\documentclass[a4paper,fleqn,usenatbib]{mnras}

\usepackage[T1]{fontenc}
\usepackage{ae,aecompl}

\usepackage{graphicx}	
\usepackage{amsmath}	
\usepackage{amssymb}	

\title[Crystalline silicates in the ISM and DEYSOs]{Absorption at 11~$\mu$m in the interstellar medium and embedded sources: evidence for crystalline silicates}

\author[C. M. Wright et al.]{
Christopher M. Wright,$^{1}$\thanks{E-mail: c.wright@adfa.edu.au (CMW)}
Tho Do Duy,$^{1,2}$
and Warrick Lawson$^{1}$
\\
$^{1}$School of Physical, Environmental and Mathematical Sciences, UNSW Canberra, PO Box 7916, Canberra BC 2610, Australia \\
$^{2}$Department of Physics, International University - Vietnam National University HCM, Block 6, Linh Trung, Thu Duc, \\
Ho Chi Minh City, Viet Nam
}

\date{Accepted XXX. Received YYY; in original form ZZZ}

\pubyear{2015}

\begin{document}
\label{firstpage}
\pagerange{\pageref{firstpage}--\pageref{lastpage}}
\maketitle

\begin{abstract}
An absorption feature is occasionally reported around 11~$\mu$m in astronomical 
spectra, including those of forming stars. Candidate carriers include 
water ice, polycyclic aromatic hydrocarbons (PAHs), silicon carbide, crystalline 
silicates or even carbonates. All are known constituents of cosmic dust in one 
or more types of environments, though not necessarily together. In this paper we 
present new ground-based 8--13~$\mu$m spectra of one evolved star, several embedded 
young stellar objects (YSOs) and a background source lying behind a large column 
of the interstellar medium (ISM) toward the Galactic Centre. Our observations, 
obtained at a spectral resolution of $\sim$~100, are compared with previous lower 
resolution data, as well as data obtained with the Infrared Space Observatory (ISO)
on these and other targets. By presenting a subset of a larger sample our aim 
is to establish the reality of the feature and subsequently speculate on its carrier. 
All evidence points toward crystalline silicate. 
For instance, the 11~$\mu$m band profile is well matched with the emissivity of 
crystalline olivine. Furthermore, the apparent association of the absorption feature 
with a sharp polarisation signature in the spectrum of two previously reported 
cases suggests a carrier with a relatively high band strength compared to amorphous 
silicates. If true, this would either set back the evolutionary stage in which silicates 
are crystallised, either to the embedded phase or even before within the ISM, or 
else the silicates ejected from the outflows of evolved stars retain some of their 
crystalline identity during their long residence in the ISM.
\end{abstract}




\section{Introduction}

The composition and evolution of cosmic dust is of great astrophysical 
interest as it from these tiny, sub-micron sized seeds that planets grow. With 
their enhanced wavelength coverage over the ground-based atmospheric windows 
at 2.9-3.4, 8-13 and 16-23~$\mu$m, the Infrared Astronomical Satellite (IRAS), 
Infrared Space Observatory (ISO) and Spitzer space telescope provided great 
impetus and progressively larger strides in the study and understanding of 
cosmic dust. This has been inclusive of ices and refractory species like 
silicates, amongst other less abundant components (e.g. \citet{molster2010mineralogy}, 
\citet{henning2010cosmic}, \citet{gibb2004interstellar}). Of particular note has 
been the 'crystalline revolution', beginning with ISO, in which routine 
detection of crystalline silicates and even the study of their specific 
mineralogies has occurred. 

Before the space-based spectrometers the existence of such crystalline silicates 
had been proposed in only a few sources. For the massive embedded YSO 
AFGL~2591 it was based on the presence of a 'shoulder' or 'inflection' around 
11~$\mu$m in its conventional absorption spectrum, along with an accompanying 
polarisation signature (\citet{aitken1988infrared}; \citet{wright1999mid}). For 
other sources it was based on a similarly placed emission feature in the spectra 
of several comets, e.g. Comet Halley (\citet{bregman1987halley}; 
\citet{campins1989identification}), and the debris disk around $\beta$ Pictoris 
(\citet{aitken1993bpic}; \citet{knacke1993silicates}). 

Through necessity these earlier identifications were typically based on the 
presence of only a single spectroscopic feature, whilst ISO and Spitzer covered 
the location of several other cosmic dust bands in the mid- and far-IR which 
could obviously strengthen identification of a candidate carrier. In so doing 
it was discovered that crystalline silicates exist around many different types 
of astrophysical sources, including dust factories (i.e. winds of evolved stars 
wherein dust condenses) and repositories (i.e. circumstellar disks around 
T Tauri and Herbig stars). The 11~$\mu$m and accompanying spectral features 
were predominantly in emission -- indicating a temperature of several hundred 
Kelvin -- such that the dust was obviously located in close proximity to the 
central star, perhaps the inner regions of the disk and/or above it within a 
disk 'atmosphere' (\citet{chiang1997rtdisk}; \citet{calvet1992rtdisk}). 

Few examples of 11~$\mu$m absorption were found, where the dust would be much 
colder, less than $\sim$ 100~K, and located in the outer disk or envelope. For 
instance, \cite{demyk1999chemical} concluded that crystalline silicates comprised 
no more than 1-2\% of the silicates in the envelopes of two massive embedded 
YSOs, AFGL7009S and IRAS19110+1045. Further, no feature was found in the ISM, 
where according to some models of cosmic dust evolution (e.g. \citet{jones2011dust}) 
it resides during the interval between its ejection from evolved stars and eventual 
deposition into a star forming region. For instance, based on the lack of an 
11~$\mu$m absorption feature \cite{kemper2005erratum} and \cite{kemper2004absence} 
placed an upper limit mass fraction of 2.2\% on crystalline silicates, with a most 
likely value of around 1\%, along the $\sim$~8~kpc path to the Galactic Centre, which 
intersects both diffuse (atomic) and dense (molecular) clouds. See also 
\cite{li2007crystallinity}, who -- using the same spectrum -- raise the upper limit 
to 3--5\% by assuming that the component in molecular clouds grows a water ice mantle, 
the broad 11--13~$\mu$m librational band of which effectively masks (or washes out) 
the narrower 11~$\mu$m crystalline silicate band. [Curiously, \cite{min2007shape} 
also used the very same spectrum to infer the presence of SiC, which has a feature 
around 11.3~$\mu$m.] 

Several scenarios have been put forward to explain the lack of a crystalline 
component in cold silicate dust. In one model the silicates condense as 
partially crystalline in the outflows of evolved stars, but are completely 
amorphised in the ISM by such processes as cosmic ray irradiation on a 
timescale as short as 70~Myr (e.g. \citet{bringa2007energetic}). Another 
instead proposes that the lifetime of dust -- against destructive processes 
like sputtering and shattering in interstellar shocks -- is only about 
$4\times10^{8}$ years, less than the $\sim$ $2\times10^{9}$ year cycling 
time between ejection and deposition (\citet{draine2009asp}). In this model 
the dust in the ISM is not stardust, but is predominantly made in the ISM, 
having re-condensed as entirely amorphous behind shock fronts. Obviously in 
both scenarios the ISM silicate dust population is amorphous, and thus so are 
the silicates eventually deposited into a molecular cloud, the gravitational 
collapse of which forms a new generation of stars. Consequently, the 
crystalline silicates seen around newly formed stars must have been annealed, 
probably within their inner disks when exposed to temperatures of 
$\sim$ 1000~K (\citet{van2004building}). They are then seen in emission.

In those cases where 11~$\mu$m absorption has been detected, either from 
ground- or space-based facilities, its identification has in many instances 
been ambiguous. See for example \cite{boogert2004spitzer} and \cite{kessler20058}. 
For instance, a potential carrier is water ice, which has a relatively strong 
and broad libration band centred between $\sim$ 12 and 13~$\mu$m for its 
crystalline and amorphous end members respectively (\citet{maldoni1998study}). 
On the basis of accompanying strong 3.1~$\mu$m water ice absorption, such an 
identification was made by \cite{soifer19814} and \cite{roche1984oh} for the 
OH/IR stars OH~231.8+4.2 and OH~32.8-0.3 respectively. For similar reasons 
\cite{demuizon1986evidence} also ascribed water ice to the feature in the IRAS 
spectra of two additional OH/IR stars, as well as the embedded YSO AFGL~4176. 
On the other hand, \cite{smith1990absorption} found no corresponding feature 
of water ice at 3.1~$\mu$m in the spectrum of the OH/IR star OH~138.0+7.3, 
and suggested instead that the 11~$\mu$m absorption could be explained by 
annealed (i.e. crystalline) silicate. 

Another potential carrier could be hydrocarbons, known to have a strong 
emission feature at 11.25~$\mu$m in the presence of ultraviolet radiation. 
In this context, \cite{bregman2000discovery} identified absorption centred 
at 11.25~$\mu$m in the embedded YSO MonR2~IRS3 with a C-H out-of-plane 
vibrational mode of PAH molecules, based on an accompanying PAH absorption 
at 3.25~$\mu$m.

More recently, with the aid of the longer wavelength coverage of ISO and/or
Spitzer, \cite{demyk2000structure} and \cite{de2014problematically} found that 
the dominant contributor of 11~$\mu$m absorption in their respective samples 
of OH/IR stars is crystalline forsterite. For a sample of protostars 
\cite{riaz20092mass} instead suggest that water ice is the dominant component. 
On the other hand, \cite{spoon2006detection} and \cite{poteet2011spitzer} were 
able to firmly identify 11.1~$\mu$m absorption with crystalline silicate -- notably 
the Mg end member forsterite -- in the Ultraluminous Infrared Galaxy (ULIRG) 
IRAS08572+3915 and the envelope of the Class~0 YSO HOPS-68 respectively. Even 
more recently, \cite{fujiyoshi2015mid} detected absorption bands of both 
crystalline olivine and pyroxene, as well as SiC, in the Subaru/COMICS 
8--13~$\mu$m spectrum of the Class~I YSO SVS13.

The review of literature described above suggests that a discrete feature 
around 11~$\mu$m is much rarer in absorption than it is in emission, especially 
in the spectra of young stars. And where such a band is inferred its 
identification is problematic, especially if only seen in isolation within 
the 8--13~$\mu$m atmospheric window. But is this really the case, or is its 
rarity instead due to insufficient signal-to-noise and/or an inappropriate 
observational approach? We have attempted to answer this question by conducting 
a mid-IR spectroscopic survey of a select sample of targets, motivated 
principally by the existence of an inflection at 11~$\mu$m in low resolution 
(R $\sim$ 40) spectra of many objects in the mid-IR polarisation atlas of 
\cite{smith2000studies}. 

In this paper we present selected ground-based results of a much larger body 
of work, which is still being worked upon. Here we include 8--13~$\mu$m spectra 
of the cold silicate dust in the envelopes or disks of several massive embedded 
YSOs as well as the path to the Galactic Centre. As a 'control', or 'template', 
we include the OH/IR star and dust factory AFGL~2403, confirmed to have 
crystalline silicates by \cite{de2014problematically}. These data are supported
and complemented by ISO observations of the same and other targets from 10 to 
45~$\mu$m, taken with the Short Wavelength Spectrometer (SWS). Our study is the 
first dedicated and systematic search for, plus statistical investigation of, 
the 11~$\mu$m absorption feature in these source types. For this paper we 
concentrate on the main phenomenological findings with some modelling of specific 
cases. We will present a full description of the sample and a complete discussion 
of the results and associated modelling in a forthcoming paper (Do Duy et al., 
in preparation). 

\section{Observations}

The 8--13~$\mu$m spectra were obtained from 21/08/2005 to 27/01/2007 using the 
facility T-ReCS (\citet{telesco1998gatircam}) and Michelle (\citet{glasse1997michelle}) 
mid-infrared long-slit spectrometers at the Gemini-S and -N telescopes 
respectively, under Gemini programmes GS-2006B-Q-81 and GN-2005B-Q-83. The 
slit width was 0.7 arcsec with T-ReCS and 0.4 arcsec with Michelle, providing 
a spectral resolving power of $\sim$~100. Standard chopping and nodding was 
implemented, with the throw chosen on the basis of the source extension. The 
data was reduced using in-house IDL codes, with the spectrum extracted by
summing the pixels across the spatial profile. Whilst not an ideal technique, for 
these bright sources there is little loss in S/N compared to optimised extraction 
methods, or Gaussian and Moffat function fits which were also tested. A standard 
star well-matched in airmass was used to correct for telluric features and provide 
the absolute flux calibration. Wavelength calibration was performed using 
telluric features in both the target and standard star spectra, and/or features 
in the filter transmission profiles. 

Complementary ISO and low resolution data was taken from the ISO Highly 
Processed Data Product archive and \cite{smith2000studies} respectively. 
Table~\ref{TAB:Observations} provides some specific observational details.
The number in brackets after the SWS01 designation refers to the
speed with which the 2.4--45.2~$\mu$m spectrum was taken, which in turn
determines the spectral resolution and signal-to-noise. Speed 1 is fastest 
and least sensitive and speed 4 is the slowest and most sensitive 
(\citet{leech2003iso}). To produce the ISO spectra we have taken the Frieswijk 
de-fringed highly processed data products for the SWS01 Astronomical Observing
Template (AOT), sigma-clipped them about a chosen S/N ratio, and then binned or 
smoothed them in wavelength bins appropriate for the respective SWS01 speeds. 
For SWS06 AOTs we have used the latest pipeline Auto-Analysis Result (AAR) 
product, sigma-clipped and then binned at a resolution more coarse than the 
fringe period.

\begin{table*}
\caption{Table of new Gemini observations, plus supporting ground-based and ISO data}
\begin{tabular}{lllllll}
\hline
 Object  &   Date      & Instrument &  Chop/Nod        & Standard  & Airmass & ISO ID   \\ 
         &             &            &  throw           & star      & Src/Std &          \\ \hline
 AFGL~2403 & 28 Sep 2006 & T-ReCS   &  8$''$ N-S       &  $\gamma$ Aql & 1.62/1.42 &     \\
           &             & SWS01(1) &                  &               & & 32000603 \\ 
           &             & SWS01(1) &                  &               & & 50200604 \\ 
 AFGL~2789 & 05 Sep 2005 & Michelle &  8$''$ N-S       &  $\eta$ Peg   & 1.30/1.22  &    \\ 
           &             & SWS01(2) &                  &               & & 26301850 \\ 
 AFGL~2136 & 15 Oct 2006 & T-ReCS   &  15$''$ 31.0$^{\circ}$ & $\lambda$ Sgr & 1.42/1.49 &   \\
           &            & SWS01(3) &                   &              & & 33000222 \\ 
           &            & SWS06    &                   &              & & 31101023 \\
 W3~IRS5~NE  & 24 Sep 2005 & Michelle &  8$''$ 36.4$^{\circ}$  & BS168    & 1.35/1.28 &  \\ 
           &             & SWS01(3) &                  &              & & 42701302 \\ 
 SgrA IRS3 & 21 Aug 2005 & Michelle &  15$''$ N-S            & $\lambda$ Sgr & 1.52/1.42 &    \\ 
 \hline
 \multicolumn{7}{|c|}{Other supporting ground-based and ISO data} \\
 AFGL~2591$^a$ & 26 Jun 1986 & UCLS-lo & 25$''$ N-S    &  $\beta$ Peg   & &  \\
 AFGL~2591$^b$ & 29-30 Sep 1987 & UCLS-hi & 24$''$ E-W &  $\beta$ Peg   & &  \\
               &             & SWS01(1) &              &               & & 02800433 \\ 
               &             & SWS01(3) &              &               & & 35700734 \\ 
 AFGL~4176$^b$ & 21 Jan 1989 & UCLS-hi &  24$''$       &  $\alpha$ Cen  & &  \\
 AFGL~4176$^b$ & 18 May 1992 & UCLS-hi &  20$''$ N-S   &  $\alpha$ Cen  & &   \\
               &             & SWS01(1) &              &               & & 11701311 \\ 
               &             & SWS06    &              &               & & 30601344 \\ 
 IRAS13481$^c$ & 19 Jan 2006 & TIMMI2  &  10$''$ N-S   &  $\lambda$ Vel  & 1.35/1.04 &          \\ 
 \hline
  \multicolumn{7}{|c|}{Other supporting ISO data} \\
  IRAS19110  &           & SWS01(2) &                 &               & & 49900902 \\
  W28~A2     &           & SWS01(1) &                 &               & & 09901027 \\
  Sgr~A$^{*}$ &          & SWS01(4) &                 &               & & 09401801 \\
             &           & SWS01(3) &                 &               & & 13600935 \\
  Sgr~A SW   &           & SWS01(4) &                 &               & & 09401905 \\
  Sgr~A NE   &           & SWS01(4) &                 &               & & 09500203 \\
  GC Pistol  &           & SWS01(4) &                 &               & & 84101302 \\
  Orion IRc2 &           & SWS01(1) &                 &               & & 68901006 \\
  Orion Pk1  &           & SWS01(4) &                 &               & & 68701515 \\
  Orion Pk2  &           & SWS01(4) &                 &               & & 83301701 \\
  Orion Bar  &           & SWS01(2) &                 &               & & 69501409 \\
 OH26.5+0.6  &           & SWS01(2) &                 &               & & 33000525 \\
 OH32.8-0.3  &           & SWS01(2) &                 &               & & 32001560 \\
 AFGL~230    &           & SWS01(2) &                 &               & & 78800604 \\
 HD100546    &           & SWS01(4) &                 &               & & 27601036 \\
 HD45677     &           & SWS01(4) &                 &               & & 71101992 \\
 HD44179     &           & SWS01(4) &                 &               & & 70201801 \\
 IRAS02575   &           & SWS01(2) &                 &               & & 86300968 \\
 IRAS10589   &           & SWS01(2) &                 &               & & 26800760 \\
 S106        &           & SWS01(2) &                 &               & & 33504295 \\
\hline
\end{tabular}
 \begin{list}{}{}
  \item $^a$ Previously published in \protect\cite{aitken1988infrared}; 
        $^b$ Previously presented in \protect\cite{wright1994thesis}; 
        $^c$ Previously published in \protect\cite{wright2008mid}
 \end{list}
\label{TAB:Observations}
\end{table*}

\section{Results}

\subsection{Spectra}

\begin{figure*}
\includegraphics[scale=0.67]{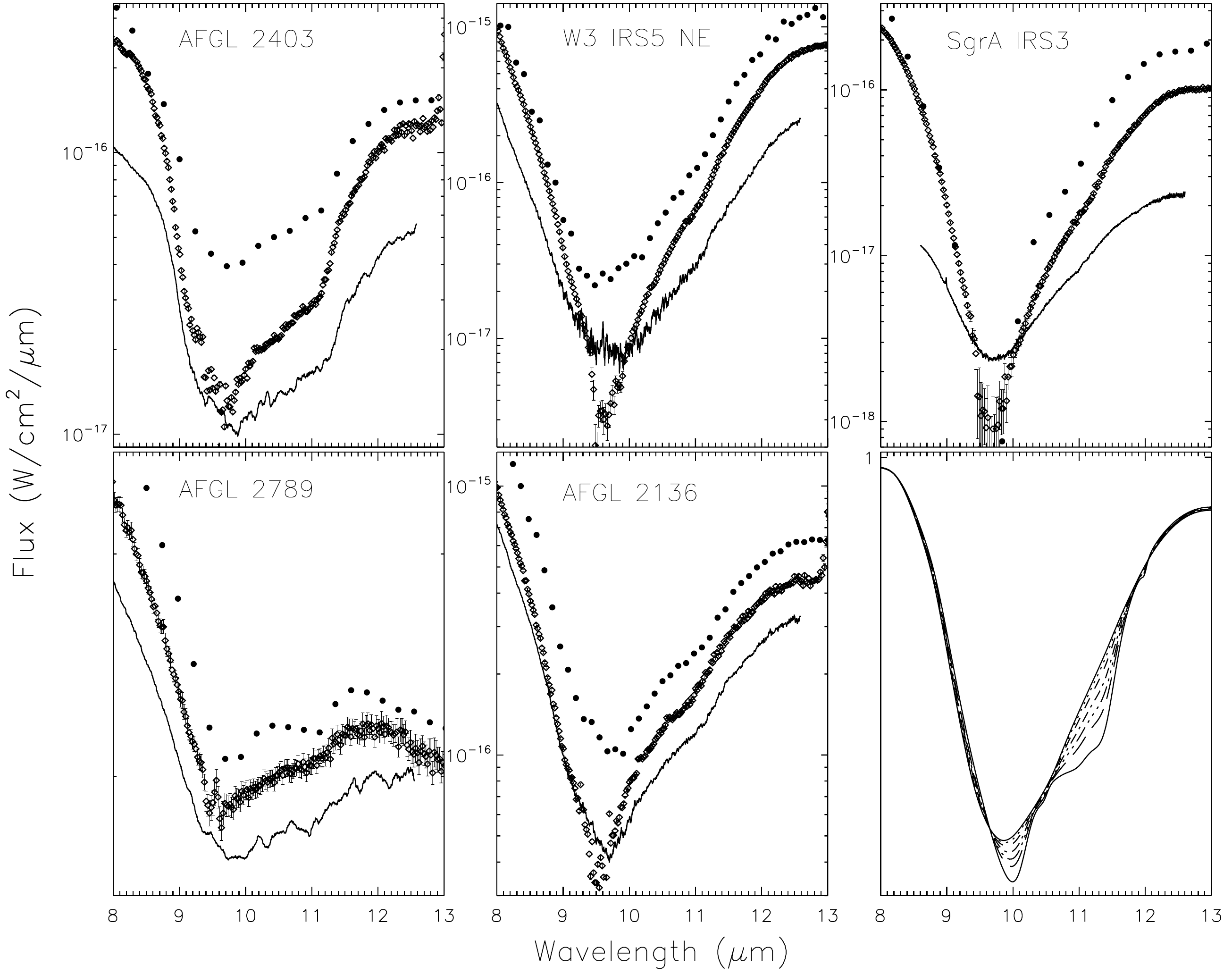}
\caption{Gemini 8--13~$\mu$m spectra of the five targets listed in 
Table~\ref{TAB:Observations}. The W3~IRS5 spectrum is of the
slightly brighter NE component of this close double, also called MIR1 in 
\protect\cite{vandertak2005subarc}. For comparison lower spectral resolution data 
(solid circles) are also provided, obtained with the UCL Spectrometer (UCLS) 
and previously presented in the spectral atlas of \protect\cite{smith2000studies}, 
scaled by factors of 0.4, 1.3, 1.0, 0.9 and 1.5 for AFGL~2403, W3~IRS5, 
SgrA~IRS3, AFGL~2789 and AFGL~2136 respectively. Also shown is the higher 
spectral resolution data (solid lines) from ISO, being the Highly Processed 
Data Products from the ISO data archive, scaled by 0.20, 0.20, 0.01, 0.9, 0.8 
respectively for AFGL~2403, W3~IRS5, SgrA~IRS3, AFGL~2789 and AFGL~2136. 
The final panel instead shows a series of EMT models for amorphous olivine with 
increasing crystalline olivine content, using a CDE. See text for details.}
\label{FIG-Spectra}
\end{figure*}

Figure~\ref{FIG-Spectra} shows the reduced Gemini 8--13~$\mu$m spectra of our 
targets, including the control source AFGL~2403, three YSOs and SgrA~IRS3. Along 
with the well known deep amorphous silicate absorption centred around 9.7~$\mu$m 
there is also a shallow feature around 11~$\mu$m, which is relatively deeper in 
AFGL~2403. The inflection seen at R $\simeq$ 40 in the UCLS spectra presented in
\cite{smith2000studies}, shown also in Figure~\ref{FIG-Spectra} as solid circles, 
is resolved here into a bona fide absorption band. For comparison we also show 
the ISO spectra of each object, noting however that they may contain relatively 
narrow artefacts around 9.35, 10.1 and 11.05 $\mu$m (with FWHM of 0.3, 0.1 and 
0.1~$\mu$m respectively) introduced by imperfect correction for the Relative 
Spectral Response Function (RSRF) of the ISO--SWS. See \citet{leech2003iso}. 

The Gemini spectrum of AFGL~2789 (V645 Cyg) is consistent with those 
previously published by \cite{hanner19988} and \cite{bowey2003galactic} at 
lower spectral resolution, inclusive of the abrupt 'jump' in flux around 
11~$\mu$m. Also, the Gemini spectrum of AFGL~2136 is consistent with the 
similar resolution 8.2--11.0~$\mu$m segment presented by 
\cite{skinner1992methanol}, inclusive of the rather sharp minimum around 
9.7~$\mu$m. 

For the relatively isolated and point-like YSOs AFGL~2136 and AFGL~2789 
(\citet{monnier2009mid}) all three of their spectra are in reasonable 
agreement in both level and shape. For the OH/IR star AFGL~2403 the shapes 
are consistent but the flux levels are notably different for all three 
spectra, which is possibly due to intrinsic variability for this type of 
source (\citet{jiminez2006ohir}, \citet{smith2003ohir}, \citet{glass2001ohir}, 
\citet{herman1985ohir}). W3~IRS5 is a mid-IR double source 
(\citet{vandertak2005subarc}), separated by about 1.1 arcsec along a 
position angle of $\sim$ 37$^{\circ}$ and embedded within more diffuse 
emission. The Gemini-Michelle spectrum presented here is of the slightly 
brighter NE component, which \citet{vandertak2005subarc} call MIR1, 
whilst the UCLS and ISO observations included both sources as well as 
the extended emission. This probably explains the slightly different 
fluxes, increasing from the Gemini to UCLS to ISO spectra in accordance 
with the increasing beam size of the respective observations. It probably
also at least partly accounts for the apparent difference in the silicate depth 
between the Gemini and other spectra.

Perhaps the best demonstration of the advantages of 8--13 $\mu$m narrow slit 
absorption spectroscopy over broad beam observations is the Galactic Centre data set. 
Clearly there is a very large difference in the depth of the silicate feature between
the Gemini and ISO data sets. There were two observations available in the ISO archive, 
one centred on IRS7 and the other on Sgr~A$^{*}$, which are very consistent with each 
other (see Appendix A). They have been coadded for Figure~\ref{FIG-Spectra}. The 
Sgr~A$^{*}$ spectrum was first presented by \cite{lutz1996sws} and subsequently by 
\cite{kemper2004absence}, who -- along with \cite{min2007shape} and 
\cite{li2007crystallinity} -- concluded from its seemingly smooth and featureless 
profile was entirely due to amorphous silicate, and thereby placed limits on other 
possible constituents. 

As well as varying amounts of extinction across the centre of the Galaxy 
(e.g. \citet{schodel2010peering}; \citet{scoville2003hubble}), even on spatial 
scales smaller than the $14'' \times 20''$ ISO beam, within that beam there 
are multiple mid-IR sources as well as extended emission comprising the N-S arm 
and E-W bar of the mini-spiral. Obviously such a complicated source structure 
will impact on the observed spectrum, e.g. partially 'filling in' the silicate 
absorption feature. Our Gemini observations are instead much closer to the ideal 
'pencil beam' absorption experiment, and thus well suited to revealing trace 
mineralogical structure. 

Another contributor to the aforementioned silicate depth difference, and the
probably related narrowness of the minimum of W3~IRS5 as well as AFGL~2136,
is the presence of NH$_{3}$ and/or CH$_{3}$OH ices at 9.0 and 9.7 $\mu$m respectively.
This is almost certainly the case for methanol for AFGL~2136, based on the work 
of \cite{skinner1992methanol} and \cite{gibb2004interstellar}. Neither ice material has 
been identified in 3--10~$\mu$m ISO spectroscopy of W3~IRS5, e.g. \cite{dartois2001search}, 
\cite{gibb2004interstellar} and \cite{gibb2001searching}, or 3~$\mu$m ground-based 
spectroscopy of \cite{brooke1996study}. But our Gemini spectra of both the NE and 
especially SW components (to be presented in Do Duy et al., in preparation) have a very 
similar shape between 9 and 10~$\mu$m to those of W33A and NGC7538~IRS9, two 
ice-rich deeply embedded YSOs with confirmed detections of NH$_3$ and CH$_3$OH 
(\citet{gibb2000w33a}, \citet{lacy1998ammonia}).

Such ices would be unlikely in the case of AFGL~2403, whilst for SgrA~IRS3 their
contribution would be very small, if at all existent (based on the relatively small optical
depth of the 3~$\mu$m water ice feature toward the Galactic Centre, compared to 
YSOs, to be discussed in a following section). But we note that their spectra in 
Figure~\ref{FIG-Spectra} also show evidence for either a discrete feature at 
9.7--9.8~$\mu$m (AFGL~2403), or again a narrow minimum of the 8--13~$\mu$m 
absorption band (SgrA~IRS3). The feature in AFGL~2403, as well as another around
9.3~$\mu$m (probably from crystalline enstatite), are more or less replicated in the 
ISO spectrum so are likely to at least be partially real. For SgrA~IRS3 the silicate
depth is in good agreement with that of \cite{pott2008enigma}, obtained at lower 
spectral resolution (R $\sim$ 30) but higher spatial resolution (mid-IR interferometry).

Unfortunately there are also potential artefacts that could produce a very
deep and/or narrow minimum of the silicate band. One is that telluric ozone at 
9.6~$\mu$m can make interpretation in this part of the spectrum problematic,
such that some authors choose not to even show this segment of their data. But as 
seen in Table~\ref{TAB:Observations} our target and standard star airmasses are 
well matched. For example, there is no residual water vapour features at 11.7~$\mu$m 
or 12.5~$\mu$m in Figure~\ref{FIG-Spectra}, and the division of the standard spectrum 
into the source spectrum has not produced large 'up-down'-type artefacts that could 
occur if the two spectra were not well aligned. Thus we do not expect a significant 
contribution from poor ozone correction to the apparent depth of the silicate feature 
in our spectra. As some evidence of this the spectrum of a second position -- which 
we call IRSX and will discuss in a following section -- obtained from the same 
Gemini-Michelle observation as SgrA~IRS3 shows no anomalous structure around 
the ozone wavelength. This strongly suggests that a reliable correction has been 
achieved (see also Appendix A).

The other artefact is difficult to quantify. As noted by \cite{roche2015ngc4418} 
and \cite{roche2006circinus} the T-ReCS and Michelle detectors suffer crosstalk 
between their different readout channels, especially prominent for bright sources 
(\citet{sako2003detector}). Whilst obvious in imaging observations it is less so for
spectroscopy, but can potentially diminish the signal along the spectral direction,
and so perturb the level and shape of the silicate minimum. This will be discussed in 
more detail in our following paper with a larger sample (Do Duy et al., in preparation).

\subsubsection{Models} 

Pre-empting the discussion to follow later, the final panel in 
Figure~\ref{FIG-Spectra} shows a series of models containing an increasing 
quantity of crystalline olivine inclusions within an amorphous silicate 
matrix. Effective Medium Theory (EMT) has been used, wherein an 'average' 
or effective dielectic function -- equivalently and otherwise referred to 
here as refractive indices or optical constants -- can be derived from the
optical constants of two or more constituent materials. See \cite{bohren1983light} 
for general details. 

For Figure~\ref{FIG-Spectra} we have used the Maxwell-Garnett (MG) mixing rule, 
which requires defining so-called matrix (or host) and inclusion materials, 
here being amorphous and crystalline silicates respectively, as well as the 
volume fraction occupied by the inclusions. Although the generalised MG 
formula can accommodate spheroidally shaped inclusions this introduces an 
extra free parameter which is unconstrained by any observations of which we 
know. Thus the version we use assumes spherical inclusions.

Different optical constants for the amorphous silicate have been tested, 
including 'astronomical silicate' of \cite{draine2003scattering} and olivine 
from \cite{dorschner1995steps}. The olivine species with equal iron and 
magnesium content, i.e. MgFeSiO$_{4}$, from Dorschner et al. is used for the 
models in Figures~\ref{FIG-Spectra}, \ref{FIG-Polyfits} and \ref{FIG-NewDepths}. 
This has also been used by different authors in their own studies of cosmic 
dust, e.g. toward the Galactic Centre by \cite{kemper2004absence} and 
\cite{min2007shape}. 

Similarly, various crystalline silicate optical constants have been trialled, 
such as those of crystalline olivine from \cite{mukai1990optical}, crystalline 
Mg$_{1.9}$Fe$_{0.1}$SiO$_{4}$ from \cite{fabian2001steps} and crystalline 
forsterite from \cite{sogawa2006infrared} and \cite{suto2006low}. Those of 
Mukai \& Koike are used for Figures~\ref{FIG-Spectra}, \ref{FIG-Polyfits} 
and \ref{FIG-NewDepths}, but our results are qualitatively (though not 
necessarily quantitatively) similar irrespective of the specific combination 
of optical constants used (Do Duy et al., in preparation). Models with a 
volume fraction of crystalline olivine of f=0.01, 0.025, 0.05, 0.075, 0.10, 
0.15 and 0.20 are shown in the final panel in Figure~\ref{FIG-Spectra}.

Absorption cross sections C$_{\rm abs}$ are calculated in the Rayleigh 
approximation, i.e. the grain size is much smaller than the wavelength. This 
is almost certainly a valid assumption in our case even for grain sizes up 
to about a micron (\citet{somsikov1999rayleigh}) in size, let alone for the 
0.1~$\mu$m grains typically inferred for the ISM (\citet{mathis1977mrn}).
Given the Rayleigh approximation is valid for the entire grain then of course
it is also valid for the EMT inclusions. 

Calculations assume a single spheroidal shape, e.g. oblate with a principal axis 
ratio of 2:1, or a continuous distribution of ellipsoids (CDE, in our case actually
spheroids). The latter is used for Figure~\ref{FIG-Spectra}, comprising both oblate 
and prolate particles, from an axis ratio of 1:1 (i.e. a sphere) up to 5:1, all 
with equal probability. What is actually plotted in Figure~\ref{FIG-Spectra} however 
is not the absorption cross section, but instead the quantity exp(--C$_{\rm abs}$) 
which 'mimics' an absorption spectrum. 

We have ran tests for different types of CDEs, e.g. with Gaussian weights and
different maximum axis ratio, and oriented spheroids as well as randomly oriented 
ellipsoids (as given in \cite{min2003shape}). Results are qualitatively similar 
(Do Duy, in preparation), but the single shape or oriented spheroids are potentially 
more realistic. This is because all of the targets presented here (apart from 
AFGL~2403) show mid-IR polarisation (\citet{smith2000studies}). This is a certain 
sign that at least some of the dust grains along the path to each object are aligned, 
probably with their short axes along the ambient magnetic field direction 
(\citet{lazarian2007tracing}). Obviously this also argues against using any kind 
of model which assumes spherical dust grains.

\subsection{Extracting the 11~$\mu$m feature and its optical depth}

At least two approaches can be made to extract the 11~$\mu$m feature and its 
optical depth. For instance, the amorphous silicate profile can firstly be 
extracted by fitting a Planck function B($\lambda$,T) to the 8 and 13~$\mu$m 
points to determine a colour temperature T$_{8/13}$. Subsequently the optical 
depth $\tau_{\lambda}$ is calculated from 
$F_{\rm obs} = B(\lambda,T_{8/13})\times$exp($-\tau_{\lambda})$, 
where $F_{\rm obs}$ is the observed flux. This is not an entirely physical 
approach as it assumes that the dust has zero emissivity at 8 and 13~$\mu$m. 
Although these wavelengths are certainly near or even at the edges of the 
amorphous silicate Si--O stretching band, cosmic dust still retains some 
emissivity there, as beautifully demonstrated in Figure 10 of \cite{fritz2011line}. 
This shortcoming can be alleviated by scaling the 
fluxes by a factor equal to an assumed emissivity at these wavelengths, e.g. 
that of the Trapezium region in Orion. This has historically been used to 
model in a straightforward way the 8--13~$\mu$m spectra of objects within 
molecular clouds and star forming regions (e.g. \citet{gillett19758}, 
\citet{hanner19988}, \citet{smith2000studies}). Thereafter a new T$_{8/13}$ 
and amorphous silicate profile can be determined.

\begin{figure*}
\includegraphics[scale=0.67]{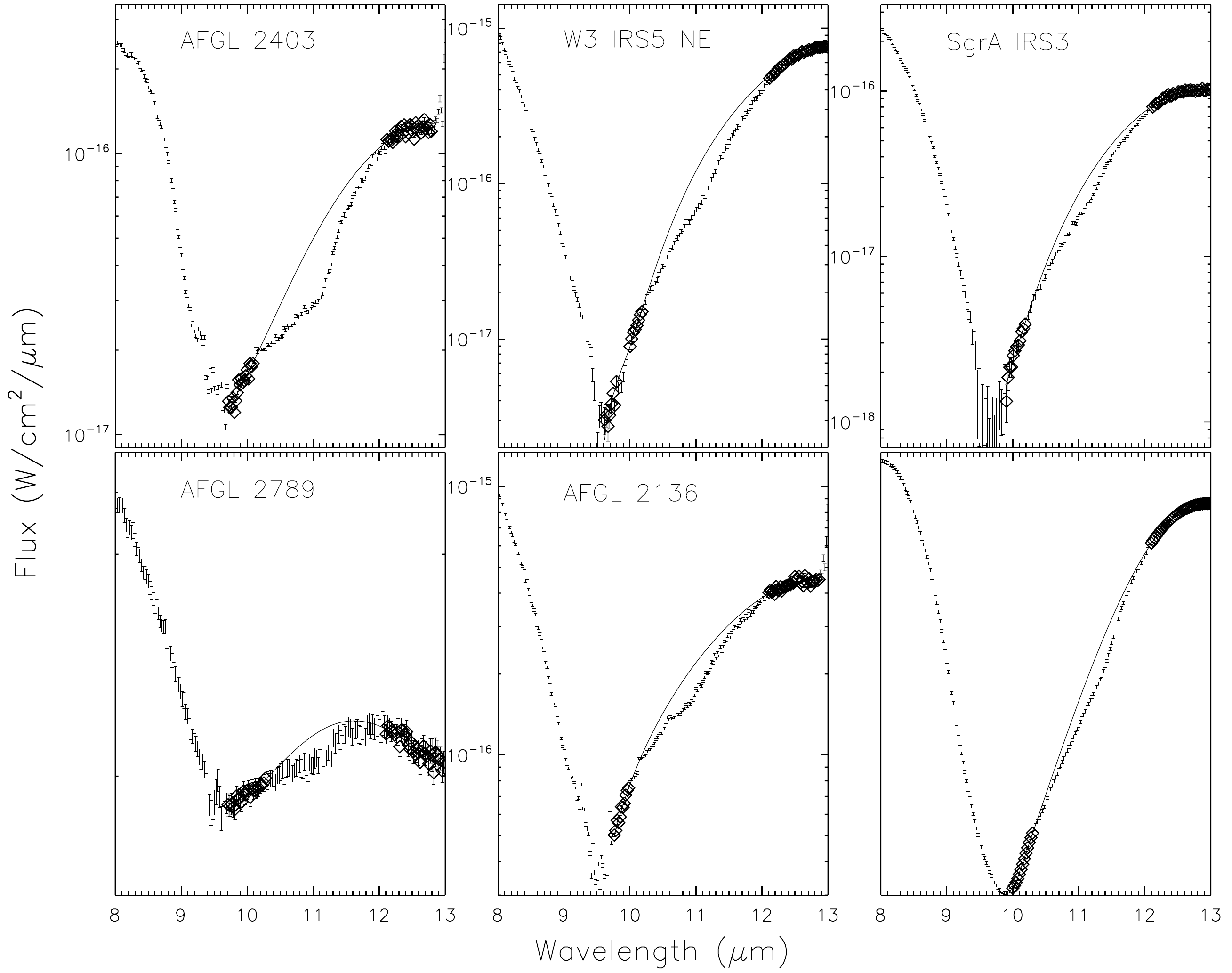}
\caption{Polynomial fits to 10-13 $\mu$m portion of the Gemini spectra in 
Figure~\ref{FIG-Spectra}, as well as a representative EMT model, in this 
case oblate 2:1 with a volume fraction of crystalline olivine of 0.05. The 
W3~IRS5 spectrum is of the slightly brighter NE component of this close 
double, also called MIR1 in \protect\cite{vandertak2005subarc}.}
\label{FIG-Polyfits}
\end{figure*}

However, this does not help with another assumption implicit in this approach, 
namely that the warm dust emitting behind the absorbing column is optically thick, 
and thus can be approximated as a black-body. This will be true in many cases 
(e.g. \citet{smith2000studies}) but will not always be true, in which case the 
underlying emitting dust would have a silicate emission feature and the 
extracted optical depth under-estimated (although the relative magnitude of 
the under-estimate will decrease with increasing absorption depth). A powerful
demonstration of how different a real continuum can be to a polynomial or even
Planck-like contunuum connected between observed fluxes can be seen in Figure~10 
of \cite{fritz2011line}. They determined the 1--19~$\mu$m extinction to the 
Galactic Centre from hydrogen recombination lines, and subsequently deduced
the unextinguished (overlying) spectrum. Nowhere do the unextinguished and 
observed spectra equal each other. 

Even so, this approach has the advantage of simplicity and consistency, and is 
commonly used (e.g. \citet{de2014problematically} and \citet{de2010determining}, 
but who instead used a linear interpolation across 8 to 13~$\mu$m rather than 
a black-body fit). After extracting the 9.7~$\mu$m feature the 11~$\mu$m absorption 
profile can then be extracted by fitting a low order polynomial from around 
10~$\mu$m to 12--13~$\mu$m, masking out the data in between these wavelengths. 
Optical depths calculated in this way, and especially the relation between the 
9.7 and 11 $\mu$m depths for the entire sample, will be presented in Do Duy 
et al. (in preparation). 

In this paper however we use a simpler approach which is less susceptible to the 
above-mentioned assumptions, but provides no information on the amorphous silicate 
band. In this approach a polynomial is fit to the observed spectrum between the 
ranges of about 9.8--10.3~$\mu$m and 12.1--13~$\mu$m, the precise ranges being 
dependent on the data quality (e.g. signal-to-noise and/or other instrumental 
or telluric artefacts). These ranges form a 'local' or 'quasi' continuum and are 
chosen to be short enough to be as free as possible from potential (strong) cosmic 
dust spectral features but long enough to adequately constrain the polynomial fit. 
We recognise that real information can be lost (or perhaps even false information 
injected) with any method of removing a continuum, as cautioned by 
\cite{jones2014physical}, which is why we perform the same steps on our model.  

Polynomial fits are shown in Figure~\ref{FIG-Polyfits} for the same five targets 
as in Figure~\ref{FIG-Spectra}. A sample model treated in precisely the same way, 
in this case for a crystalline olivine volume fraction of 0.05, is shown in the 
last panel. We note here that broadly equivalent approaches were used by 
\cite{poteet2011spitzer} and \cite{spoon2006detection} to extract their 11~$\mu$m 
absorption features.

\begin{figure*}
\includegraphics[scale=0.75]{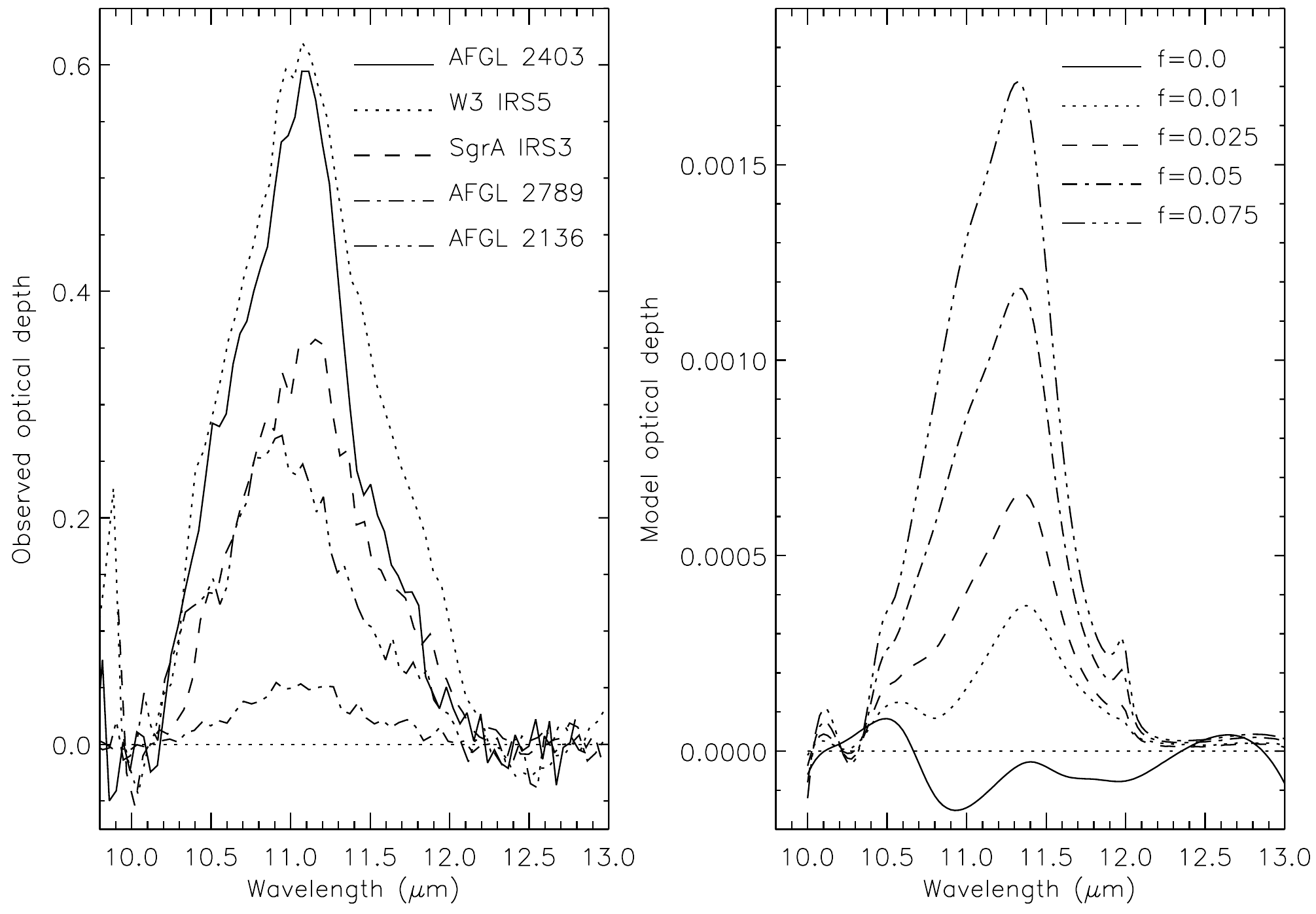}
\caption{In the left and right hand panels are shown the optical depth 
profiles around 11~$\mu$m extracted from the Gemini observations of 
Figure~\ref{FIG-Spectra}, and EMT models of oblate 2:1 grains with varying
volume fraction of crystalline olivine inclusions. The same technique has 
been used for both the observations and models. The observations
have been averaged in two pixel wide bins for plotting purposes.}
\label{FIG-NewDepths}
\end{figure*}

The 11~$\mu$m feature profile, and its optical depth $\tau$, is subsequently 
calculated by extrapolating the polynomial across the interval and then 
deriving $\tau$ using a similar equation to that above, in this case 
$F_{\rm obs} = F_{\rm cont} \times$ exp($-\tau_{\lambda})$, 
where $F_{\rm cont}$ is the local continuum given by the polynomial. The left 
hand panel of Figure~\ref{FIG-NewDepths} shows for the same five sources in 
Figs.~\ref{FIG-Spectra} and \ref{FIG-Polyfits} the 11~$\mu$m feature extracted 
in this way, whilst the right hand panel shows the model treated in precisely 
the same fashion for volume fractions of crystalline silicate of 0.0, 0.01, 
0.025, 0.05 and 0.075. That no 11~$\mu$m feature is 'recovered' for the 
purely amorphous silicate lends credibility to the approach.  

\section{Discussion}

\subsection{Central wavelength and profile of the 11~$\mu$m feature}

The central wavelength of the 11~$\mu$m absorption feature is 
11.10$\pm$0.10~$\mu$m for all objects. Whilst that for AFGL~2136 appears to 
be below 11~$\mu$m in Figure~\ref{FIG-NewDepths} this is likely to be an 
artefact introduced by noise and/or the defringeing process necessary for 
some T-ReCS spectra. The corresponding feature extracted from its ISO spectrum 
in Figure~\ref{FIG-Spectra} is fully consistent with being centred at 
11.1~$\mu$m. Such a central wavelength was also found for the features 
discovered by \cite{poteet2011spitzer} and \cite{spoon2006detection} in 
a Class~0 YSO and a ULIRG respectively.

\begin{figure}
\includegraphics[scale=0.445]{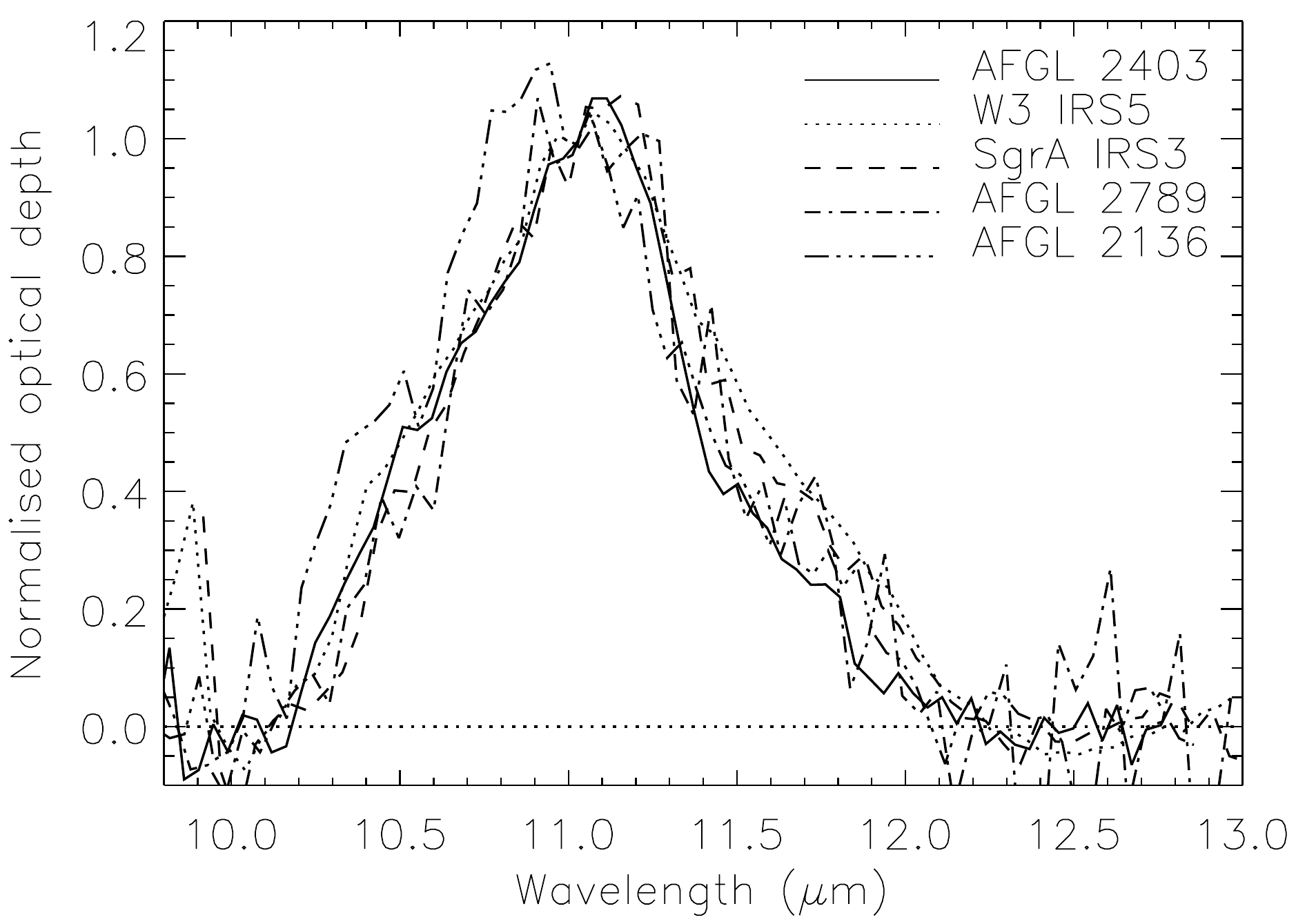}
\caption{Normalised profiles of the 11.1~$\mu$m absorption feature extracted from 
the Gemini spectra. Each profile has been divided by a peak value given by the mean 
between 10.9 and 11.2~$\mu$m. The observations have been averaged in two 
pixel wide bins for plotting purposes.}
\label{FIG-NormTau}
\end{figure}

Figure~\ref{FIG-NormTau} displays all the features again, but this time 
normalised to their respective peaks. Like the central wavelength the profile 
is also remarkably similar for each source, noting that they represent a 
range of different environments from a dust factory (AFGL~2403) to the ISM 
(SgrA~IRS3) to dense molecular clouds or even circumstellar envelopes/disks 
(W3~IRS5, AFGL~2136). The profile is not symmetric about the peak, dropping 
essentially monotonically on the short wavelength side, whilst on the long 
wavelength side the drop is much steeper to around 11.5~$\mu$m at which 
point it becomes more gradual. 

That the profile is so similar for sources from low (AFGL~2789) to 
high (W3~IRS5) extinctions strongly suggests that our technique to extract the 
11~$\mu$m band is not influenced by potential crosstalk of the T-ReCS and Michelle
detectors. Indeed, as seen in Figure~\ref{FIG-Polyfits} we have not used the deepest
part of the silicate band -- where such crosstalk might be expected to be most 
severe -- for the polynomial fit for SgrA~IRS3 and AFGL~2136.

\subsection{Possible carriers of the 11.1~$\mu$m absorption}

Several potential candidates exist for the carrier of the 11.1~$\mu$m absorption 
feature reported here, including hydrocarbons, water ice, silicon carbide (SiC), 
carbonates and crystalline silicates. All have been identified as components of 
cosmic dust in one or more types of environments through astronomical spectra 
and/or as pre-solar inclusions within meteorites or interplanetary dust 
particles (IDPs), e.g. \cite{boogert2015observations}, \cite{zinner2013laboratory}, 
\cite{bradley2010astromineralogy} and \cite{draine2003interstellar}. Considering
all of the above candidate species, we present arguments below which we believe 
strongly supports a crystalline silicate identification. If nothing else, the 
data itself, plus modelling and other plausibility arguments, are more consistent 
with crystalline silicate than any other candidate. 

To assist with the discussion below we list in Table~\ref{TAB:OpDep} for each 
target the optical depths at various wavelengths for which a discrete spectral 
feature has been detected. We have included four other sources in the table, 
namely a second Galactic Centre position plus the deeply embedded YSOs (DEYSO) 
AFGL~2591, AFGL~4176 and IRAS13481-6124. For convenience we call the 
Galactic Centre source SgrA IRSX, the 8--13~$\mu$m spectrum of which was 
obtained from the same Gemini-Michelle observation as SgrA~IRS3. Its position 
is several arcsec south of IRS3, within the E-W bar of the SgrA mini-spiral. The 
three DEYSOs have previously been identified to have an 11~$\mu$m absorption 
band by \cite{aitken1988infrared}, \cite{wright1994thesis} and \cite{wright2008mid} 
respectively. These objects will be discussed more fully in following subsections (see 
for instance Figures~\ref{ISO-Spectra_10-12um}, \ref{FIG-AFGL2591} and 
\ref{FIG-IRAS13481}).

\begin{table*}
\scriptsize
\caption{Optical depths, $\tau_\lambda$}
\begin{tabular}{llllllllll}
\hline
 Object    &           & \multicolumn{2}{c}{H$_2$O ice} & \multicolumn{4}{c}{Hydrocarbons}  & Unknown & Silicate \\
           &   11.1    &    3.0         &     6.0       &   3.4     & 3.47       &  6.2     &  7.25    & 6.85    &  9.7     \\ \hline 
 AFGL~2403 & 0.60/0.55  &   nd        &  nd, <0.02      &   nd, <0.05   &  nd, <0.05 &  $\leq$0.04 & nd, <0.005 & 0.045@6.6~$\mu$m   & 2.7-4.4 \\ 
 AFGL~2789 & 0.05/0.04  &  nd, <0.05  &  $\leq$0.02  &  nd, <0.005 & nd   &   nd, <0.01 &  $\leq$0.005  & nd  & {\bf 0.7}-2.3 \\ 
           &            &             &              &             &  $\leq$0.02@3.53~$\mu$m &         &               &  0.035@6.6~$\mu$m  &               \\ 
 AFGL~2136 & 0.27/0.28  &   2.72-3.60    & 0.23-0.30    & $\leq$0.12 &  0.10-0.14 & 0.03-0.06 &  0.04    & 0.18-0.27   & {\bf 3.5}-5.1 \\ 
 W3~IRS5   & 0.60/$\geq$0.45  &   2.78-3.48    & 0.26-0.33    & nd, <0.03  & 0.133-0.14 & <0.02     &  0.04    & 0.22-0.28   & {\bf 5.6}-7.4 \\ 
 SgrA~IRS3 & 0.35       & $\leq$0.3-0.63 &     --       & 0.19-0.31 &  0.12-0.20 & 0.18      &   --     & 0.16        & $\geq$7 \\
 SgrA~IRSX &  0.15/0.20 &   0.50         &  0.10@6.1~$\mu$m  & 0.20      & 0.05 & 0.05      &  0.03    & 0.05        & 3.6    \\     
 AFGL~2591 & 0.27       &   0.69-0.92    &    0.12      & $\leq$0.03 &  0.045     & <0.02    & $\leq$0.03 & 0.04-0.17 & {\bf 2.8}-4.4 \\ 
 AFGL~4176 & 0.22/0.20  &   0.3-0.5      &    0.03      &  0.05     &  0.05      &  <0.02    & $\leq$0.01 & 0.02@6.7$\mu$m      & {\bf 3.1}-4.8 \\
 IRAS13481 & 0.10       &   --           &   --         &  --       &  --        &  --       &   --     &     --      &  $\geq$1      \\ 
\hline
\end{tabular}
 \begin{list}{}{}
  \item \noindent Note 'nd' means 'not detected', but a 1$\sigma$ upper limit may 
        be given.  
        The two values x/y for $\tau_{11.1}$ refer to those determined from the 
  			Gemini and ISO spectra respectively. In the case of W3~IRS5 the
			ISO value is a lower limit since 11.25~$\mu$m PAH emission perturbs
			the extracted profile. \\ 
        Optical depths at 9.7~$\mu$m are mainly taken from
        \protect\cite{wright1994thesis} and \protect\cite{smith2000studies}, 
        with the two values being appropriate for optically thick (i.e. 
        featureless blackbody-like) and optically thin underlying emission. 
        The value in bold face is the preferred figure based on the $\chi^{2}$ 
        of the fit. Otherwise, for SgrA $\tau_{9.7}$ is from 
        \protect\cite{roche1985investigation} and for IRAS13481-6124 
        $\tau_{9.7}$ is from Do Duy et al. (in preparation). \\
  			References for optical depths of the other species are:\\
  			AFGL~2136: \protect\cite{gibb2004interstellar}, 
        \protect\cite{schutte2003origin}, \protect\cite{dartois2002combined}, 
        \protect\cite{keane2001ice}, \protect\cite{dartois2001search}, 
        \protect\cite{brooke1999new}, \protect\cite{schutte1996discovery},        
        \protect\cite{willner1982infrared} \\ 
        W3~IRS5: \protect\cite{gibb2004interstellar}, 
        \protect\cite{keane2001ice}, \protect\cite{brooke1996study}, 
        \protect\cite{allamandola1992infrared}, \protect\cite{smith1989absorption}, 
        \protect\cite{willner1982infrared} \\ 
  			SgrA~IRS3: \protect\cite{pott2008enigma}, 
  			\protect\cite{moultaka2004dust}, \protect\cite{chiar2002hydrocarbons}, 
  			\protect\cite{tielens1996infrared}, \protect\cite{pendleton1994near}, 
  			\protect\cite{sandford1991interstellar} \\ 
  			SgrA~IRSX: Apart from the first figure for $\tau_{11.1}$ from this work, all 
  			other values are from ISO spectroscopy, and hence 'averaged' across a field 
  			of view containing most or all of the mini-spiral; \protect\cite{gibb2004interstellar} \\
        AFGL~2591: \protect\cite{gibb2004interstellar}, 
        \protect\cite{dartois2001search}, \protect\cite{brooke1999new}, 
        \protect\cite{smith1989absorption}, \protect\cite{willner1982infrared};
        the 6--7~$\mu$m region is heavily influenced by H$_2$O gas phase lines 
        (\protect\citet{helmich1996hotwater}) \\
        AFGL~4176: \protect\cite{persi1986infrared} and our own analysis of the
        ISO--SWS01 spectrum; the 6--7~$\mu$m region is heavily influenced by 
        H$_2$O gas phase lines (\protect\citet{vandishoeck1996hotwater}) \\
        AFGL~2403 and AFGL~2789: Our own analysis of the ISO--SWS01 spectra
 \end{list}
\label{TAB:OpDep}
\end{table*}

\subsubsection{Water ice}

Water ice possesses a librational band, the peak wavelength of which varies 
between about 12 and 13~$\mu$m for its crystalline and amorphous phases 
respectively (e.g. \citet{mastrapa2009optical}; \citet{maldoni1998study}). 
Its astronomical identification has historically been extremely difficult, 
with only a handful of good cases, e.g. the embedded YSO AFGL~961 
(\citet{cox1989line}; \citet{smith2011librational}, though see also 
\citet{robinson2012water}) and a few low mass YSOs such as HH46~IRS 
in \cite{boogert2008c2d}. Its detailed study has thus been restricted, due 
in part to its broadness and overlap with the minimum between the 10 and 
20~$\mu$m silicate bands. 

\begin{figure*}
\hspace*{0.75cm}
\includegraphics[scale=0.70]{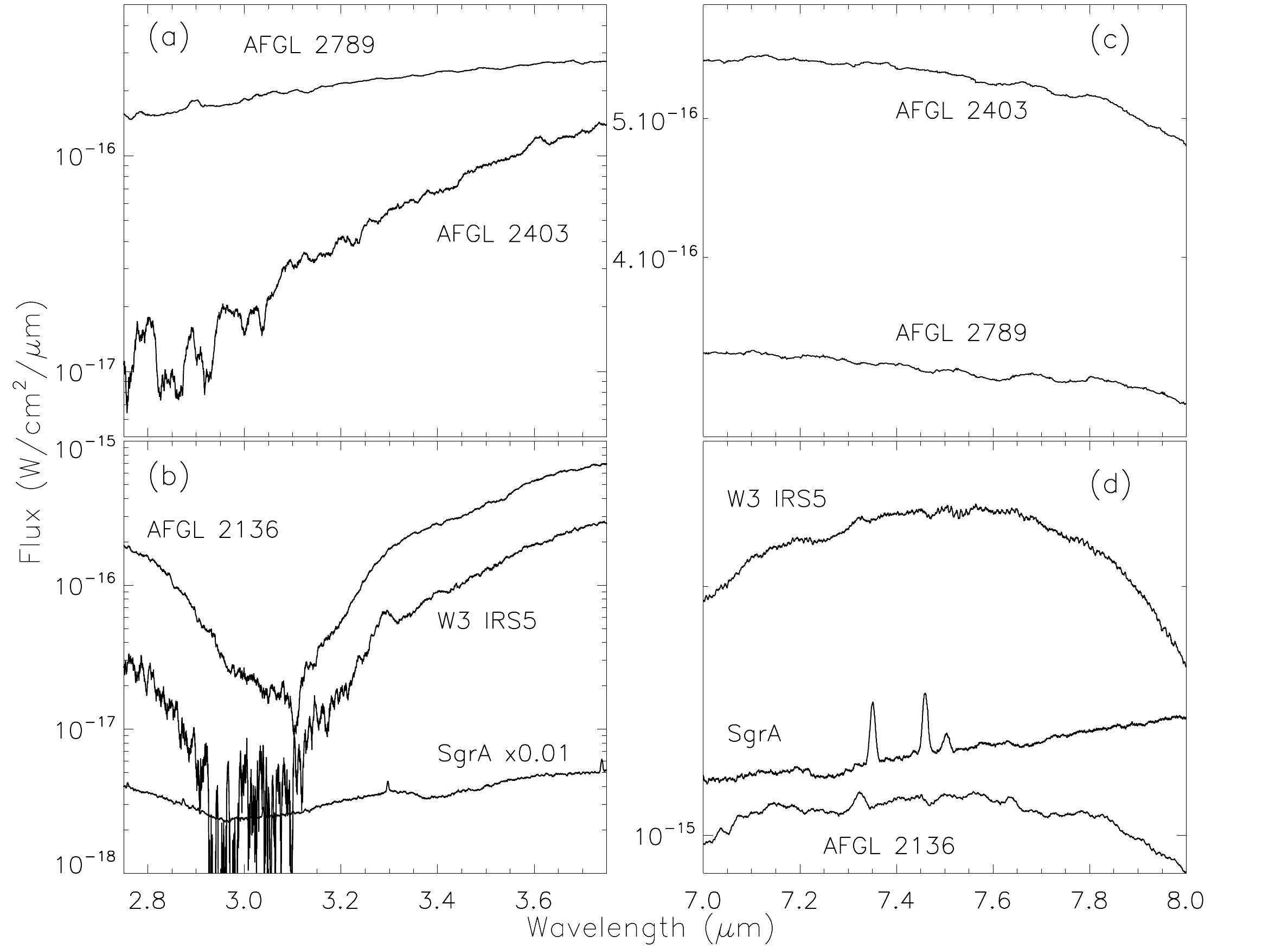}
\caption{Sigma-clipped and smoothed ISO spectra in the region of 
a) the 3.05~$\mu$m water ice band and b) the 7.25~$\mu$m hydrocarbon band. 
In (a) a hydrocarbon feature at 3.4~$\mu$m is seen toward the Galactic Centre, 
and perhaps also for AFGL~2136. Whilst the 3.4~$\mu$m band is certainly detected 
in small-beam and narrow-slit spectra of SgrA~IRS3, the 3~$\mu$m absorption is 
relatively weak or non-existent (see text for details). That there is probably 
no water ice band in AFGL~2403 is demonstrated by the fact that the signal is 
flat from 3.1~$\mu$m onwards, unlike the cases of W3~IRS5 and AFGL~2136. In (b) 
there is a known 7.25~$\mu$m hydrocarbon feature toward the Galactic Centre, 
and probably also in the spectra of W3~IRS5 and AFGL~2136, but not toward 
AFGL~2789 or AFGL~2403. Probable 7.7~$\mu$m methane ice absorption is detected 
in AFGL~2136, and possibly W3~IRS5 and SgrA, but no 7.7~$\mu$m feature is seen 
in AFGL~2789 and AFGL~2403.}
\label{FIG-ISOicehycs}
\end{figure*}

In centrally heated dust shells it is also susceptible to radiative transfer 
effects, such that cool dust emission can 'fill-in' and essentially mask the 
water ice feature, as shown by \cite{robinson2013models} and 
\cite{robinson2010librational}. For instance, whilst some of their models did 
result in a clearly identifiable water ice signature, even resembling the 
feature we observe (e.g. Fig. 14 in \citet{robinson2010librational}), they 
predict unrealistic levels of absorption within the intrinsically much stronger 
3.05~$\mu$m water ice band, certainly much higher than seen in our targets 
(Table~\ref{TAB:OpDep} and Figure~\ref{FIG-ISOicehycs}). Further, the overall 
appearance of the $\geq$ 10~$\mu$m portion of mid-IR spectra of YSOs with 
possible librational band absorption in \cite{boogert2008c2d} is much flatter 
than we see in our two Gemini-observed and bona-fide embedded YSOs 
AFGL~2136 and W3~IRS5, as well as AFGL~2591, AFGL~4176 and IRAS13481 
to be discussed in a following sub-section. These considerations, plus the 
difference between the expected and observed central wavelengths, already 
argues against a water ice explanation.

Even so, water ice absorption is definitely identified in a few of our objects 
at 3.05~$\mu$m (\citet{gibb2004interstellar}; \citet{smith1989absorption}) and 
6.0~$\mu$m (\citet{keane2001ice}). But in neither AFGL~2403 nor AFGL~2789 is 
it detected (though we note for AFGL~2403 there is little or no continuum 
below 3~$\mu$m in the ISO data which water ice could absorb against). See 
Figures \ref{FIG-ISOicehycs} and \ref{FIG-ISO6p85}. Thus, in at least these 
two sources a water ice carrier for the 11.1~$\mu$m absorption can almost 
certainly be ruled out.

For the Galactic Centre source SgrA~IRS3 there is conflicting evidence whether 
it has 3~$\mu$m water ice absorption. Within a broad beam, such as that of 
ISO, definite absorption is detected (e.g. \citet{chiar2000composition} and 
Figure~\ref{FIG-ISOicehycs}), but several authors, including 
\cite{chiar2002hydrocarbons}, \cite{moultaka2005vlt} and \cite{moultaka2004dust} 
have shown that the water ice column varies significantly, up to a factor of 5, 
over relatively small spatial scales of 0.5--2~pc, and certainly within the ISO 
beam size. This has been attributed by \cite{chiar2002hydrocarbons} to the clumpy 
nature of the molecular clouds within the Galactic Centre region. 

The spectrum of IRS3 in \cite{willner1982three} in a 2.5 arcsec beam has a 
3.4~$\mu$m hydrocarbon absorption feature (see next section) but no strong 
ice absorption (as stated by the authors). They instead infer that it has 
probable H$_2$O gas phase bands from the stellar atmosphere, with peak 
absorption occurring near 2.9~$\mu$m. Indeed, it can be seen in the ISO 
spectrum in Figure~\ref{FIG-ISOicehycs} that the 'ice' feature in SgrA occurs 
at $\sim$ 2.95~$\mu$m, significantly shorter than that of the YSOs AFGL~2136 
and W3~IRS5. 

The spectrum of SgrA IRS3 in \cite{pendleton1994near}, obtained in a 
2.7~arcsec aperture, has a smoothly rising continuum from 2.9--3.6~$\mu$m, 
apart from 3.4~$\mu$m hydrocarbon absorption, unlike the nearby sources IRS7 
and IRS6E which have clear absorption around 3~$\mu$m. The spectra of IRS3 
and IRS7 in \cite{moultaka2004dust}, taken in a 0.6~arcsec slit, are very 
similar to those of \cite{pendleton1994near}, but that of IRS3 in 
\cite{chiar2002hydrocarbons}, also in a 0.6~arcsec slit, is very different. 
The work of \cite{moultaka2005vlt} appears to resolve the discrepancy, 
showing that IRS3 is coincident with a region of much reduced H$_2$O 
absorption. Thus, by analogy with AFGL~2403 and AFGL~2789 it appears highly 
unlikely that water ice could be responsible for the 11.1~$\mu$m absorption 
seen in SgrA~IRS3.

Finally, assuming the same carrier is responsible for all the 11.1~$\mu$m features 
we have detected then the clear lack of a correlation between $\tau_{11.1}$ and 
$\tau_{3.0}$ or $\tau_{6.0}$ in Table~\ref{TAB:OpDep} almost certainly rules out 
water ice as the carrier. Note for instance the discrepancies between 
$\tau_{3.0}$/$\tau_{11.1}$ for AFGL~2591 and AFGL~4176 (as well as SgrA~IRS3) and 
the other two YSOs AFGL~2136 and W3~IRS5. 

\subsubsection{Hydrocarbon}

Our observed central wavelength is inconsistent with that expected from 
isolated (gas-phase) PAHs, for which the typically observed peak wavelength 
is at 11.22--11.25~$\mu$m, at least in the case of emission 
(e.g. \citet{verstraete2001aromatic}; \citet{witteborn1989new}). The central 
wavelength may change in the case of absorption when the PAH or related 
hydrocarbon is embedded in or on a host matrix or perhaps as some kind of 
mantle constituent, e.g. together with water ice. For example, 
\cite{bernstein2005mid} find that in a water ice matrix PAH bands in the 
11--13~$\mu$m region can shift by $\pm$~5-10~cm$^{-1}$. So a gas-phase 
band at 11.25~$\mu$m could feasibly occur in the range of 
11.25$\pm$0.13~$\mu$m. But \cite{bregman2000discovery} identify 11.25~$\mu$m 
absorption in the embedded YSO MonR2~IRS3 with a C--H out-of-plane vibrational 
mode of PAH molecules, a wavelength obviously inconsistent with our data. 

As shown by \cite{witteborn1989new} for the four sources they studied, the 
11.25~$\mu$m PAH band does have an asymmetric shape, with a long wavelength 
wing, and is thus broadly consistent with our feature in Figure~\ref{FIG-NormTau}. 
But assuming our quasi-continuum-subtracted profiles in Figure~\ref{FIG-NormTau} 
are a true representation of the feature profile then its FWZI is inconsistent 
with the much narrower PAH emission band, which only extends between about 
11.0--11.6~$\mu$m in the four targets of \cite{witteborn1989new}. Furthermore, 
such a PAH feature would likely be accompanied by other bands, particularly an 
in-plane bending mode at 8.6~$\mu$m of comparable integrated strength and an 
even stronger C--C mode at 7.7~$\mu$m. No sign of 8.6~$\mu$m absorption or 
emission is seen in our data (Figure~\ref{FIG-Spectra}), whilst that at 
7.7~$\mu$m in Figure~\ref{FIG-ISOicehycs} for AFGL~2136 and possibly 
W3~IRS5 (potentially explaining the apparent 'early' onset of its 9.7~$\mu$m 
silicate absorption band) is almost certainly due to methane (CH$_4$) ice as 
described in \cite{gibb2004interstellar}.  

Other hydrocarbon absorption features include those of aliphatic groups at 
3.38, 3.42, 3.47, 6.85 and 7.25~$\mu$m, and aromatic groups at 3.3 and 
6.2~$\mu$m, and have been identified in absorption spectra along several 
sightlines through the ISM, eg. \cite{chiar2013structure} and references 
therein. This includes the Galactic Centre (e.g. Figure~\ref{FIG-ISOicehycs} 
and Figure~\ref{FIG-ISO6p85} here, as well as \citet{chiar2002hydrocarbons}, 
\citet{chiar2000composition} and \citet{tielens1996infrared}), but to our 
knowledge no such feature around 11~$\mu$m has been postulated, let alone 
identified. For the Galactic Centre sightline the 3.38, 3.42 and 3.47~$\mu$m 
features appear as a triplet of comparable strengths, whilst for 
YSOs only a broad feature centred near 3.47~$\mu$m is typically detected 
(\citet{brooke1999new}; though the ISO spectrum of AFGL~2136 does appear 
to have a discrete but weak 3.4~$\mu$m feature in Figure~\ref{FIG-ISOicehycs},
confirmed after extracting an optical depth spectrum from 3.2 to 3.7~$\mu$m). 
The 3.2--3.8~$\mu$m long wavelength wing, peaking at around 3.3~$\mu$m and 
which almost ubiquitously accompanies the 3~$\mu$m water ice feature in 
molecular clouds, is also commonly attributed to a 'continuum' of hydrocarbon 
absorption, e.g. \cite{gibb2004interstellar} and \cite{smith1989absorption}.
   
Absorption at 7.25~$\mu$m is probably also present in the ISO spectra of W3~IRS5 
and AFGL~2136, but not toward AFGL~2403 and AFGL~2789 (Figure~\ref{FIG-ISOicehycs}). 
Since neither the 7.25 nor 7.7~$\mu$m features are seen toward these latter two 
objects, nor features at 3.4, 6.2 and 6.85~$\mu$m, then a hydrocarbon carrier 
for their 11.1~$\mu$m absorption can almost certainly be ruled out.

Once again, assuming the same carrier is responsible for all the 11.1~$\mu$m 
features we have detected then (despite the low number statistics) the lack of 
a correlation between $\tau_{11.1}$ and any of $\tau_{3.4}$, $\tau_{3.47}$, 
$\tau_{6.2}$, $\tau_{6.85}$ or $\tau_{7.25}$ in Table~\ref{TAB:OpDep} almost 
certainly rules out hydrocarbons as the carrier. Note for instance the 
discrepancies between $\tau_{3.47}$/$\tau_{11.1}$ for AFGL~2591 and the other 
two YSOs AFGL~2136 and W3~IRS5. Further, with larger sample sizes 
\cite{brooke1999new} and \cite{brooke1996study} find that the 3.47~$\mu$m 
hydrocarbon feature does correlate with the 3~$\mu$m water ice band, and 
\cite{thi2006vlt}, \cite{smith1989absorption} and \cite{willner1982infrared} 
find that the 3.2--3.8~$\mu$m long wavelength wing also correlates with the 
ice band. Since there is no obvious correlation between the 11.1~$\mu$m and 
water ice bands (see previous subsection) then it is highly unlikely a 
correlation could exist between the 11.1~$\mu$m feature and these other bands. 
We note here that many of the objects in the aforementioned works are common 
to our larger sample, so these correlations will be studied in more detail 
in Do Duy et al. (in preparation).

\subsubsection{Carbonates}

Carbonates have been identified in interplanetary dust particles (IDPs) 
and meteorites, principally via bands at around 6.8 and 11.4~$\mu$m 
(e.g. \citet{bradley1992combined}; \citet{sandford1986acid}; 
\citet{sandford1985laboratory}). To our knowledge there has been no 
pre-solar carbonate grain detected, i.e. one with an isotopic anomaly 
compared to our solar system (e.g. \citet{zinner2013laboratory}). Carbonates 
have also been tentatively identified in the far-infrared spectra of a few 
extra-solar-system objects, including calcite (CaCO$_{3}$, near 93~$\mu$m) 
and dolomite (CaMg(CO$_{3})_{2}$, near 62~$\mu$m) in two planetary 
nebulae by \cite{kemper2002detection} and \cite{kemper2002mineral}, and 
calcite in the solar-type protostar NGC1333 IRAS4 by 
\cite{ceccarelli2002discovery}.  

If carbonates were responsible for our 11.1~$\mu$m absorption feature 
then they would have to be Mg-rich (i.e. MgCO$_3$) as it is only for the 
Mg cation that the band occurs at 11.10~$\mu$m. For other abundant and 
likely cation metals Ca and Fe the feature occurs at 11.33 and 11.42~$\mu$m 
respectively (\citet{lane1997thermal}). Otherwise dolomite at 11.19~$\mu$m 
is just within our 0.1~$\mu$m uncertainty band.

\begin{figure}
\includegraphics[scale=0.43]{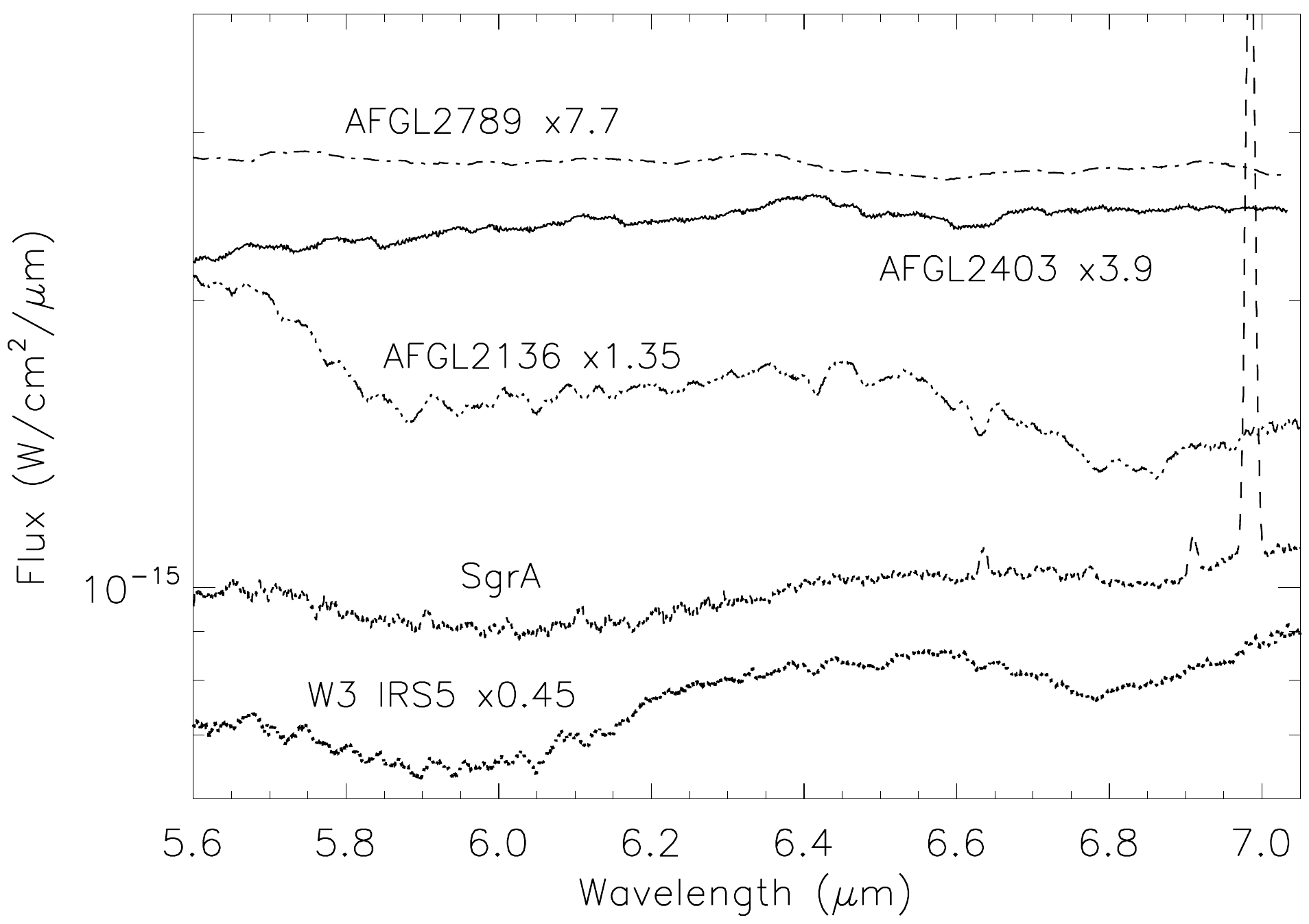}
\caption{Sigma-clipped and smoothed ISO spectra in the region of the 6.0 and 
6.85~$\mu$m water ice and 'unidentified' bands respectively. Relatively sharp
features in the spectrum of AFGL~2136, e.g. at $\sim$ 6.4 and 6.6~$\mu$m, arise 
from hot gas-phase water (\protect\cite{vandishoeck1996hotwater}).}
\label{FIG-ISO6p85}
\end{figure}

The 6.8~$\mu$m carbonate band is intrinsically several times stronger than the 
11.1--11.4~$\mu$m band, providing a potentially critical diagnostic constraint. 
As seen in Figure~\ref{FIG-ISO6p85} three of our targets do have an 
absorption feature centred around 6.8~$\mu$m. In fact, this feature is almost 
ubiquitous in the spectra of both high and low mass YSOs (\citet{boogert2008c2d}; 
\citet{gibb2004interstellar}), and is seen in the ISM toward the Galactic Centre 
(\citet{chiar2000composition}; \citet{tielens1996infrared}). At least for the 
YSOs it is assessed to be made up of two components, differing in their 
volatility, centred around 6.75 and 6.95~$\mu$m (\citet{boogert2008c2d}, 
\citet{keane2001ice}). Whilst several candidates exist for the feature(s), 
positive identification of either component remains a mystery 
(\citet{boogert2015observations}), and according to \cite{boogert2008c2d} 
the carrier cannot be the same for the YSOs and the ISM. We refer to the 
aforementioned papers for a discussion of the relative merits of each 
candidate. However, \cite{keane2001ice} rule out a carbonate interpretation 
based on the overall shape of the observed band being poorly fit by 
carbonates, although they also use the perceived lack of an accompanying 
11.4~$\mu$m feature to support their case, which we have shown is possibly 
incorrect for many sources.  

As for the cases of water ice and hydrocarbons, Figure~\ref{FIG-ISO6p85} shows 
that neither AFGL~2789 nor AFGL~2403 have a feature around 6.8~$\mu$m, so that 
a carbonate carrier for their 11.1~$\mu$m absorption feature can almost certainly 
be ruled out. Further, assuming the same carrier is responsible for all the 
11.1~$\mu$m features we have detected then the lack of a correlation between 
$\tau_{11.1}$ and $\tau_{6.85}$ in Table~\ref{TAB:OpDep} almost certainly rules 
out carbonates as the carrier. Note for instance that the higher quality ISO 
measurement of $\tau_{6.85}$ for AFGL~2591 --  given in \cite{gibb2004interstellar} 
and compared to the much lower spectral resolution data of 
\cite{willner1982infrared} -- is up to a factor of $\sim$ 5 lower than that of any other 
object, yet their $\tau_{11.1}$ are broadly similar.

\subsubsection{Silicon carbide (SiC)}

Silicon carbide has a lattice mode, the central wavelength of which occurs -- on 
average -- at 11.15$\pm$0.05~$\mu$m in emission (and occasionally in absorption) 
in astronomical spectra of carbon stars (\citet{clement2003new}). Along with the 
agreement in central wavelength with our feature, the FWZI's are also reasonably 
consistent. Thus, SiC could be a candidate for the absorption band we observe in 
our small sample of Figure~\ref{FIG-Spectra}. Pre-solar SiC has been found in 
meteorites, suggesting some must survive after being formed in C-star outflows. 
But it has not so far been unambiguously detected in the ISM (\citet{whittet1990sicism}), 
although \cite{min2007shape} inferred a fractional abundance of 2.6--4.2\%, with 
9--12\% of the available Si in SiC grains, based on a shoulder around 11~$\mu$m 
in the extinction curve toward the Galactic Centre. Such a shoulder was also detected 
in the VLTI MIDI study of SgrA~IRS3 by \cite{pott2008enigma}, who state it occurs 
at 11.3~$\mu$m and also tentatively assign it to SiC.

To our knowledge only a single detection has been claimed for the presence of 
SiC in the disk or envelope of a young star, namely SVS13 (\citet{fujiyoshi2015mid}). 
But its spectrum looks markedly different to those we present here, and the SiC 
identification was based largely on a unique mid-IR polarisation signature 
(\citet{fujiyoshi2015mid}; \citet{smith2000studies}; \citet{wright1999mid}). 
Once again we defer a full discussion of the SiC possibility to a subsequent 
paper describing the full sample (Do Duy et al. in preparation). For now we 
instead use polarisation data in the following section to argue against SiC 
being the carrier.

\subsection{A crystalline silicate carrier}

Given the similarity of central wavelength and band profile for all five sources 
in Figures~\ref{FIG-Spectra} to \ref{FIG-NormTau}, as well as three other YSOs 
to be presented in this section -- AFGL~2591, AFGL~4176 and IRAS13481-6124 -- we 
believe it is very likely that the same carrier is responsible for their 
11.1~$\mu$m absorption features. Further, the above discussion highlights 
that -- of the several possible carriers -- water ice, hydrocarbons and carbonates 
can almost certainly be excluded in the cases of AFGL~2789 and AFGL~2403 given 
the lack of concomitant features in those spectra. Moreover, given the absence 
of a correlation between the depths of the 11.1~$\mu$m feature and 3--8 $\mu$m 
bands of water ice, hydrocarbons and carbonates, a strong argument exists that 
none of these materials can be the 11.1~$\mu$m carrier in any of the objects. 
Crystalline silicates and perhaps SiC therefore remain the only options. We 
will show later that SiC in at least three sources is highly unlikely, based
on polarisation considerations.

\begin{figure*}
\includegraphics[scale=0.675]{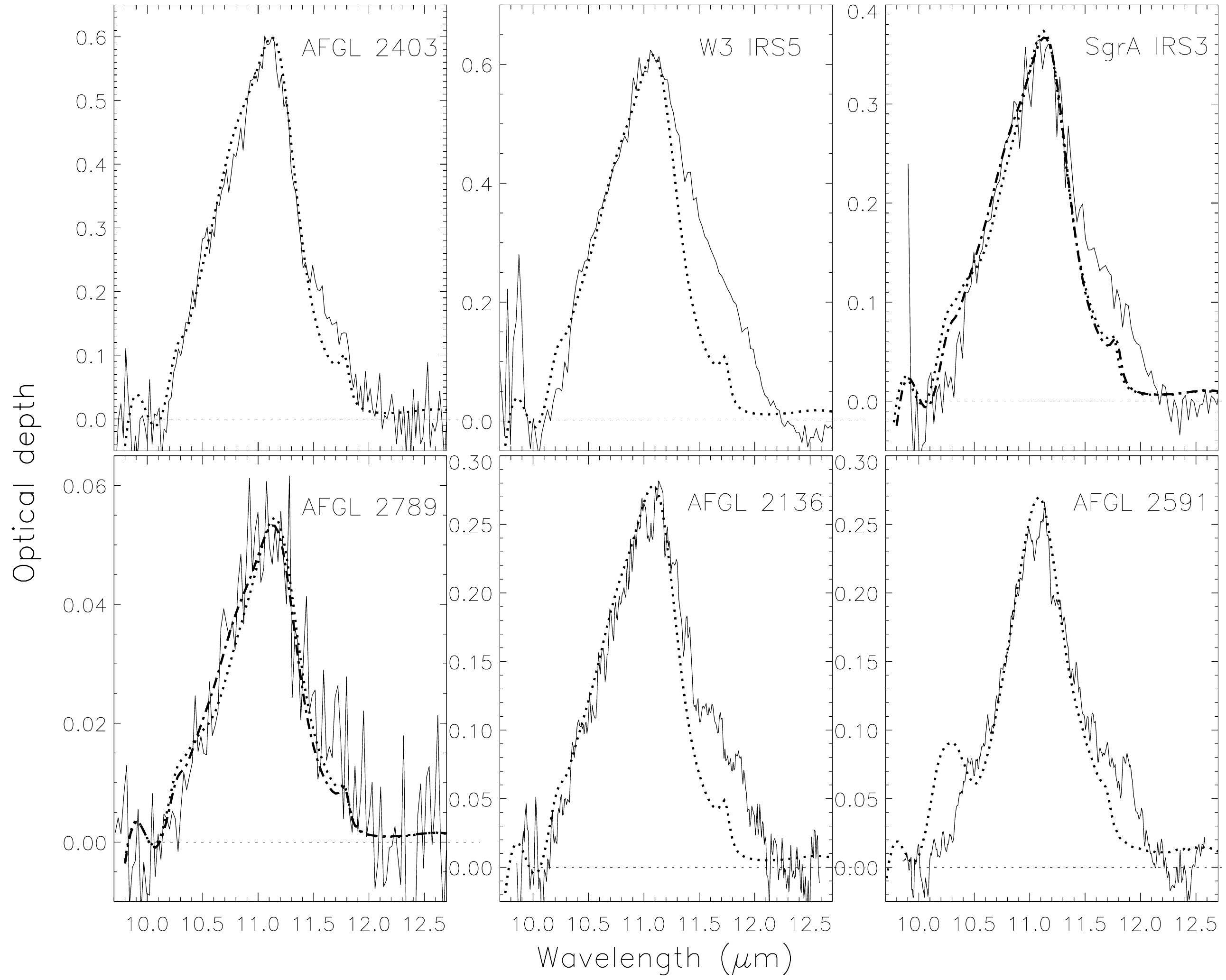}
\caption{Comparison between observed and model 11.1~$\mu$m feaures, the latter 
extracted using the same polynomial technique as described in the text. The fractions 
of crystalline olivine in each model are 0.075 for AFGL~2403, 0.05 in W3~IRS5 
and AFGL~2136, 0.025-0.05 (dotted/dashed) for SgrA~IRS3 and AFGL~2789 and 
0.01 for AFGL~2591, using the optical constants of \protect\cite{mukai1990optical} 
along with those of \protect\cite{dorschner1995steps}. The model feature has been 
shifted by -0.2~$\mu$m for AFGL~2403, AFGL~2789 and SgrA~IRS3, -0.25 for 
W3~IRS5 and AFGL~2136, and -0.29~$\mu$m for AFGL~2591, consistent with the 
findings of \protect\cite{tamanai200610}. This shift aligns the primary peak but not 
the secondary peaks at 10.5 and 11.9~$\mu$m, which apparently do not shift between
aerosol and matrix-embedded particles in the work of \protect\cite{tamanai200610}.
The ISO data used for AFGL~2136 and AFGL~2591 have been averaged in 20
pixel wide bins for plotting purposes.}
\label{FIG-Obs-XMods}
\end{figure*}

If any trend can be seen in Table~\ref{TAB:OpDep} it is that $\tau_{11.1}$ 
increases with increasing $\tau_{9.7}$, e.g. the respective values from 
AFGL2789 to IRAS13481-6124 to SgrA~IRSX to the four other YSOs as well 
as SgrA~IRS3 (excluding AFGL~2403 given its status as a dust factory). 
Interestingly, \cite{alexander2003isocam} find a correlation between the 
depth of a feature at 11.2~$\mu$m and the depth of the 9.7~$\mu$m silicate 
band in their sample of ISOCAM spectra of YSOs in the Corona Australis, 
$\rho$ Ophiuchus, Chamaeleon I and Serpens molecular clouds. Whilst they 
do not show a correlation plot, they state that the proportionality is 
negative, which we presume to mean that the 11.2~$\mu$m depth decreases 
as the 9.7~$\mu$m depth increases (or vice-versa). This then leads them 
to identify the 11.2~$\mu$m feature as an emissive shoulder on the silicate 
feature, rather than an independent feature of some other species. This 
seems highly unlikely to us, as many of their spectra resemble those 
presented here, e.g. HH100~IR in their Figure 4 and which is part of our 
larger sample to be presented in Do Duy et al. (in preparation).

The putative $\tau_{9.7}$ -- $\tau_{11.1}$ correlation that we find does not 
necessarily mean that the 11.1~$\mu$m feature must originate from a silicate. 
But it does mean that whatever carrier is responsible always occurs together 
with silicates. Moreover, given no other known cosmic dust constituent seems 
able to account for the 11.1~$\mu$m feature there is a strong implication that 
it must itself be a silicate band.

The central wavelength of 11.10$\pm$0.10~$\mu$m is consistent with crystalline 
silicate, especially the magnesium-rich olivine end-member forsterite. Admittedly, 
at first sight the observed central wavelength appears to be different to that 
typically quoted of $\sim$ 11.3~$\mu$m for crystalline forsteritic olivine of 
\cite{fabian2001steps} and others (e.g the models in Figure~\ref{FIG-NewDepths}). 
But as shown by \cite{tamanai200610} this is likely to be a result of the 
conditions under which the laboratory data were taken. They showed that for 
free-flying, aerosol crystalline forsterite the primary bands occur at around 
9.85 and 11.1~$\mu$m, shifted shortward by $0.20\pm0.05$~$\mu$m from the band 
position when the forsterite is embedded on a KBr pellet. Thus, the wavelength 
of the astronomical feature we detect and that expected for terrestrial 
crystalline forsterite are the same. On the other hand, weaker bands at 10.1, 
10.4 and 11.9~$\mu$m show little or no shift. That the main bands shift and 
the minor bands do not shift obviously makes comparing complete forsterite 
profiles of laboratory and observed spectra problematic.

The feature in our 'template' or 'control' target AFGL~2403 almost certainly 
arises from crystalline forsterite, or at least olivine with a higher Mg than 
Fe content. This is because accompanying detections of both the 33.6 and 
69~$\mu$m forsterite bands were made by \cite{de2014problematically}. Accepting 
this to be the case then the similarity of the band profile -- central wavelength 
and overall shape -- in the other sources suggests the same carrier.

As seen in Figure~\ref{FIG-Obs-XMods} the observed and model profiles broadly 
resemble each other. For the model we have used a single oblate 2:1 shape as it 
appears to best match the mid-IR polarisation profile of the diffuse interstellar 
medium dust (\citet{wright2005mid}, \citet{wright2002mineralogy} and in preparation; 
see also \citet{draine2006x} and \citet{hildebrand1995shapes}). The model profiles 
have been shifted shortward by 0.2--0.3~$\mu$m, in accordance with the results 
of \cite{tamanai200610} and which nicely aligns the peak wavelengths at 11.1~$\mu$m. 
Unfortunately such a bodily shift of the profile also moves the 10.5 and 11.9~$\mu$m 
sub-bands, which as noted above is not replicated in the results of \cite{tamanai200610}.

Whilst we have not attempted to optimise the comparison between the model 
and observed profiles in Figure~\ref{FIG-Obs-XMods}, the crystalline olivine 
volume fraction is around 7.5\% in AFGL~2403, and the others vary between 1\% 
and 5\%. Assuming the density of the crystalline and amorphous silicates are 
the same then these figures are also their mass fractions. The value for 
AFGL~2403 is in reasonable agreement with the abundance (mass fraction) of 
11$\pm$3\% and 8$\pm$2\% inferred by \cite{de2014problematically} from the 
11~$\mu$m and 69~$\mu$m bands respectively. The value of 2.5--5\% for 
SgrA~IRS3 is larger than the best-fit mass fraction of 1.1\% for the ISM path to the 
Galactic Centre of \cite{kemper2005erratum} or 0.6--1.5\% of \cite{min2007shape}, 
but our 'minimum' value is in reasonable agreement with the firm upper limit of 
2.2\% of \cite{kemper2005erratum}. Our range is also in reasonable agreement
with the limit of 3--5\% postulated by \cite{li2007crystallinity}. 

As a caveat on the above figures we note that they assume the optical properties 
of specific silicates, i.e. the crystalline olivine of \cite{mukai1990optical}
mixed with amorphous olivine MgFeSiO$_4$ of \cite{dorschner1995steps}. A different 
set of refractive indices, which may well have used a different technique in their 
determination, or have a different Mg/Fe ratio, for either one or both of the 
amorphous and crystalline components, may change these estimates. 

\begin{figure}
\includegraphics[scale=0.40]{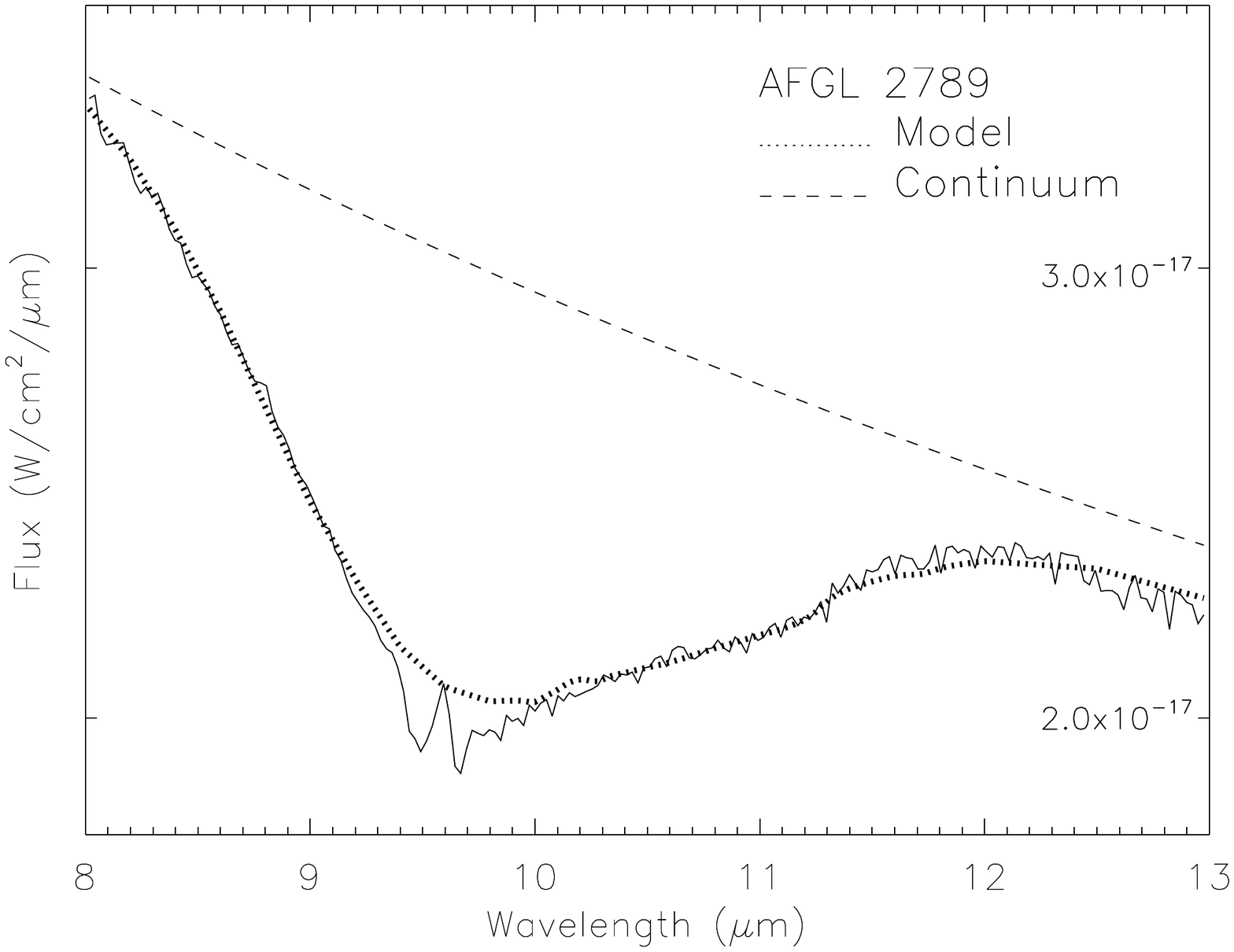}
\caption{Comparison between observations of AFGL~2789 (solid line) and a 
representative model (dotted line). The modelling method is adapted from 
that of \protect\cite{hanner19988} and \protect\cite{hanner199510}, which 
finds its roots in \protect\cite{gillett19758}. In this case it includes three 
separate populations of dust, resulting in mass fractions of 0.1~$\mu$m sized 
grains of amorphous olivine, amorphous pyroxene and crystalline forsterite of 
58\%, 39\% and 3\% respectively. Optical constants for olivine and pyroxene 
are taken from \protect\cite{dorschner1995steps} and for forsterite from 
\protect\cite{fabian2001steps}. The wavelength range 9.2--10.0~$\mu$m has 
been excluded from the fit due to the imperfect telluric correction around 
the 9.6~$\mu$m ozone band.} 
\label{FIG-AFGL2789Mod}
\end{figure}

As an example, when the crystalline component was changed to the forsteritic 
olivine Mg$_{1.9}$Fe$_{0.1}$SiO$_4$ sample of \cite{fabian2001steps} we were 
not able to obtain as good a match to the extracted optical depths. The model 
profile remained too narrow compared to the observations even up to a crystalline 
fraction of 7.5\%. A proper model fit to the entire observed spectrum, as 
opposed to our relatively simplistic approach using a single extracted feature, 
is probably required in these cases. An example of this is demonstrated in 
Figure~\ref{FIG-AFGL2789Mod} for AFGL~2789. In this case our inferred value 
for the crystalline olivine fraction of 2.5-5\% in Figure~\ref{FIG-Obs-XMods} -- using 
\cite{mukai1990optical} optical constants -- is consistent with the value of about 
3\% obtained from a preliminary model and which uses \cite{fabian2001steps} optical 
constants. This model will be described in detail in Do Duy et al. (in preparation). 

\subsubsection{Other signatures of crystalline silicate in the 8--13~$\mu$m region?}

Our (re-)discovery of the 11.1~$\mu$m feature, and its likely association 
with crystalline silicate (specifically forsterite), motivated us to search 
for other spectral features which could strengthen this identification. Within 
the 8--13~$\mu$m window discrete features might be present at around 10 and 
11.9~$\mu$m, with perhaps other shoulders or inflections in between, as 
suggested by the models in the final panel of Figure~\ref{FIG-Spectra}. 

\begin{figure}
\includegraphics[scale=0.40]{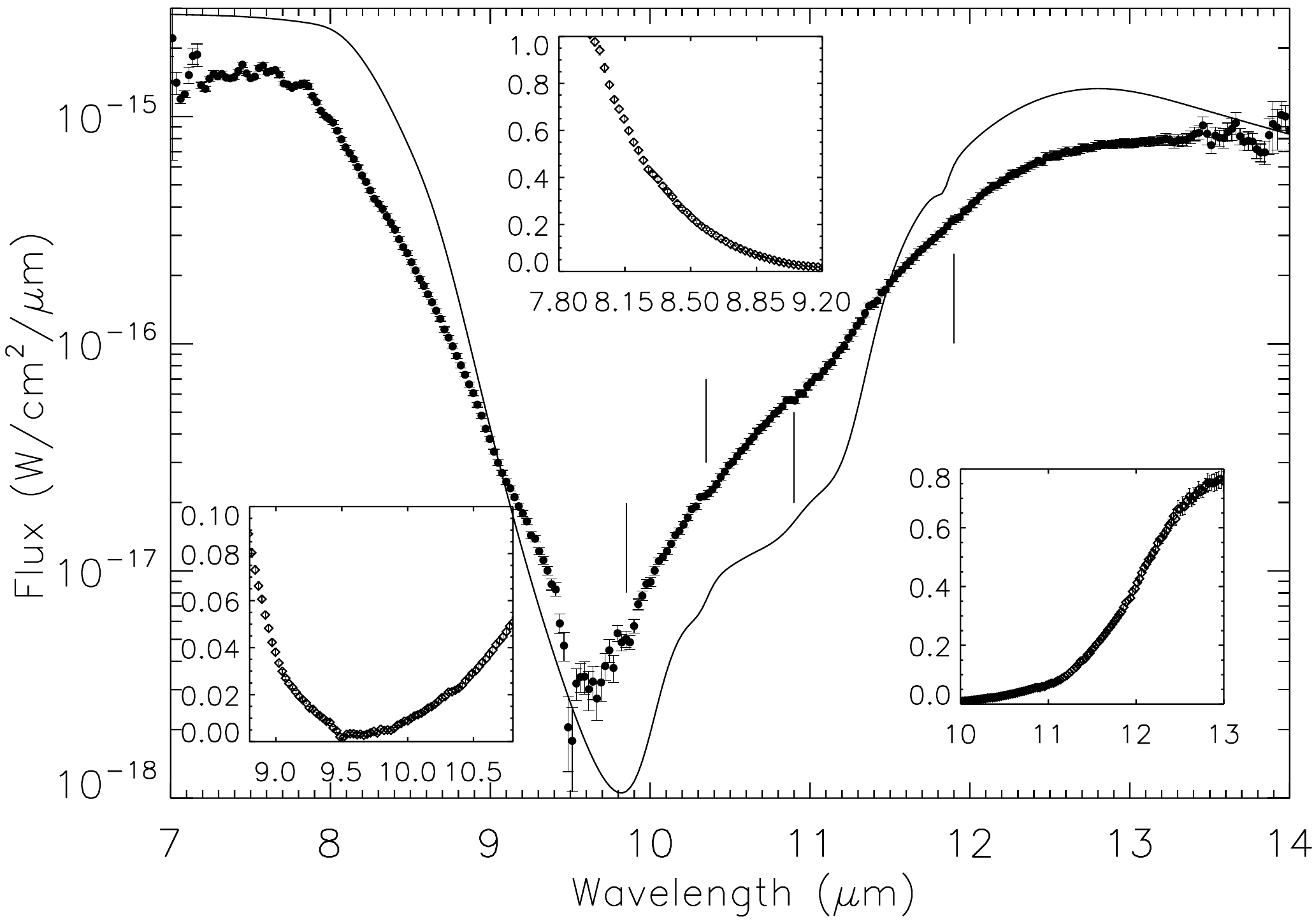}
\caption{Spectrum of W3~IRS5~NE reproduced from Figure~\ref{FIG-Spectra}, along 
with a representative model containing a volume fraction of 0.20 of crystalline olivine
from \protect\cite{mukai1990optical}. Insets show zooms, on a linear flux scale and in
units of 10$^{-15}$ W/cm$^{2}$/$\mu$m, of selected wavelength intervals. The zoom 
around 8.3~$\mu$m is included since both \protect\cite{fujiyoshi2015mid} and 
\protect\cite{poteet2011spitzer} detected a feature at this wavelength in the YSOs SVS13 
and HOPS-68 respectively. In Fujiyoshi et al.'s model it came from annealed SiO$_2$, 
whilst Poteet et al. did not mention it in their paper. The vertical lines guide the eye 
to possible correspondences between the observations and model. Note that the model 
has been bodily shifted by 0.15~$\mu$m to shorter wavelengths, in line with the finding 
of \protect\cite{tamanai200610} that the main fortsterite peaks, but not the minor peaks, 
shift between free-flying and matrix-embedded measurements. Consequently the
11.9~$\mu$m features do not precisely align in the plot.}
\label{FIG-W3IRS5}
\end{figure}

Given its large optical depth and good signal-to-noise our best ground-based 
spectrum for searching for other features is probably that of W3~IRS5~NE. This is 
presented again in Figure~\ref{FIG-W3IRS5}, along with a model with a relatively 
large crystalline olivine volume fraction in order to emphasise the features. 
The model is shifted by 0.15~$\mu$m shortward, consistent with the work of 
\cite{tamanai200610}, and is not meant to be a model for W3~IRS5~NE, but merely 
to guide the eye to possible similarities. 

The spectrum of W3~IRS5 does -- at least qualitatively -- display features, 
or perhaps better described as perturbations, that are tantalisingly similar 
to those expected from a mixture of amorphous and crystalline olivine. These are 
marked in Figure~\ref{FIG-W3IRS5} with vertical bars. For instance, there appears 
to be a very weak 11.9~$\mu$m feature. Also, in the middle of the broad 11~$\mu$m 
band there is a slope change in the model spectrum which is potentially reflected 
in the data. Similar such 'features' are also seen between 9.8--10.5~$\mu$m in both 
the model and data. 

\subsubsection{Crystalline silicate feature at 11.85~$\mu$m}

Admittedly the existence of features other than at 11.1~$\mu$m in our Gemini 
spectra is not entirely conclusive. But at R $\simeq$ 100 our ground-based spectra 
barely have the spectral resolution to detect the aforementioned features. Therefore 
we utilised the ISO--SWS database, for which R is about an order of magnitude higher 
and which also allows a search for crystalline features at longer wavelengths, 
e.g. 20--45~$\mu$m. 

\begin{figure}
\includegraphics[scale=0.40]{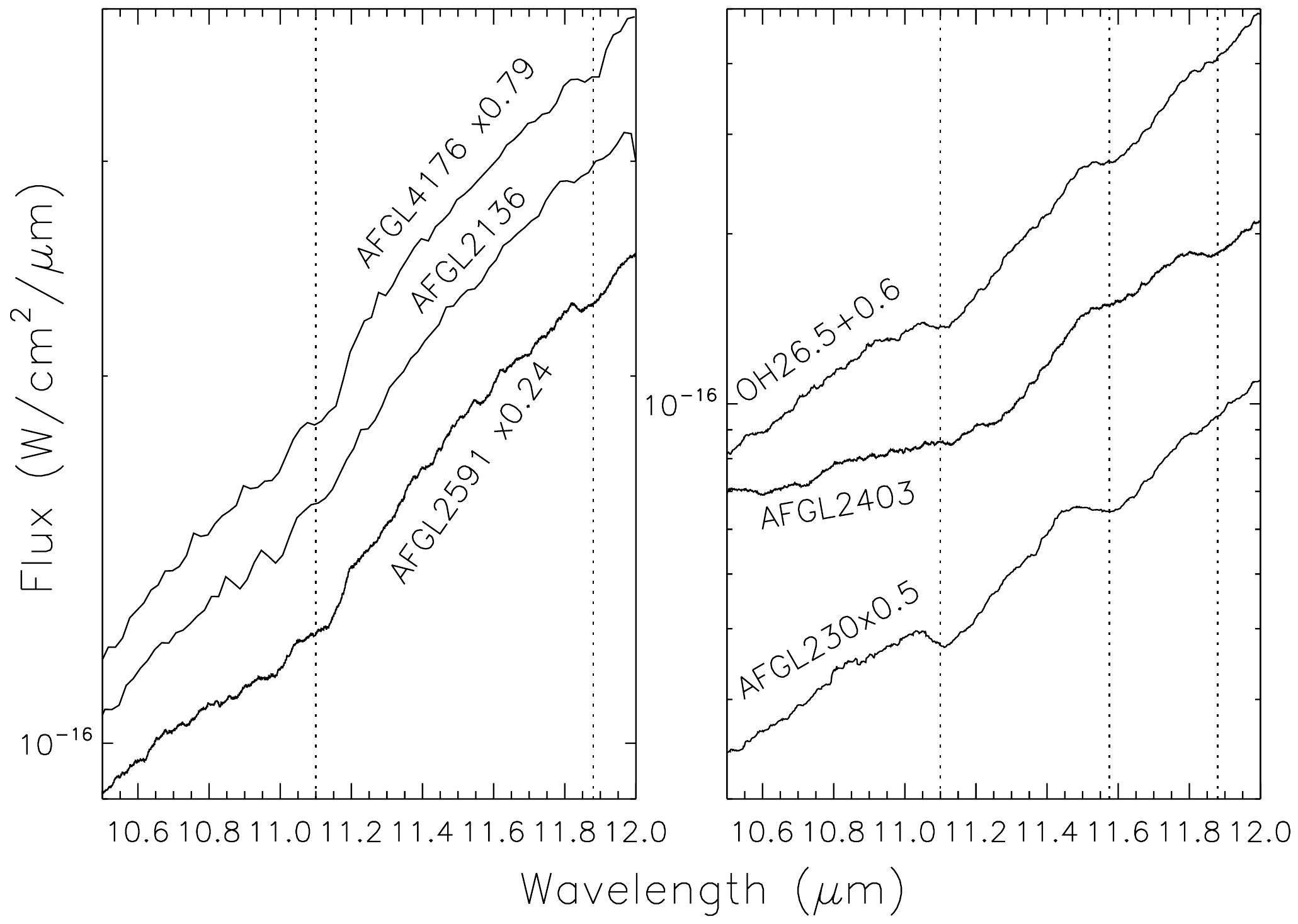}
\caption{10.5--12.0~$\mu$m ISO--SWS spectra of three YSOs (left hand panel) 
and three OH/IR stars (right hand panel). In all three YSOs, all of which 
have a clear 11.1~$\mu$m absorption feature, there is also an apparent 
absorption band centred around 11.85~$\mu$m. That both features also exist 
in the OH/IR stars, known sources of crystalline silicates, suggests a similar 
interpretation for the YSOs. See also \protect\cite{sylvester19992} for 
detailed analysis of the OH26.5 and AFGL~230 ISO 2--200~$\mu$m spectra.}
\label{ISO-Spectra_10-12um}
\end{figure}

The left hand panel of Figure~\ref{ISO-Spectra_10-12um} shows the SWS06 spectra 
of the massive embedded YSOs AFGL~2136 and AFGL~4176, as well as the SWS01 
spectrum of AFGL~2591. For comparison the SWS01 spectra of the OH/IR stars 
AFGL~2403, OH26.5+0.6 and AFGL~230 (OH127.8+0.0) are shown in the right hand 
panel. In much the same way that AFGL~2403 is used as a template for the 11.1~$\mu$m 
feature, OH26.5+0.6 and AFGL~230, as known sources of crystalline silicates, also act 
as templates for other potential features. This includes crystalline enstatite in 
the case of AFGL~230, and which we discuss in a little more detail in Appendix B. 

All the OH/IR template sources clearly show the 11.1~$\mu$m forsterite feature. But 
in addition they possess a feature around 11.6~$\mu$m, most prominent in AFGL~230 
and which can be identified with crystalline enstatite. Furthermore, OH26.5 and 
AFGL~2403 also show a band at $\sim$~11.85~$\mu$m. Similarly, the three YSOs possess 
such an 11.85~$\mu$m feature. Notably, the extracted 11~$\mu$m band for AFGL~2591, 
and probably also for AFGL~2136, in Figure~\ref{FIG-Obs-XMods} shows this feature, 
as would be expected. We also find the feature in the SWS spectrum toward the Galactic 
Centre (Figure~\ref{IRS3X-ISOstar}), but given the special status of this ISM path we 
reserve its discussion to a later section (also see Appendix A). 

Since the relevant band 2C of the SWS is not documented to have a feature in its 
RSRF at 11.85~$\mu$m, unlike the case at 9.35, 10.1 and 11.05~$\mu$m, we assess 
that it is a real spectral feature in these targets. Paradoxically the non-detection 
of an 11.85~$\mu$m feature in the ISO spectrum of W3~IRS5 supports this contention, 
but which we attribute to the complicated source structure within the large ISO 
beam (e.g. its binary nature and extended mid-IR emission; \cite{vandertak2005subarc}). 

Finally, although not noted by the respective authors, we point out that a band at 
this wavelength appears in the crystalline silicate-rich spectra of the ULIRG 
IRAS08572+3915 in \cite{spoon2006detection} (their Figure~2) and the Class~0 
protostar HOPS-68 in \cite{poteet2011spitzer} (their Figure~2).

\subsubsection{Crystalline silicate features at 20--30~$\mu$m}

\begin{figure*}
\includegraphics[scale=0.72]{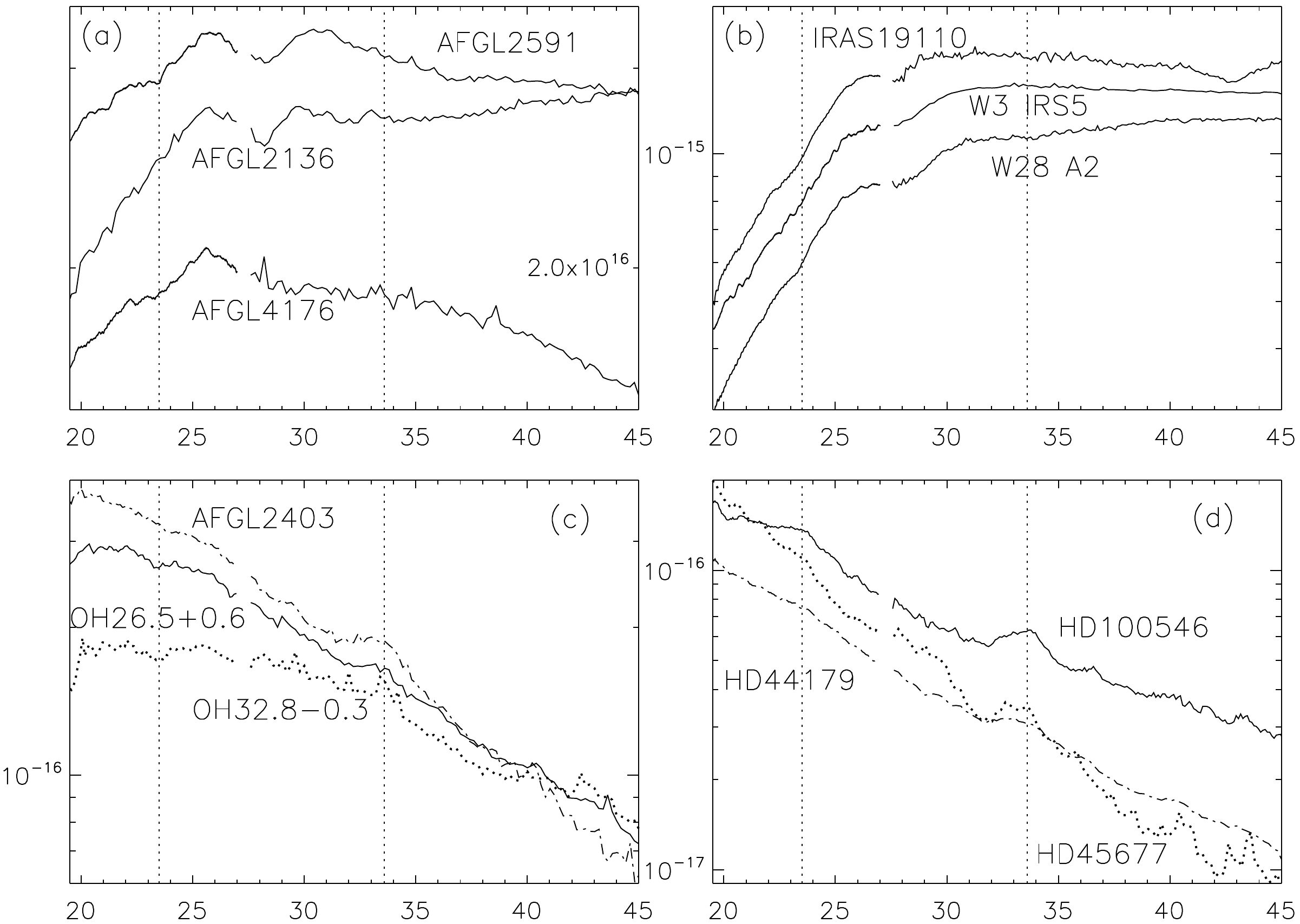}
\caption{19.5--45~$\mu$m ISO--SWS01 spectra of a selection of targets. The vertical 
axis is flux measured in units of W/cm$^2$/$\mu$m and the horizontal axis is wavelength 
in microns. Included in each panel are three embedded YSOs in (a), another three YSOs 
in (b) but which might have a more complex emission structure within the relatively 
large ISO beam (e.g. the binary nature of W3~IRS5 and associated extended emission), 
three OH/IR stars in (c), two Herbig Be stars in (d) and the pre-planetary nebula 
HD44179 (the Red Rectangle) also in (d). Vertical lines mark the wavelengths of 
crystalline silicate features at 23.5 and 33.6~$\mu$m. The SWS spectrum of HD100546 
has previously been studied in detail by \protect\cite{malfait1998spectrum}, and 
those of HD45677 and HD44179 by \protect\cite{molster2002crystalline}. Since the 
ISO--SWS aperture size increased from 14$\times$27 to 20$\times$33 arcsec between 
bands 3D and 4, and the sources may be marginally extended at these wavelengths, 
there could be a small flux mismatch such that the band 4 data had to be scaled 
to match band 3D. Scaling factors applied to each source are the following: 
AFGL~2136 band 4 - 0.70; 
AFGL~4176 band 3D - 0.52, band 4 - 0.61; 
AFGL~2591 band 3D - 0.53, band 4 - 0.42; 
W3~IRS5 band 4 - 0.77; 
W28~A2 band 3D - 0.71, band 4 - 0.75; 
IRAS19110 band 3D - 3.73, band 4 - 3.39; 
OH26.5 band 3D - 0.95; 
OH32.8 band 3D - 3.21, band 4 - 2.94;
AFGL~2403 band 3D - 3.36, band 4 - 3.21;
HD100546 band 4 - 0.85;
HD45677 band 3D - 1.52, band 4 - 1.69;
HD44179 band 3D - 0.29, band 4 - 0.26.}
\label{Spectra_20to45um}
\end{figure*}

Our identification of probable crystalline silicates in our sample of YSOs 
(as well as the ISM toward the Galactic Centre, see Appendix A) is further 
strengthened when consideration is made of the 20--45~$\mu$m interval.
Figure~\ref{Spectra_20to45um} shows the ISO SWS01 spectra of several of 
our targets, plus others, in this spectral range. The data was treated in 
a similar manner to that previously described, but with the following 
additional considerations. 

Firstly, data from band 3E, with the relatively narrow wavelength interval of 
27.5--29~$\mu$m, was completely neglected. It is notoriously unreliable in its 
spectral shape, severely affected by fringes, and in most cases can at best 
only be used to provide a flux (see \citet{leech2003iso}). This means that 
there is a small gap in our spectra, but which is partially filled by the 
overlapping of band 4 down to around 28~$\mu$m. 

Secondly, we neglected the band 3D data beyond 27.0~$\mu$m because of the well 
documented blue leak, in which around 10\% of the 14~$\mu$m flux leaks to the 
$\geq$~27~$\mu$m region (see \citet{leech2003iso}). Thirdly, the band 4 data 
was corrected for the related effects of delayed responsivity from 40--45~$\mu$m 
and memory effects from 28--33~$\mu$m, which for relatively strong sources can 
cause a large difference in the spectral shape in these regions between the up 
and down scans. See Appendix C for a brief description. Notably band 3D is immune 
to such effects, and the up/down scans overlay almost precisely for the objects 
considered here. 

Included in Figure~\ref{Spectra_20to45um} are a variety of sources, comprising six 
YSOs in (a) and (b), three OH/IR stars in (c), two Herbig Be stars in (d) and one 
pre-planetary nebula (PPN, HD44179 also known as the Red Rectangle) also in (d). 
As previously mentioned the OH/IR stars are established sources of crystalline 
silicates, and the same is true for the Herbig stars and PPN objects 
(e.g. \citet{molster2002crystalline}). Thus, they are included here as templates 
in the study of the YSOs, few of which have previously been inferred to possess 
crystalline silicates (e.g. \citet{demyk1999chemical}).

All of the YSOs have at least one, and in several cases two, absorption features 
in the 20--30~$\mu$m interval, one at about 23.5~$\mu$m and the other around 
28~$\mu$m. The 23.5~$\mu$m band is visible as a shallow feature in AFGL~2591 and 
AFGL~4176, or as a shoulder (or inflection) in IRAS19110+1045 (G45.07+0.13) and 
W28~A2 (G5.89-0.39). These latter two also possess the 11.1~$\mu$m absorption 
feature in their ISO spectra (see Appendix B, where we also speculate on the 
presence of crystalline enstatite in these two YSOs). \cite{demyk1999chemical} 
and \cite{dartois1998detection} previously detected the 23.5~$\mu$m feature in 
IRAS19110, as well as another YSO AFGL~7009S, but neither pursued an analysis. 
Along with W28~A2 we have found it in several other YSOs. 

The 23.5~$\mu$m band has a corresponding absorption feature in the three OH/IR 
stars, previously presented in \cite{sylvester19992} for OH26.5 and OH32.8, and 
a corresponding emission feature in the two Herbig Be stars and one PPN. It has 
been detected in many other sources in both ISO and Spitzer spectra of OH/IR and 
other evolved stars (e.g. \citet{jiang2013crystalline}; \citet{molster2002crystalline}),
predominantly as an emission feature, as well as in Herbig and/or T Tauri star 
disks (\citet{juhasz2010dust}; \citet{watson2009crystalline}; 
\citet{sargent2009dust}; \citet{meeus2001iso}). 

In all these cases the 23.5~$\mu$m feature is universally identified as a crystalline 
forsteritic band, based on its similarity to a feature seen in laboratory measurements 
of magnesium rich crystalline olivines (\citet{pitman2010infrared}; 
\citet{sogawa2006infrared}; \citet{suto2006low}; \citet{koike2003compositional}; 
\citet{jager1998steps}; \citet{mukai1990optical}). To our knowledge there is no 
documented problem with the RSRF of the SWS band 3D, and so we favour a 
crystalline olivine interpretation in the much younger YSOs -- still in their embedded 
phase -- as well. A detailed discussion is deferred to a later paper (Do Duy et al., 
in preparation), but in Appendix B we show the feature extracted in a similar manner 
to that for the 11.1~$\mu$m band, as well as comparison to a representative 
amorphous+crystalline silicate model.

An absorption feature at around 28~$\mu$m is also evident in the YSOs in 
Figure~\ref{Spectra_20to45um}, being most apparent in AFGL~2591 and AFGL~2136.
The fact that this feature occurs across two separate bands of the SWS has 
both good points and bad points. For instance, that both the long and short 
wavelength sides of bands 3D and 4 respectively dip down provides a level of 
confidence that they trace a real spectral feature. This is despite the central 
wavelength being part of the 'missing' band 3E, and that the larger band 4 
aperture size may include more extended emission. On the other hand, one must 
always be wary about features at the band edges given that the RSRF of band 4 
does decrease relatively steeply from about 30~$\mu$m to 29~$\mu$m, and that we 
are utilising data beyond the nominal 29~$\mu$m minimum 'valid' wavelength for 
band 4 (\citet{leech2003iso}). 

As 'insurance' against the possibility that the 28~$\mu$m feature is an artefact 
we have examined many tens of other SWS01 spectra covering several different source 
types, spectral shapes and flux levels. We do not see a pattern that would suggest 
our 28~$\mu$m feature identification is an artefact. A few examples are included in
Figure~\ref{Spectra_20to45um}-c and -d, where there is no apparent problem with the 
RSRF. This is in the sense that for the objects in (c) and (d) their spectra continue 
to monotonically decrease in the 'transition interval' from band 3D to band 4, 
showing no 'anomalous' structure mimicking the shape of the RSRF in that region. 
Indeed, in our experience the majority of artefacts introduced by the ISO RSRF have 
their origin in relatively narrow 'downward' features which in turn mimic 'emission' 
bands in the target spectrum.   

Accepting that the 28~$\mu$m feature in our YSO sample in 
Figure~\ref{Spectra_20to45um}-a and -b is real then once again a similar band has 
previously been detected. This has typically been in emission, in spectra of both 
dust factories (outflowing winds of evolved stars) and dust repositories (disks 
around young stars). These respectively include old stars in perhaps all post-main 
sequence evolutionary phases (\citet{jiang2013crystalline}; \citet{gielen2007dust}), 
and cicumstellar disks of Herbig Ae/Be stars (\citet{juhasz2010dust}) and T Tauri 
stars (\citet{watson2009crystalline}). Again it is universally attributed to 
crystalline silicate based on its similarity to a laboratory band.
 
As already noted there is no 28~$\mu$m feature in the other sources of 
Figure~\ref{Spectra_20to45um}, either in emission or absorption (except perhaps 
for HD45677 in emission). But in the case of the OH/IR stars the crystalline 
silicate features appear to switch from absorption at 23.5~$\mu$m to emission at
33.6~$\mu$m. It is thus natural to conclude that the 28~$\mu$m feature is likely 
to be self-absorbed in these sources, due to radiative transfer effects within 
their circumstellar shells, and is hence difficult or impossible to distinguish 
against the continuum without extremely good signal-to-noise. 

Finally, we note that 23.5 and 28~$\mu$m absorption bands have been identified in 
the crystalline silicate-rich spectra of at least one other embedded YSO, the 
Class~0 object HOPS-68 by \cite{poteet2011spitzer}. 

\subsubsection{Crystalline silicate and other features at 30--45~$\mu$m}

This brings our discussion to the conspicuous absence of a 33.6~$\mu$m feature 
in the YSOs of Figure~\ref{Spectra_20to45um}, despite the presence of this 
feature in absorption in HOPS-68 and in emission in all the other sources of 
Figure~\ref{Spectra_20to45um}-c and -d. Our current explanation is that this 
band is self-absorbed in most YSOs, in a similar fashion to the 28~$\mu$m 
band of some OH/IR stars. 

\begin{figure}
\includegraphics[scale=0.6425]{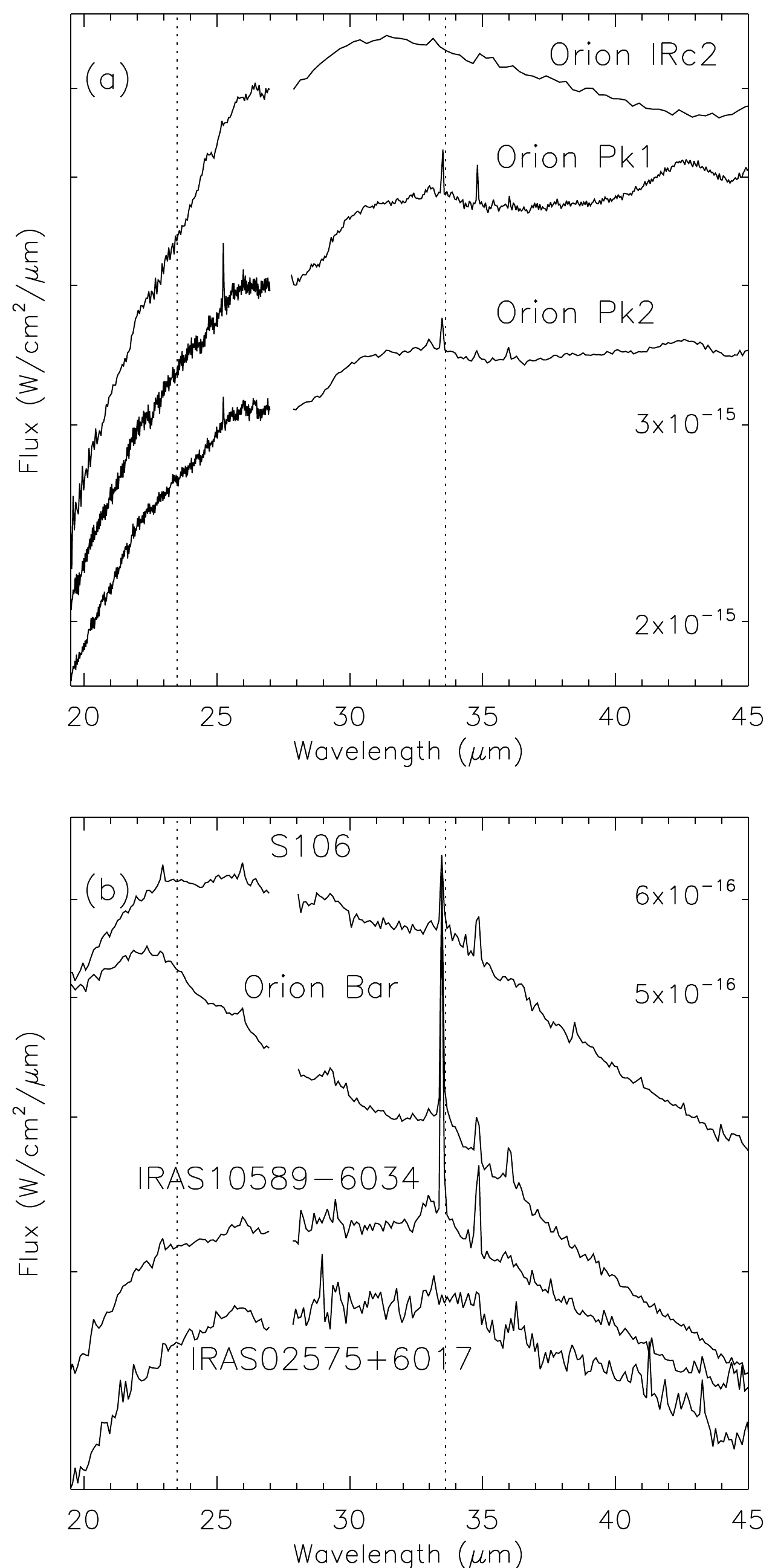}
\caption{(a) 19.5--45~$\mu$m ISO--SWS01 spectra of three positions in Orion, 
centred on IRc2, Pk1 and Pk2. The IRc2 and Pk1 spectra have previously been 
presented in \protect\cite{gibb2004interstellar} and 
\protect\cite{rosenthal2000iso}, though without discussion of the presence or 
otherwise of crystalline silicates or the 43~$\mu$m water ice feature. 
(b) 19.5-45~$\mu$m ISO--SWS01 spectra of several HII regions. That of S106 has 
previously been presented by \protect\cite{vandenancker2000iso}, the two IRAS 
sources by \protect\cite{peeters2002iso} and the Orion Bar by 
\protect\cite{cesarsky2000silicate}.
Scaling factors applied to each source are the following: 
Orion Pk2 band 4 - 0.52;
Orion Pk1 band 3D - 1.56, band 4 - 0.99;
Orion IRc2 band 3D - 0.53, band 4 - 0.42;
Orion Bar band 3D - 0.40;
S106 band 3D - 1.27, band 4 - 0.58;
IRAS02575 band 3D - 4.82, band 4 - 3.86;
IRAS10589 band 3D - 2.46, band 4 - 1.77.}
\label{ISO-HII-20to45um}
\end{figure}

A radiative transfer study, varying parameters such as the central source 
temperature and luminosity, circumstellar envelope density and temperature 
structure, and the overall opacity and dust composition, is required to test 
this hypothesis. Whilst beyond the scope of the present paper we have begun 
work on this study, but note here that this feature is also not apparent in 
the ULIRG IRAS08572+3915 of \cite{spoon2006detection}, despite the crystalline 
silicate bands at lower wavelengths being clearly detected.

As some indication that spectral bands in the 30--45~$\mu$m region can go 
into emission, absorption or disappear depending on specific parameters of the 
source, we can look to the case of IRAS19110+1045. Whilst this YSO has a 
3.1~$\mu$m water ice absorption band, as well as features of other ice species, 
in the 2--10~$\mu$m interval, they are in no way abnormal or atypical compared 
to the other embedded YSOs measured with ISO. Yet it is the only one with an 
absorption feature at $\sim$~43~$\mu$m, seen in panel (b) of 
Figure~\ref{Spectra_20to45um} and discussed in detail by 
\cite{dartois1998detection}, who attributed it to a lattice mode 
of crystalline water ice.

Similarly, in the case of the Orion YSO cluster, bright in the mid-IR and including 
the BN~Object and IRc2, no 43~$\mu$m ice band can be discerned, as seen in panel (a) 
of Figure~\ref{ISO-HII-20to45um}, and consistent with the independent (but higher 
S/N) SWS06 spectrum presented in \cite{vandishoeck1998iso}. Yet only an ISO beamsize 
or so away to the SE and NW along the outflow axis (\citet{allen1993explosive}), at 
the shock positions known as Pk1 and Pk2 respectively (\citet{beckwith1978observations}), 
the band appears prominently in emission.  
 
To potentially further alleviate a concern that the 33.6~$\mu$m crystalline 
silicate feature may be absent in star forming environments we show spectra 
of several HII regions in panel (b) of Figure~\ref{ISO-HII-20to45um}. In all 
of them, representing a range of excitation conditions and flux levels, there 
is a possible feature, or indeed a plateau of emission, extending between 
about 32 and 37~$\mu$m. It bears a remarkable similarity to the complex of 
crystalline silicate bands seen for instance in the OH/IR stars, Herbig Be
stars and PPN in (c) and (d) respectively of Figure~\ref{Spectra_20to45um}.

Features similar to these, and interpreted as evidence for Mg-rich crystalline
silicates, were first identified in ISO--SWS spectra of star forming regions 
(HII regions, photodissociation regions or PDRs) by \cite{jones1999infrared} in 
M17 and \cite{cesarsky2000silicate} in Orion. However, a question was raised over 
their reality by \cite{molster2005crystalline}, who claim them to be artefacts. See 
also \cite{peeters2005high} who cites a paper in preparation by Kemper et al., but
which has not so far been published to our knowledge. 

There are certainly good reasons to be careful in interpreting the presence 
of features in ISO--SWS band 4 spectra, given the aforementioned responsivity 
and memory effects. That the 'average' dust temperature is such that the 
Planck-like spectra of most of these targets turns over in the 30--40~$\mu$m 
region (see for example the spectral atlas of \citet{peeters2002iso}) also makes 
feature identification and extraction more complicated. 

Further, in this particular range it is feasible that not only a high continuum 
flux but also the bright emission line intensity (e.g. [S~III] at 33.481, [Si~II] 
at 34.815 and [Ne~III] at 36.014~$\mu$m) could perturb the shape of the spectrum. 
This would probably depend alot on the speed with which the particular SWS01 
observation was conducted, being more likely for speed 1 than speed 4. In the 
many tens of SWS01's we have looked at -- across a broad range of source types, 
flux levels and speeds -- we have not seen an obviously attributable such effect, 
or at least not one which is sufficiently broad to 'mimic' a several micron wide 
(FWZI) emission plateau. 

Finally, there are features in the RSRF of band 4 at $\sim$ 31, 33 and 36.5~$\mu$m 
which could feasibly conspire to create the observed feature (we note no in-orbit 
band 4 RSRF was ever derived, and hence it relies on laboratory data; 
\citet{leech2003iso}). But the only one which we know as having been documented 
to appear in fully reduced spectra is that at 33~$\mu$m, seen for instance in the 
SWS06 data of Orion IRc2 in \cite{wright2000iso}. 

We have looked into all of the above-mentioned band 4 issues, and remain 
confident in the reality of the solid-state dust features seen in 
Figure~\ref{ISO-HII-20to45um}. This includes the 43~$\mu$m water ice band 
at the Orion shock positions in Figure~\ref{ISO-HII-20to45um}-a, and the 33.6 and 
36~$\mu$m features in several HII regions in Figure~\ref{ISO-HII-20to45um}-b.

\subsection{A polarisation perspective}

Detection of a polarisation signature from the 11.1~$\mu$m absorption band could 
potentially provide a valuable constraint on its carrier. Linear polarisation via 
dichroism obviously requires grains to be non-spherical with a particular axis 
mutually aligned along a common direction (e.g. \citet{aitken1989spectropolarimetry}). 
In most alignment mechanisms it is the short axis of spinning grains which becomes 
aligned along the direction of an ambient magnetic field (see 
\citet{lazarian2007tracing}). 

\begin{figure}
\includegraphics[scale=0.43]{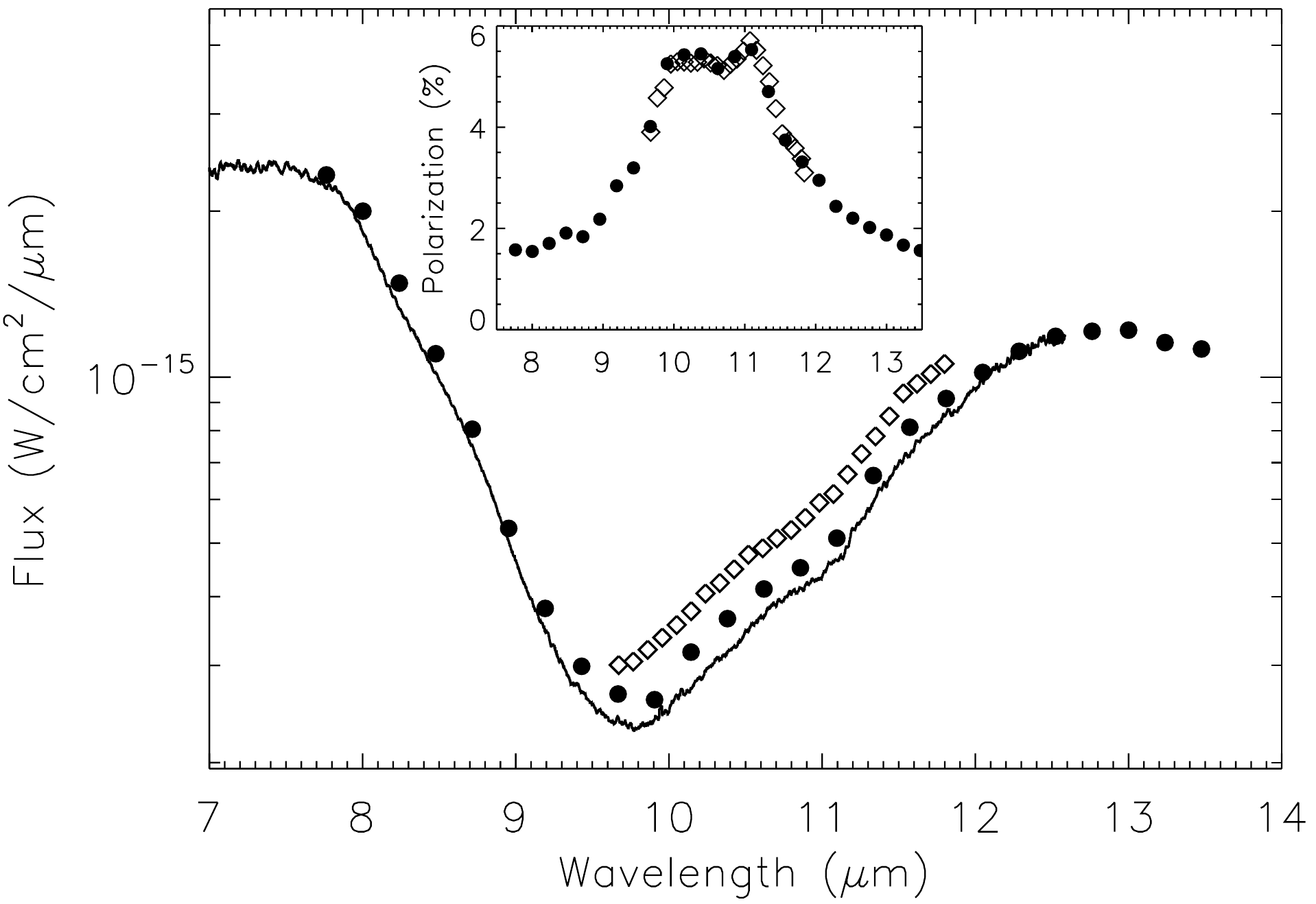}
\caption{The main plot shows the UCLS low and high resolution spectra of AFGL~2591 
(solid circles and open diamonds respectively), whilst the ISO spectrum is shown 
as a solid line. Note how an inflection around 11~$\mu$m at low resolution is 
resolved into a bona fide absorption band at higher resolution. The inset 
displays the low (solid circles) and high (open diamonds) resolution polarisation 
spectrum, containing the sharp polarisation feature around 11.1~$\mu$m first discovered 
by \protect\cite{aitken1988infrared}.}
\label{FIG-AFGL2591}
\end{figure}

Assuming grains to be spheroidal -- making the problem more tractable -- cross
sections for absorption of radiation along the major and minor axes peak at 
different wavelengths (\citet{draine1984optical}). The polarisation cross section 
C$_{\rm pol}$ is formed by a subtraction of these cross sections, in a sense 
magnifying small differences between them, whilst the absorption cross section 
C$_{\rm abs}$ is formed from a sum (\citet{lee1985infrared}). Since the cross 
sections are also sensitively dependent on the grain dielectric function, a 
unique identifier of the responsible material, then spectropolarimetry -- especially 
across a resonance -- becomes a powerful probe of dust grain mineralogy, more 
so than conventional spectroscopy alone.

\begin{figure*}
\includegraphics[scale=0.75]{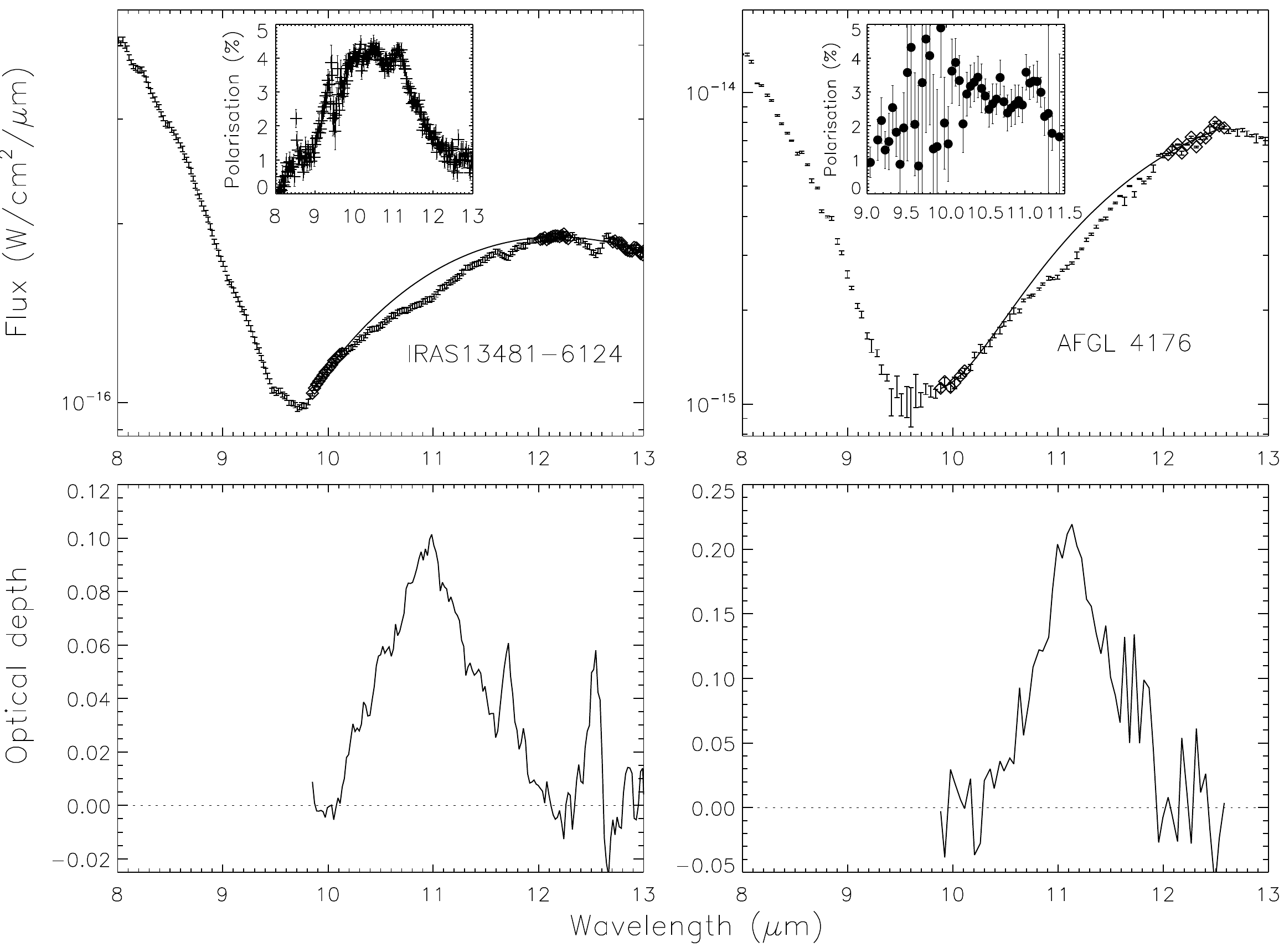}
\caption{The top panel shows the observed spectra of IRAS13481-6124 (left) and 
AFGL~4176 (right), previously presented in \protect\cite{wright2008mid} and 
\protect\cite{wright1994thesis} respectively, along with an extrapolated 
polynomial across the 11.1~$\mu$m feature. The insets show the polarisation 
data, demonstrating a positive detection of a feature around 11.1~$\mu$m in 
IRAS13481 and a tentative detection in AFGL~4176. The IRAS13481 data was 
obtained with TIMMI2 in January and June 2006 at the ESO 3.6~m telescope, 
whilst the conventional spectrum of AFGL~4176 was obtained with the UCLS on 
the ANU 2.3~m telescope in January 1989 and the polarisation data acquired 
with the UCLS on the AAT in May 1992. The bottom panel shows the optical 
depth spectra of the extracted 11.1~$\mu$m feature, which is obviously very 
similar in central wavelength and profile to those in 
Figure~\ref{FIG-Obs-XMods}. The narrow features around 11.7 and 12.5~$\mu$m 
in IRAS13481 are due to telluric absorption bands.}
\label{FIG-IRAS13481}
\end{figure*}

\subsubsection{Polarisation-to-absorption ratio considerations}

At a spectral resolution R $\simeq$ 40 \cite{aitken1988infrared} first found a 
polarisation signature likely to be associated with 11.1~$\mu$m absorption in the 
massive embedded YSO AFGL~2591. They interpreted the feature to be due to a 
structured -- as opposed to amorphous -- silicate produced during an annealing 
episode. This was based principally on the feature having a polarisation-to-absorption 
ratio, or p/$\tau$, of around 0.05 compared to a value of $\sim$~0.02 for the 
amorphous silicate feature, as well as preliminary modelling of the 8--13~$\mu$m 
spectrum using the dielectric function of disordered, radiation damaged olivine. 

The quantity $p/\tau$ acts as a proxy for the material band strength (opacity or 
cross section per gram in cm$^2$/g, or C$_{\rm abs}$/V in cm$^{-1}$) as shown by 
\cite{martin1975some}. Obviously the polarisation depends on the degree of 
alignment of the grains and/or the angle to which the magnetic field is inclined 
to the plane-of-the-sky, whilst $\tau$ depends on neither. In the case of AFGL~2591 
the comparison of $p/\tau$ for each dust component relies on them being similarly 
aligned, a safe assumption given the constancy of the observed polarisation position 
angle presented in \cite{aitken1988infrared}. 

\cite{wright1999mid} extended the modelling of AFGL~2591 to more realistic 
optical properties, and found a reasonable match between observed and modelled 
8--13~$\mu$m spectra using a mixture of amorphous and crystalline silicate, 
represented by the \cite{draine1984optical} 'astronomical silicate' and 
crystalline olivine from \cite{mukai1990optical}. The volume fraction of 
crystalline olivine inclusions occupying the amorphous silicate matrix was 
around 17.5\%. But as will be shown later this possibly represents an absolute 
maximum, and could be a factor of around two lower for a different set of 
optical constants.

Higher resolution (R $\simeq$ 100) observations of the AFGL~2591 11.1~$\mu$m 
polarisation feature obtained with the UCLS instrument, shown in 
\cite{wright1994thesis} and here in Figure~\ref{FIG-AFGL2591}, confirmed its 
reality and revealed its profile. Similar resolution observations obtained 
with Michelle on UKIRT also showed it, presented in \cite{wright2005mid}. 
These demonstrated that the polarisation maximum was shifted to a longer 
wavelength, by 0.05--0.10~$\mu$m, than the extinction maximum, a direct 
prediction of polarisation by dichroism.

These higher resolution data sets also established that the true p/$\tau$ of 
the feature is more like about 0.1, twice that inferred by \cite{aitken1988infrared} 
and suggesting an even higher band strength and thus more structured material. 
For instance, if the band strength of the 'astronomical silicate' of 
\cite{draine1984optical} is 10$^{4}$~cm$^{-1}$ then that of the 11.1~$\mu$m 
carrier is of the order of 5$\times$10$^{4}$~cm$^{-1}$, or 3000 and 15000~cm$^2$/g 
respectively if expressed as an opacity. 

Interestingly, at 11~$\mu$m the crystalline forsterite of \cite{fabian2001steps} 
has a band strength of around 20000~cm$^{2}$/g along the $x$ and $y$ axes, in good 
agreement with the observed value. On the other hand, the \cite{mukai1990optical} 
crystalline olivine band strength is only $\sim$~5500~cm$^{2}$/g. This difference 
already suggests that the two sets of optical data would require different amounts 
of the crystalline component to match the data.

For SiC the band strength is instead around 3500~cm$^{2}$/g and just over 
5000~cm$^{2}$/g for the \cite{pegourie1988optical} and \cite{laor1993spectroscopic} 
varieties respectively, whilst it ranges between 40000--90000~cm$^{2}$/g for the 
samples of \cite{pitman2008optical} and Choyke \& Palik in \cite{palik1985hocs}. 
These SiC values of either several thousand or several tens of thousand cm$^{2}$/g 
are obviously too low and too high respectively compared to the AFGL~2591 data, and 
make SiC of any variety an unlikely carrier of the 11~$\mu$m feature in at least 
AFGL~2591. Furthermore, the silicate--SiC polarisation models presented in 
\cite{fujiyoshi2015mid} for the YSO SVS13 show that SiC actually broadens the 
8--13~$\mu$m polarisation profile, as well as shifts the peak to longer wavelengths, 
with an increasing SiC contribution. It is never able to reproduce the sharp peak 
of the relatively narrow 11.1~$\mu$m polarisation feature seen in AFGL~2591.

Since the discovery of \cite{aitken1988infrared} it took 20 years until a second 
positive detection of an 11.1~$\mu$m polarisation feature was made by 
\cite{wright2008mid}, again in a massive embedded YSO called IRAS~13481-6124. 
Before that the only other possible case was AFGL~4176, also a DEYSO, presented 
in \cite{wright1994thesis}. See Figure~\ref{FIG-IRAS13481}. In these two objects 
the respective $p/\tau$ for the 10 and 11.1~$\mu$m bands are $\sim$ 0.01 and 0.05 
for AFGL~4176, and $\leq$0.04 and $\sim$ 0.10 for IRAS13481.

A caveat on these $p/\tau$ comparisons at 9.7 and 11.1~$\mu$m, alluded to earlier, 
is that $p$ obviously originates only from aligned dust, whilst $\tau$ could have 
contributions from both aligned and unaligned dust. The comparisons above would not
change if there was an unpolarised component to the spectrum only so long as the 
mineralogy of the unaligned and aligned dust was the same. However, if there was an 
unpolarised component from amorphous silicates, then $(p/\tau)_{9.7}$ would decrease 
faster than $(p/\tau)_{11.1}$. The latter could then appear to be (unreasonably) 
much larger than the former, and a direct comparison misleading concerning their 
relative band strengths. From the modelling presented in the next subsection this 
could be the case for AFGL~2591, although it does not change our assessment of 
crystalline silicates. On the other hand, if there was a polarised component 
from amorphous silicates, then $(p/\tau)_{9.7}$ would be unchanged whilst 
$(p/\tau)_{11.1}$ would decrease. In this case the latter could then appear to 
be (unreasonably) much lower than it would otherwise be (even as low as zero), 
and again a direct comparison misleading concerning their relative band strengths. 
This situation is also considered in the following subsection, potentially 
explaining why an 11.1~$\mu$m polarisation signature is not seen in the majority 
of sources in the mid-IR polarisation atlas of \cite{smith2000studies}.

\subsubsection{Modelling the polarisation spectrum}

The original modelling of the AFGL~2591 polarisation by \cite{aitken1988infrared} 
and subsequently \cite{wright1999mid} used optical constants for both amorphous 
and crystalline silicates that are somewhat outdated and/or cannot provide 
mineralogical information. For example, the crystalline olivine of 
\cite{mukai1990optical} was not identified with a particular Fe/Mg ratio. 

Further, the astronomical silicate of \cite{draine1984optical} 
was constructed to fit the 10~$\mu$m emission spectrum of the Trapezium region of 
Orion, and its 20-to-10~$\mu$m band ratio was defined from an 'average' of the 
emission of dust shells around oxygen-rich evolved stars. Whilst of 
significant utility in radiative transfer modelling of spectral energy distributions, 
due to its broad wavelength coverage and consistency with causality via the 
Kramers-Kroenig relations, it has not proved as successful in modelling the 
detailed shape of specific 10 and 20~$\mu$m silicate bands. This is especially the
case in polarisation, even for the BN Object in the same cloud, Orion, 
as the Trapezium (\citet{wright2005mid}, \citet{wright2002mineralogy}; 
\citet{hildebrand1995shapes}; \citet{o1994effect}; \citet{henning1993porous}; 
\citet{aitken198910}, \citet{aitken1988infrared}; \citet{lee1985infrared}, 
\citet{draine1984optical}; but see also \citet{fujiyoshi2015mid} and 
\citet{wright1999mid} where it is mixed with SiC to provide a good match to 
the Class~I YSO SVS13 polarisation).

We have thus calculated new models using laboratory-based refractive indices
of silicates with well defined mineralogical properties. For the crystalline 
component the \cite{fabian2001steps} forsteritic olivine with formula 
Mg$_{1.9}$Fe$_{0.1}$SiO$_4$ is used. We have already noted previously the 
difference in band strength between this and the \cite{mukai1990optical} sample. 
This may partly stem from the fact that the Mukai \& Koike and Fabian et al. 
optical constants are respectively derived from transmission and polarised 
reflection measurements. Originating from these different techniques is that the 
Mukai \& Koike data consists of only a single set of optical constants (actually 
oscillator parameters), whilst those of Fabian et al. comprise three data sets, 
corresponding to the vibrational directions parallel to the crystallographic 
axes $x$, $y$ and $z$. \cite{sihvola1994anisotropic} and \cite{sihvola1994dielectric} 
present generalised formulae for the effective dielectric function in the case 
of randomly oriented and spherical biaxial crystallite inclusions -- of which 
forsterite is a member given its orthorhombic structure.

For the amorphous component the olivine with formula Mg$_{0.8}$Fe$_{1.2}$SiO$_4$ 
from \cite{dorschner1995steps} was initially used, since it provides a very good 
match -- in terms of both wavelength of peak polarisation ($\lambda_{\rm P,max}$)
and FWHM -- to the polarisation profile of dust in the diffuse ISM, as shown in 
\cite{wright2002mineralogy} and \cite{wright2005mid}. It was also used by 
\cite{mathis1998silicates} in his models of the diffuse ISM silicate features,
and their consistency with observed heavy element abundances. The scenario we
envisaged was that this would be the form of the bulk of the silicate dust 
deposited in the molecular cloud from which AFGL~2591 formed, and which would 
subsequently be processed.

However, whilst this combination could provide a good match to the 10~$\mu$m
polarisation spectrum of AFGL~2591 it also produced a 20~$\mu$m polarisation 
far in excess (about a factor of 2) of that observed by \cite{aitken1988infrared}. 
Given that the position angles within the 10 and 20~$\mu$m windows are the 
same, and spectrally constant, then pure dichroic absorption is almost certainly 
the sole operating mechanism. In other words it is unlikely that a 'crossed 
polariser' effect is diluting the 20~$\mu$m polarisation. Therefore the 
discrepancy between the observed and model 20~$\mu$m polarisation must originate 
in the optical constants of the amorphous silicate, which indeed has a relatively 
high 20-to-10~$\mu$m band ratio. In addition to this problem, the combination 
also had an internal inconsistency in the difference between the Fe/Mg ratios 
of the two components.

Other laboratory-derived optical constants for amorphous olivine exist in the
publications of \cite{day1979mid}, \cite{scott1996ultraviolet} and 
\cite{jager2003steps}, specifically including the magnesium end-member forsterite, 
Mg$_2$SiO$_4$. Additionally, \cite{day1981infrared} provides refractive indices 
for the iron end member Fe$_2$SiO$_4$. Interestingly, these works plus others 
like \cite{dorschner1995steps} show a strong trend whereby the 20-to-10~$\mu$m 
band ratio increases with increasing iron content. This suggests that perhaps 
the amorphous host (matrix) component Mg$_{0.8}$Fe$_{1.2}$SiO$_4$ contains too 
much iron. On these bases we elected to use the optical constants of a Mg-rich 
amorphous silicate, specifically those of \cite{day1979mid} which gave a FWHM and 
$\lambda_{\rm P,max}$ more consistent with observations.

\begin{figure*}
\includegraphics[scale=0.625]{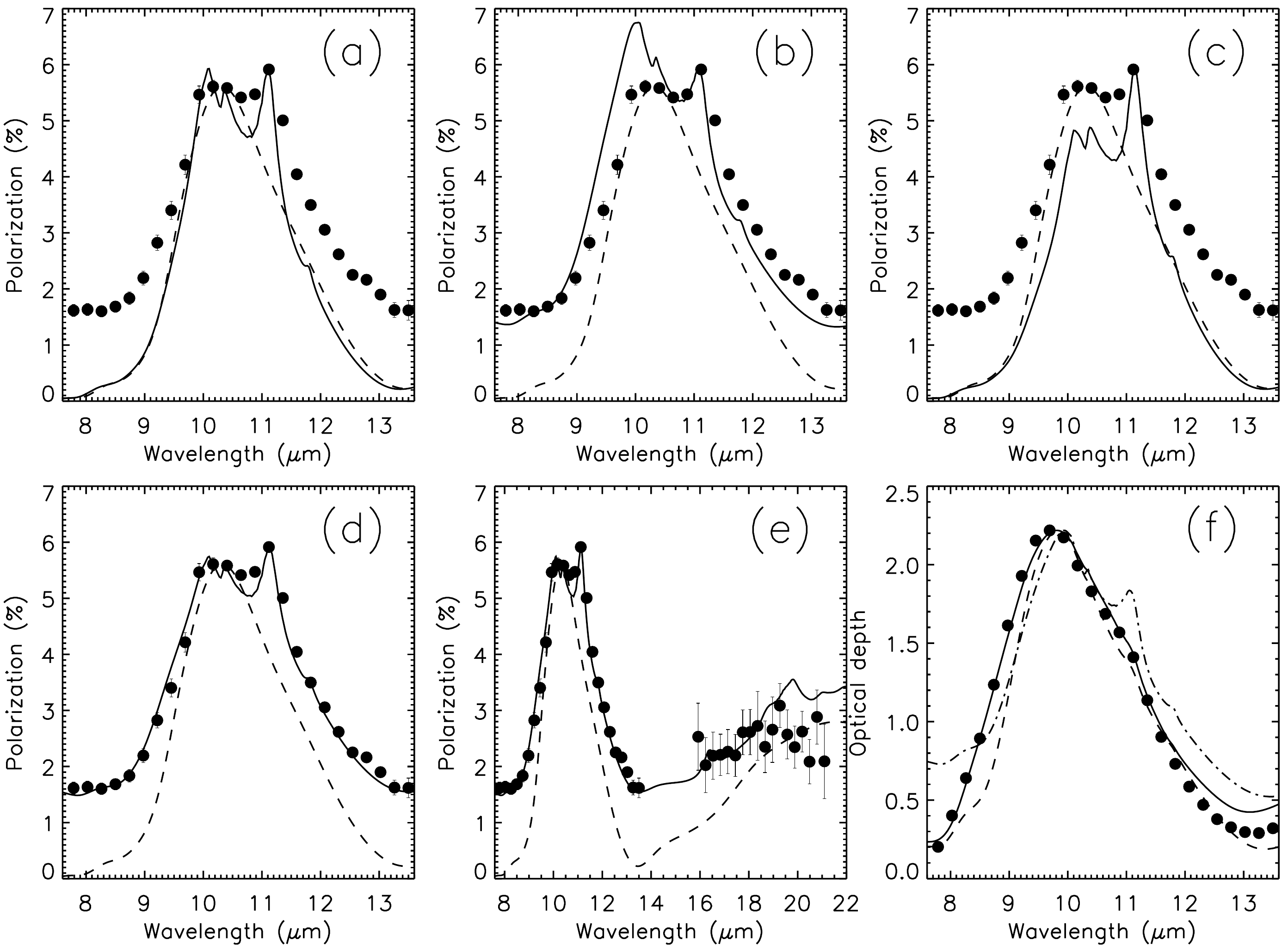}
\caption{a-e: Models for the polarisation of AFGL~2591, using EMT and a mixture 
of amorphous and crystalline silicates, and assuming the Rayleigh approximation.
The grains are assumed to be prolate with a principal axis ratio of 2:1. 
(a) amorphous+0.1$\times$(crystalline forsterite), 
(b) amorphous+0.1$\times$(crystalline forsterite)+0.24$\times$(graphitic carbon),
(c) amorphous+0.1$\times$(crystalline forsterite)+0.40$\times$vacuum, 
(d) amorphous+0.1$\times$(crystalline forsterite)+0.24$\times$(graphitic carbon)+
0.40$\times$vacuum, where the numbers refer to the volume fraction of that
particular component of the grain.
(e) Same as (d) but extended to the 20~$\mu$m band. In all cases the dashed line
is for the amorphous component only. 
(f) Models for the optical depth spectrum, where the dot-dashed curve is the 
absorption cross section for the same model in (d) and (e), whilst the solid 
curve has an additional but unpolarised amorphous component. Consistent with the 
work of \protect\cite{tamanai200610} these partially crystalline models have 
been shifted by $-0.1~\mu$m. See text for further details.}
\label{FIG-AFGL2591-Mods}
\end{figure*}

Figure~\ref{FIG-AFGL2591-Mods} shows the observed AFGL~2591 polarisation (a-e) 
and optical depth $\tau$ (f) spectra against calculations of the polarisation 
and absorption cross sections respectively. The optical depth has been extracted 
using the method outlined in Section 3.2, also used by \cite{fujiyoshi2015mid} 
for the Class~1 YSO SVS13. The peak optical depth and 8--13~$\mu$m colour 
temperature of around 2.3 and 385~K agree reasonably well with the best-fit 
optically-thick two-component model in \cite{smith2000studies}.

With our selection of optical constants a reasonable match to the polarisation 
could be found using a mildly prolate-shaped grain with a crystalline silicate 
inclusion volume fraction of 0.10 (Figure~\ref{FIG-AFGL2591-Mods}-a), and in 
the interests of clarity we show only this model. We do not claim that this is 
the shape of the grains within the AFGL~2591 envelope, as there is certainly 
sufficient flexibility in the model's input parameters (i.e. shape, principal 
axis ratio, optical constants, EMT mixing rule, inclusion volume fraction) to 
provide an equivalently good match for oblate grains, or even a CDE.

For instance, a finding of prolate grains would need to be tempered by the fact 
that i) oblate grains better match the ISM 8--13~$\mu$m profile (Wright et al., 
in preparation), as well as the BN Object in the Orion molecular cloud 
(\citet{hildebrand1995shapes}; \citet{lee1985infrared}; \citet{draine1984optical}, 
and ii) perfectly aligned prolate grains are only half as efficient polarisers as 
oblate grains of the same principal axis ratio (\citet{hildebrand1995shapes}, 
\citet{hildebrand1988magnetic}. The latter potentially places a constraint on 
the grain alignment mechanism. 

On the other hand, \cite{greenberg1996true} surmise that prolate grains, of a 
weakly-constrained though preferred axial ratio of 3:1, better fit the 8--13~$\mu$m 
BN profile, though apparently only by inclusion of an as-yet unproven organic 
refractory mantle (e.g. see \citet{li2014spectropolarimetric}; 
\citet{chiar2006spectropolarimetry}; \citet{adamson1999spectropolarimetric}). 
They also argue that elongated (prolate-like) rather than flattened (oblate-like) 
grains are a more realistic end-product of clumping. Also, \cite{siebenmorgen2014dust}
suggest that prolate silicate grains -- with axial ratio around 2:1 -- are responsible 
for the ISM visible polarisation along 4 of the 5 lines-of-sight they studied. 

Probably the most robust conclusion we can make about the grain shape is that 
they do not need to be very highly elongated or flattened. Indeed, both mildly 
prolate and oblate shapes can adequately match the 'trapezoid-like' shape of 
the AFGL~2591 polarisation profile, as well as the peak of the 11.1~$\mu$m 
feature, though obviously the model profile in Figure~\ref{FIG-AFGL2591-Mods}-a 
is too narrow. For a principal axis ratio of 2:1 the prolate grain better matches 
the ratio of the two main peaks at around 10.1 and 11.1~$\mu$m. But this ratio 
is shape-dependent, and for the same crystalline fraction of 0.10 a 6:1 oblate 
grain provides an equivalent match. For a 2:1 oblate grain the crystalline 
fraction must be increased to around 0.15 to adequately match the peak ratio 
at 10.1 and 11.1~$\mu$m.

The narrowness of the amorphous+crystalline silicate model in 
Figure~\ref{FIG-AFGL2591-Mods}-a has already been mentioned, and no amount of 
fiddling with the input parameters can broaden it and/or provide the polarisation 
observed in the 'wings' at $\leq$~8.5~$\mu$m and $\geq$~13~$\mu$m. One or more 
extra sources of emissivity are required. To identify what these could be we 
appeal to the well known and established growth-by-coagulation scenario of dust 
in molecular clouds (e.g. \citet{bromley2014challenges}; \citet{stognienko1995optical}). 
In one variant of this scheme proposed by \cite{mathis1989porous} the most 
abundant species of solid state matter -- i.e. sub-$\mu$m silicate and carbon 
particles -- stick together when they collide to form composite particles, upon 
which mantles of ices and/or hydrocarbons or organics can form. Further collisions 
continue to grow the particle, the interior of which may then contain many voids. 
Such a picture of porous (or fluffy) dust has essentially become a standard feature 
of recent cosmic dust models, capable of explaining many phenomena 
(e.g. \citet{voshchinnikov2006dust}) and perhaps finding direct support in 
observations of at least some interplanetary dust particles 
(IDPs; \citet{bradley2010astromineralogy}). 

In any model for the AFGL~2591 polarisation we can effectively rule out the 
presence of ices since the 3.1~$\mu$m water ice band is unpolarised 
(\citet{holloway2002spectropolarimetry}; \citet{hough1989spectropolarimetry}; 
\citet{dyck1980ice}; \citet{kobayashi1980narrow}). But other major 
components of the dust to be considered would be carbon, vacuum and possibly
metallic iron. None have a spectral feature in the 8--13~$\mu$m region, but instead
would form a 'continuum'. Since most other models, such as those cited above,
use silicate, carbon and vacuum, and to minimise as much as possible the number 
of free parameters, we also confine our model to these three components. Also,
the fact that metallic iron is featureless suggests it would only be a proxy for one 
or both of the also featureless carbon and vacuum components. 

The refractive indices of the carbon component are taken from \cite{jager1998optical} 
for their sample synthesized by pyrolizing cellulose material at 1000$^{\circ}$~C, 
which they describe as an ordered graphitic substance. Panels (a) to (d) of 
Figure~\ref{FIG-AFGL2591-Mods} thus shows the sequence of how such a 
multicomponent dust model is built up to match the 8--13~$\mu$m portion of 
the AFGL~2591 polarisation spectrum (solid lines), whilst panel (e) includes 
the 16-21~$\mu$m data (\citet{aitken1988infrared}; see also \citet{smith2000studies}).  
Dashed lines are for the case of purely amorphous Mg$_2$SiO$_4$ silicate with 
the same grain shape parameters. 

In constructing these models the volume fractions of carbon and vacuum are 0.24 
and 0.40 respectively. Such relatively large figures may begin to push the validity 
of EMT, so we have used the Bruggeman rule for the multi-component grain, due in 
large part to its inherent symmetry with respect to interchanging the components. 
This rule was also preferred by \cite{sokolik1999aerosol} in their modelling of 
multi-mineral aerosol aggregates. It has been shown by various authors to be more 
robust for large volume fractions of inclusions and/or a relatively large contrast 
between the refractive indices of the host and inclusion materials 
(\citet{ossenkopf1992silicates}; \citet{ossenkopf1991emt}; 
\citet{stroud1975generalized}). It has provided a good approximation to more exact 
methods like the Discrete Dipole Approximation (DDA) to calculate the optical 
properties of small particles (\citet{voshchinnikov2007emt}; \citet{perrin1990emt}), 
and even mid-IR experiments on aerosol particles with mineral compositions (e.g. 
silica, corundum, haematite, anhydrite; \citet{ruan2011aerosol}). In calculating 
such an EMT we have again used the work of \cite{sihvola1994anisotropic}, who 
presents a generalised formula encompassing several of the known mixing rules, 
including Maxwell-Garnett, Bruggemann (Polder-van Santen) and Coherent Potential.
Further, he provides an iterative solution for the effective medium of a 
multi-component mixture, negating the need to solve a cubic, quartic etc. polynomial 
for 2, 3, etc. different inclusion materials. 

In the context of the relative volume fractions, the figure of 0.40 for 
vacuum is within the likely range postulated by \cite{mathis1998silicates} 
and \cite{mathis1996dustmodels} of $\geq$ 0.25 -- based on abundance 
constraints -- and $\leq$ 0.60 -- based on the width of the visible 
interstellar polarisation curve. For the silicate and carbon components 
\cite{mathis1996dustmodels} states for his model that silicates (or Fe/Mg/Si 
oxides) comprise 77\% of the mass of the composite silicate+carbon+vacuum 
grains, with the other 23\% obviously comprising the various forms of 
carbon. Since the volume fraction of vacuum is 0.45 in his model, and
using densities of respectively 3.3 and 2.0 g~cm$^{-3}$ for silicate and
carbon, then their inferred volume fractions are $\sim$ 0.20 and 0.35. 
Thus, the volume fractions of carbon and vacuum in our AFGL~2591 polarisation 
model are in pretty good agreement with those of ISM dust from 
\cite{mathis1996dustmodels}. 

The final model in Figure~\ref{FIG-AFGL2591-Mods}-d provides quite a good match 
to the entire 8--13~$\mu$m spectrum, inclusive of its overall trapezoid shape, 
FWHM, peak ratios at 10.1 and 11.1~$\mu$m, and the short and long wavelength 
'wings'. This is apart from the region between the two peaks at around 
10.5--10.8~$\mu$m, likely a result of the laboratory and cosmic crystalline 
forsterites (unsurprisingly) not being exact analogues.

The model and observed 20-to-10~$\mu$m polarisation ratios in 
Figure~\ref{FIG-AFGL2591-Mods}-e are also in reasonable agreement. As noted by 
\cite{smith2000studies}, AFGL~2591 has the lowest such ratio amongst the 4 objects 
for which the absorptive component is well constrained (from a total of 6 with 
data in both windows). That it is also unique amongst those objects in 
having an 11.1~$\mu$m polarisation feature suggests a mineralogical relation 
between the two phenomena, possibly the high Mg/Fe ratio proposed here. 

Having said that, the 19--22~$\mu$m portion of the spectrum is problematic, 
especially in the sense of the data not showing a distinct feature at about 
19.5~$\mu$m. We have not found a model, nor can currently suggest a 
mineralogical explanation, to produce the sharp 11.1~$\mu$m feature but be 
essentially structureless in the 20~$\mu$m band. However, we note the 
relatively poor S/N at 20~$\mu$m compared to 10~$\mu$m, which results from 
the extreme difficulty in conducting spectropolarimetric observations in 
the 20~$\mu$m atmospheric window, with its multitude of telluric water vapour 
bands. New observations on a 10-m class telescope with a dual-beam instrument, 
such as CanariCam (\citet{telesco2003canaricam}), would be advantageous.

Our model also predicts quite prominent secondary polarisation peaks at around 
10.4 and 11.9~$\mu$m. Although we are very confident in our identification, 
observational confirmation of these would seal the interpretation of crystalline 
fortsteritic olivine silicate in at least AFGL~2591. This was attempted by 
\cite{wright2005mid} using Michelle on UKIRT, but the S/N was inadequate. Again, 
CanariCam on the Gran Telescopio Canarias would provide the best -- currently 
only -- opportunity to detect these features.

\subsubsection{Separate grain populations of different mineralogy?}

As seen in the dot-dash line in panel (f) of Figure~\ref{FIG-AFGL2591-Mods} the 
model which nicely matches the AFGL~2591 polarisation cannot match the optical 
depth spectrum. This is unlike the case of SVS13 in \cite{fujiyoshi2015mid} where
the two observational quantities are well represented by the same model. For 
AFGL~2591 there is clearly a large discrepancy in the amounts of crystalline 
forsterite required to account for the absorption feature in the conventional 
spectrum of AFGL~2591 and the associated polarisation signature, i.e. volume
fractions of around 1\% and 10\% respectively. Further, the amorphous forsterite 
is too narrow to account for the bulk of the AFGL~2591 absorption (dashed line). 
We hypothesise that these discrepancies are due to an unpolarised and purely 
amorphous (or much lower crystallinity) component to the absorption. The solid 
line of Figure~\ref{FIG-AFGL2591-Mods}-f shows such a model for the same grain 
shape parameters as in a-e, and where the purely amorphous component contributes 
about twice as much as the partially crystalline one. For this purely amorphous 
component we used the olivine Mg$_{0.8}$Fe$_{1.2}$SiO$_4$ of \cite{dorschner1995steps} 
for reasons already stated above.

If correct this would constitute evidence of distinct populations of grains with 
different mineralogies, although we cannot say which lies closer to the star, or 
indeed whether they are mixed within the same region. An argument against the 
latter scenario is that one population is aligned and the other not, although it 
is feasible that one might couple to the magnetic field more effectively than the
other (e.g. if the partially crystalline and Mg-rich silicate grains have a 
component of free iron). Whatever the case, for a dual-population scenario the 
'apparent' crystallinity as determined from the conventional spectrum is an 
'average' over both populations and/or along the line-of-sight, and is obviously 
lower than the 'real' crystallinity within a localised region. 

This also raises an interesting quandary when AFGL~2591 is compared with the 
few tens of other polarised YSOs in the atlas of \cite{smith2000studies}, including 
W3~IRS5, AFGL~2136 and AFGL~2789 here, as well as SgrA~IRS3. The 11.1~$\mu$m 
absorption feature in their conventional flux spectra is very similar to that 
of AFGL~2591, yet their polarisation shows no evidence of an associated 
signature. So could it be that the estimated crystalline fractions in these 
objects are in fact lower limits to what might actually exist? Perhaps a 'pocket' 
of higher crystallinity dust exists along their lines-of-sight, but which is 
either not aligned, or has a lower abundance, and thus contributes little or 
nothing to the polarisation. Or is it instead that the disk and/or envelope 
of AFGL~2591 (along with IRAS13481-6124 and perhaps AFGL~4176) contains 
dust that is more highly processed than most other embedded YSOs, perhaps from 
an annealing episode as postulated by \cite{aitken1988infrared}? 

\begin{figure}
\includegraphics[scale=0.415]{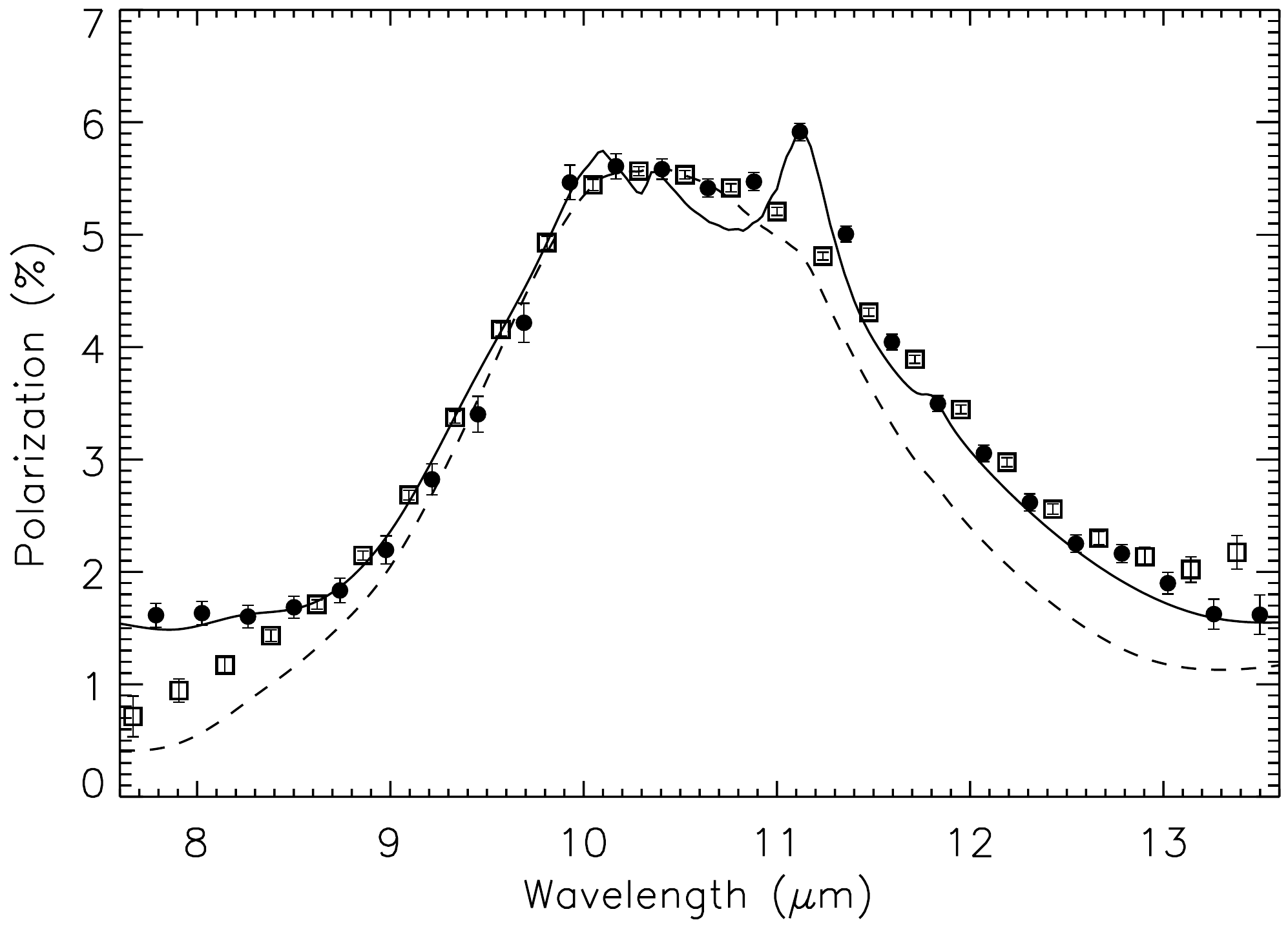}
\caption{The solid line shows the same model as in Figure~\ref{FIG-AFGL2591-Mods}-d 
and -e, whilst the dashed line includes a component of polarisation from purely 
amorphous silicate which contributes about twice as much as the partially crystalline
component. The open squares are the data for the BN Object in Orion 
(e.g. \protect\cite{smith2000studies}, \protect\cite{aitken198910}). Curiously, the 
BN data and the dashed line bear at least a qualitative similarity to each other in
the vicinity of 11~$\mu$m, namely a steepening of the profile, which may hint at a
previously unknown crystalline fraction in BN. Indeed, its conventional spectrum in
the mid-IR interferometric study of \protect\cite{boley2013midir} shows an 11~$\mu$m 
feature on some baselines, and our own larger data set shows it too (Do Duy et al., 
in preparation).}
\label{FIG-AFGL2591-Mod813um}
\end{figure}

Given the similarity of the sources in terms of their age, mass and that they 
drive an outflow, plus the almost ubiquitous presence of the 11.1~$\mu$m absorption 
band (as shown here and in Do Duy et al., in preparation), it is tempting to favour
the first scenario. Figure~\ref{FIG-AFGL2591-Mod813um} shows such an example, where
the same model is used as in Figure~\ref{FIG-AFGL2591-Mods}-f for the optical
depth, but now the amorphous component is polarised. The 11.1~$\mu$m polarisation 
feature is almost indistinguishable, and only a minor tweak to the relative 
contributions of the partially crystalline and amorphous components would make it 
disappear altogether, consistent with its non-detection in most other targets.
See the Figure~\ref{FIG-AFGL2591-Mod813um} caption for further details.
Whilst a promising explanation, a real answer to the questions posed above must 
await higher sensitivity and resolution data on more sources, hopefully to be 
obtained with CanariCam in the near future.

\subsection{Speculating on implications for the cosmic dust life-cycle}

We have presented a strong case for the widespread existence of crystalline 
silicates -- comprising up to a few or even several percent of the total 
silicate content -- in various cold astrophysical environments. Thus we can 
begin to speculate on the implications this might have for cosmic dust 
evolution. This is particularly so for the interstellar medium, represented 
here by SgrA~IRS3. 

\subsubsection{YSO envelopes or disks}

The detection of the 11.1~$\mu$m absorption band in all of our Class~1 YSOs, along 
with one or more other bands at 11.85, 23.5 and 28~$\mu$m, suggests that the hitherto 
couple of reported cases of the Class~0 YSO HOPS-68 by \cite{poteet2011spitzer} and
Class~I YSO SVS13 of \cite{fujiyoshi2015mid} are not unique or in any way peculiar. 
Hence the silicate crystallisation phase must either i) begin much earlier 
than generally accepted during YSO evolution, i.e. if not before then during the 
embedded Class~0--I rather than the protoplanetary Class~II--III phase, or ii) the 
dust originally deposited in the parent molecular cloud from the surrounding ISM was 
already partially crystalline. 

If the former case is true then there are some interesting questions which could be 
posed. For instance, is a crystallisation process still 'required' to occur in the 
disks of T Tauri and Herbig stars, as reviewed by \cite{henning2010cosmic}? If not, 
then is a radial mixing or some other transport process still needed to explain the 
existence of crystalline silicates in comets, which are thought to have formed in 
the cold, outer regions of our own protoplanetary disk? Or if a crystallisation 
process is still 'required', will it be more efficient, and/or act over a shorter 
time, if the starting point already contains crystalline seeds, from which the 
process may propagate?

On the other hand, if the latter case is true does this then raise a question over 
the efficiency of the amorphisation and/or destruction processes proposed in the 
ISM? Or if not, then how do ISM grains, i.e. those formed in situ in the ISM, 
condense and/or accrete as partially crystalline?

Answers to these questions, if indeed they become necessary, will only be possible 
once we have a firmer basis on which to conclude that ISM silicate dust does indeed 
contain a crystalline component. In the following subsections we examine the evidence 
for partially crystalline grains along the path to the Galactic Centre, and discuss 
some implications.

\subsubsection{Path to the Galactic Centre}

Along the extended path toward the Galactic Centre, passing though spiral arms of 
the Galaxy, there are multiple molecular and diffuse clouds. Of the approximately 
30 magnitudes of visual extinction typically quoted as an 'average' toward the 
Galactic Centre, \cite{whittet1997infrared} suggest that about 10 magnitudes occurs 
in molecular clouds and the remainder in diffuse clouds, where grains would 
probably be with and without ice mantles respectively. 

Before beginning specific speculation a caveat is in order, namely that we do 
not know with precision how the silicate extinction -- let alone that within the 
11.1~$\mu$m absorption band -- is distributed along the line-of-sight to SgrA~IRS3. 
Indeed, it is not even known with certainty where IRS3 is located with respect to 
the dynamical centre of our Galaxy (\citet{goto2008absorption}). 

We discuss IRS3's location in Appendix A, but for now assert that at least half 
of the total silicate absorption, including that of the 11.1~$\mu$m band, occurs 
in the ISM and is not merely local to IRS3. This is supported by the presence of 
the band in the 8--13~$\mu$m spectra of IRS1, IRS7 and IRS10 obtained with 
Subaru/COMICS and shown in \cite{okada2003mid}, although they did not discuss it. 
It is also supported by our own mid-IR spectroscopy of the Galactic Centre 
Quintuplet cluster (Do Duy et al. in preparation), as well as ISO observations of 
several other positions toward SgrA. As previously noted some of the latter data 
has already been presented by various authors, but here and in Appendix A we show 
additional data as well as provide a new analysis of the previously published 
results. Details can be found in Appendix A but here we simply quote the important 
findings.

The 11.1~$\mu$m feature can be seen in the spectrum of a position several 
arcseconds south of IRS3, within the E-W bar of the SgrA mini-spiral and 
which we call IRSX, obtained from the same Gemini Michelle observation as 
SgrA~IRS3 (Figure~\ref{IRS3X-ISOstar}-a and -c). This is similar to the 
position at which the "diffuse bar" spectrum was obtained by \cite{okada2003mid}. 
Whilst the 11.1~$\mu$m optical depth of IRSX is about half that 
of IRS3, the IRSX amorphous silicate optical depth is also roughly half 
that for IRS3. So the relative strengths of the 11.1~$\mu$m band are 
similar. Further, despite the assertion of \cite{kemper2004absence} the 
11.1~$\mu$m feature can indeed be seen in both ISO spectra centred near 
SgrA$^*$, again with a similar $\tau_{11.1}$/$\tau_{9.7}$ as for IRS3 and
IRSX. Even more telling for a crystalline silicate identification is that 
in the same ISO spectra an 11.85~$\mu$m feature is also detected (see 
Figure~\ref{IRS3X-ISOstar}-b and -d). 

Similar to the case of the embedded YSOs, a 23.5~$\mu$m feature is seen in 
the SgrA$^*$ spectrum, as well as in spectra taken at positions offset 
approximately 41 and 45 arcseconds to the SSW and NNE respectively, within 
the so-called circumnuclear disk or ring (CND/R). The same holds for a 28~$\mu$m 
band. See Figure~\ref{ISO-SgrA-20to45um}-a and -b. The two features are even 
more prominent in the spectrum of the so-called Pistol star in 
Figure~\ref{ISO-SgrA-20to45um}-c. 

All of the above leaves us in little doubt that crystalline silicates are
present in one or more clouds in the ISM along the line-of-sight toward the
Galactic Centre. Averaged along the line-of-sight the crystalline component
comprises up to a few percent of the total silicate content, as judged from the 
crude model for the 11.1~$\mu$m band of SgrA~IRS3 in Figure~\ref{FIG-Obs-XMods}.

However, the precise abundance is not very well constrained, due mostly to 
uncertainties in the dust properties. These stem from both the morphology of 
the grains, e.g. bare versus mantled silicate, as well as the optical properties
of the silicate (especially the crystalline component) and the mantle constituent(s).
For instance, \cite{li2007crystallinity} note that the original crystalline 
upper limit of 2.2\% reported by \cite{kemper2005erratum} must be raised to 
around 5\% if a water ice mantle coats the grains (assuming a mantle-to-core 
volume ration of $\sim$ 0.55). 

In addition to our rough estimate of the crystalline silicate fraction toward
SgrA~IRS3 we can try to obtain an independent estimate from its mid-IR polarisation 
spectrum presented in \cite{aitken1986infrared} and \cite{smith2000studies}. 
Unlike AFGL~2591 and IRAS13481-6124, but like most other YSOs, an 11.1~$\mu$m 
polarisation signature has not been detected despite the very clear absorption 
band in Figure~\ref{FIG-Spectra}. It is interesting to see if an upper limit 
from the polarisation agrees with the estimate from the conventional spectrum.

\begin{figure}
\includegraphics[scale=0.47]{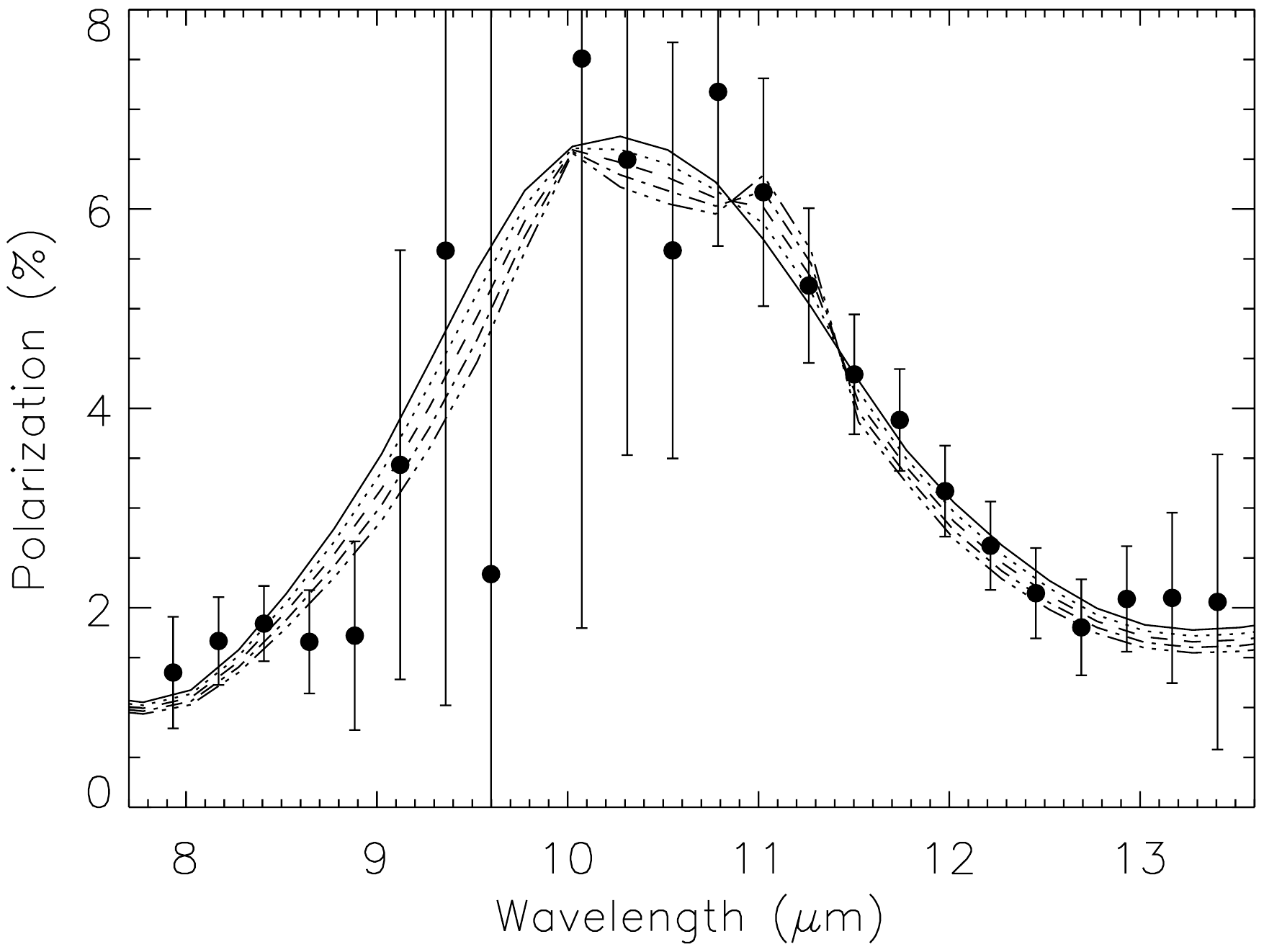}
\caption{Mid-IR polarisation spectrum of SgrA~IRS3, along with representative 
models. Note how between about 10.5 and 11.5~$\mu$m the model polarisation is 
firstly quite flat and then steepens. Neither property is seen in the data that 
can be reasonably trusted to accurately represent the true polarisation profile, 
i.e. from about 10.8~$\mu$m onwards.}
\label{FIG-SgrAIRS3-Mod813um}
\end{figure}

Figure~\ref{FIG-SgrAIRS3-Mod813um} shows the 8--13~$\mu$m polarisation spectrum
of SgrA IRS3, similar to that shown in \cite{smith2000studies}, along with several
models. Given the poor S/N in the middle of the band the models have been normalised
to the average polarisation within the 10.7--12.0~$\mu$m interval. The basis of the
models is the olivine silicate Mg$_{0.8}$Fe$_{1.2}$SiO$_4$ from 
\cite{dorschner1995steps}. However, this was too narrow compared to the data, not 
being able to fit the short and long wavelength 'wings' of the band, let alone 
produce the level of polarisation observed from 1.4--4.2~$\mu$m along the very 
nearby IRS7 sightline by \cite{adamson1999spectropolarimetric} and 
\cite{nagata1994spectropolarimetry}. Therefore, similar to the case of 
AFGL~2591, and consistent with the interstellar dust life-cycle suggested by 
\cite{jones2013evolution}, inclusions of graphitic carbon (optical constants 
from \citet{jager1998optical}) and vacuum were added using EMT, with volume 
fractions of 0.125 and 0.20 respectively. This model is shown as the solid line, 
which assumes oblate 2:1 grains (though the grain shape is unconstrained given 
the wavelength of peak polarisation is not constrained by the data). Having 
also been binned to match the approximate resolution (or sampling) of the 
data of about 0.25~$\mu$m, this model provides an acceptable match to the 
spectrum.

The dashed line and three subsequent others are for models with crystalline silicate 
inclusions with volume fractions of 0.025, 0.05, 0.075 and 0.10, using the optical 
constant of \cite{fabian2001steps}. Within the error bars we can not rule out 
any crystalline silicate fraction in this range. However, despite the poor S/N, given 
the monotonic decrease of the polarisation from about 10.8~$\mu$m onwards we believe 
that if a crystalline silicate fraction of more than about 5\% were present then it 
would have been detected. Whilst noting that this refers only to the aligned dust 
population, whereas the conventional spectrum samples all dust, at least the two 
estimates of the crystalline silicate fraction are consistent. Having said that, 
the quality of the polarisation data is inadequate, and it is hoped that CanariCam 
on the GTC can produce a more stringent constraint.

In the interests of completeness we also examined the effect SiC has on the 
polarisation, since \cite{min2007shape} infers that between 2.6 and 4.2\% of the 
dust grain mass is SiC. If this is implemented into our model then the volume
fraction would be between about 2 and 3\%. Using SiC optical constants from
\cite{laor1993spectroscopic}, preferred by \cite{min2007shape}, or those of
\cite{pegourie1988optical}, then such a volume fraction could still be consistent
with the observations. On the other hand, if the optical constants of 
\cite{pitman2008optical} or \cite{palik1985hocs} are assumed then a volume
fraction of 2--3\% significantly perturbs the polarisation spectrum, and would 
probably have been detected. 

\subsubsection{Crystalline silicates in the ISM}

It is widely believed that silicates in the ISM are either completely amorphous or 
have such a low crystalline fraction as to be undetectable (\citet{henning2010cosmic}), 
to the point where this has essentially become a basic tenet of the cosmic dust life 
cycle (e.g. \citet{jones2013evolution}). One version of this cycle asserts that the 
silicate dust ejected into the ISM by AGB stars is destroyed through processes like 
collisional fragmentation and sputtering on a timescale shorter than the ISM residence 
time. This residence time, defined as the period between when a grain is injected into 
the ISM from its parent evolved star outflow to when it is consumed in a new star, 
is generally agreed to be a few Gyr (\citet{slavin2015destruction}; 
\citet{jones2011dust}; \citet{draine2009asp}). On the other hand, the destruction 
time scale is typically thought to be around a factor of 10 lower 
(\citet{draine2009asp}, \citet{mouri2000survival}), so that something like 
$\sim$ 95\% of the silicate dust observed in the ISM was formed in the ISM. 

Precisely how silicates form in the ISM is not well understood (\citet{jones2011dust}), 
but may involve accretion of condensible atoms onto the surface of pre-existing 
particles (\citet{draine2009asp}). \cite{voshchinnikov2010accretion} claim to have
found evidence for such accretion. Such ISM silicate grains are expected to be completely 
amorphous. More recent estimates of the silicate destruction timescale 
have increased it by a factor of several, e.g. \cite{slavin2015destruction}, 
alleviating somewhat the requirement to form silicates in the ISM but not completely 
eliminating it.

Whatever the case, accepting that only 5\% of silicate dust in the ISM is stardust, 
and that at most 15\% of such stardust is crystalline (\citet{henning2010cosmic}), 
then we may expect to observe a crystalline fraction of $\leq$ 0.75\%. However, this 
does not take account of the relatively rapid amorphisation processes which would 
probably act on the dust. As reviewed by \cite{henning2010cosmic}, the approximately 
15\% crystalline mass fraction of silicate dust ejected into the ISM by AGB stars 
is likely to be amorphised through energetic particle collisions on timescales of 
tens of Myr, much shorter than both the destruction and residence timescales 
(\citet{bringa2007energetic}; \citet{jager2003steps}). 

Thus we would expect any surviving silicate stardust in the ISM to be completely 
amorphous, and any silicate dust grown in the ISM to also be completely amourphous. 
Obviously this is consistent with the previous upper limits of a few to several 
percent for the path to the Galactic Centre (\citet{li2007crystallinity};
\cite{min2007shape}; \citet{kemper2005erratum}). But it is inconsistent with 
the detection of ISM crystalline silicate reported here of up to a few percent 
for essentially the same path, i.e. to SgrA~IRS3. So detection of crystalline 
silicates toward SgrA~IRS3 and other Galactic Centre positions has a potentially 
very significant bearing on one or more of: i) the lifetime of stellar-produced 
silicate grains in the ISM, ii) amorphisation of stellar-produced silicates in 
the ISM, and/or iii) the growth process of silicate grains in the ISM.

Of course a caveat here is that, despite multiple sightlines toward the GC 
containing crystalline silicates, this could still be regarded as a special case. 
They are all along a somewhat confined path within our Galaxy that intersects 
many of the same clouds, as judged for instance by the similar velocities of 
absorption components of several molecular tracers (e.g. see discussion in 
Appendix A for references). A sensitive survey of other sightlines is clearly 
required, which we have begun and a few examples of which will be presented in 
our paper describing the whole sample (Do Duy et al., in preparation). 

Notably however, even if the GC ISM path is unique and/or the crystalline silicates 
occur within a particular cloud (e.g. within a few hundred parsecs of the GC itself), 
our results still have important implications for the cosmic dust life cycle. Any 
amount of crystallinity in these environments would be unexpected.

\section{Conclusions}

The major observational result of this paper is that an absorption feature at 
11.1~$\mu$m exists in the spectra of eight embedded Class~1 YSOs -- AFGL~2136, 
W3~IRS5, AFGL~2789, AFGL~2591, AFGL~4176, IRAS13481-6124, W28~A2 and 
IRAS19110+1045 -- and the path through the ISM to the Galactic Centre sources 
SgrA~IRS3 and a position we call (for convenience) SgrA~IRSX. Curiously, this 
feature has escaped widespread attention. This is despite it being hinted at 
in previous observations of multiple YSOs and the ISM at lower spectral 
and/or spatial resolution, and actually resolved in a Class~0 YSO by 
\cite{poteet2011spitzer}. It even appears in the mid-IR interferometric
study of many DEYSOs of \cite{boley2013midir}, including three studied 
here -- AFGL~2136, AFGL~4176 and IRAS13481-6124 -- where in some cases it 
also shows a possible spatial (baseline) dependence. To date the feature 
has not been studied in the detail provided here. 

The wavelength of this feature, and its spectral profile, is similar to that 
observed toward the OH/IR stars and dust factories AFGL~2403, OH26.5+0.6 and
AFGL~230, known to be sources of crystalline silicates. Also, other potential 
carriers for the feature in AFGL~2789 -- such as water ice, hydrocarbons and 
carbonates -- can be excluded to varying degrees of confidence (high, medium 
and high respectively) based on the lack of expected accompanying features at 
shorter wavelengths. A model of separate populations of sub-micron amorphous 
olivine and crystalline forsterite grains match very well the observed 
AFGL~2789 spectrum. Further, models of magnesium-rich (forsteritic) crystalline 
olivine mixed with amorphous olivine -- to represent a partially crystalline 
grain structure as an 'effective medium' -- match very well the central 
wavelength and spectral profile. However, this is only the case if the work 
of \cite{tamanai200610} is utilised, which shows an $\sim$~0.2~$\mu$m shift 
of the primary features to shorter wavelengths. 

Detection of a polarisation signature in the sources AFGL~2591 and IRAS13481-6124, 
and possibly AFGL~4176, suggests a carrier with a relatively high band strength, 
and this too is consistent with a crystalline silicate interpretation. The implied
volume fraction from an EMT model to the AFGL~2591 polarisation is around 10\%, 
larger than the few percent that would be inferred based solely on the absorption 
feature in the conventional spectrum of this as well as the other YSO and Galactic 
centre targets. This suggests that 'pockets' of relatively high crystallinity 
material could be hidden in at least some of the other sources, only revealed 
if the dust is aligned.   

Using the ISO Data Archive we also examined the 20-45~$\mu$m spectra of these 
and other sources. Doing so allowed us to identify features that support our
contention that the 11.1~$\mu$m feature is carried by crystalline silicate. 
For instance, we find evidence for features at 11.85, 23.5 and 28~$\mu$m in
absorption toward several of our YSO targets, as well as toward multiple 
positions within the Galactic Centre. A 33.6~$\mu$m absorption band is 
however conspicuously absent in all these spectra. But such a feature is 
detected in emission, in conjunction with another at 36~$\mu$m, in the spectra 
of several HII regions. This provides further confidence in our assessment 
that crystalline silicates are commonly, if not ubiquitously, present in the 
very early stages of star formation. 

The seemingly common existence of crystalline silicates at any abundance level 
in the ISM, and envelopes or disks of embedded YSOs, presents a challenge to models 
which attempt to explain the cosmic dust life cycle. These models use as a foundation 
that silicate dust is completely amorphous in the ISM and molecular clouds. There 
are certainly good physics-based reasons to believe this would be the case, given
the multitude of grain destruction and amorphisation processes proposed in the
literature. But observationally this appears to be no longer the situation. Of 
course, more observations are required, especially of truly diffuse ISM paths 
and ideally coupled with spectropolarimetry.

Finally, we have shown that ground-based mid-IR slit spectroscopy, at a moderate 
spectral resolution, and even of bright sources on 10~m-class telescopes, still has 
a part to play in the study of cosmic dust. We have a much larger sample of sources, 
most of which also show the 11.1~$\mu$m absorption band, e.g. well known YSOs like 
S140~IRS1, S255 IRS1 and NGC7538~IRS1, and ISM sightlines to the Galactic Centre 
Quintuplet members. The entire sample is currently under study, is producing some
interesting and surprising results, and will be presented and discussed in a 
forthcoming paper (Do Duy et al., in preparation).

\section*{Acknowledgements}

Based on observations obtained at the Gemini Observatory, which is operated by the 
Association of Universities for Research in Astronomy, Inc., under a cooperative 
agreement with the NSF on behalf of the Gemini partnership: the National Science 
Foundation (United States), the National Research Council (Canada), CONICYT (Chile), 
the Australian Research Council (Australia), Minist\'{e}rio da Ci\^{e}ncia, 
Tecnologia e Inova\c{c}\~{a}o (Brazil) and Ministerio de Ciencia, Tecnolog\'{i}a e 
Innovaci\'{o}n Productiva (Argentina).
Based on observations with ISO, an ESA project with instruments funded by ESA Member 
States (especially the PI countries: France, Germany, the Netherlands and the United 
Kingdom) and with the participation of ISAS and NASA. The version of the ISO data 
presented in this paper correspond to the Highly Processed Data Product (HPDP) set 
called "`High resolution processed and defringed SWS01s"' by W. Frieswijk et al. 
CMW acknowledges financial support from an Australian Research Council Future 
Fellowship FT100100495 and previously DP0345227. Tho Do Duy acknowledges financial 
support from a UNSW Canberra Research Training Scholarship and a UNSW Canberra 
Tuition Fee Scholarship. We thank David Aitken, Craig Smith and Patrick Roche for
obtaining much of the UCLS data used to complement the work here, and the 
reviewers for well considered and thorough reviews of both this work and an earlier
version.




\bibliographystyle{mnras}
\bibliography{CMWright_MidIR-Crystalline_references}

\begin{thebibliography}{}
\makeatletter
\relax
\def\mn@urlcharsother{\let\do\@makeother \do\$\do\&\do\#\do\^\do\_\do\%\do\~}
\def\mn@doi{\begingroup\mn@urlcharsother \@ifnextchar [ {\mn@doi@}
  {\mn@doi@[]}}
\def\mn@doi@[#1]#2{\def\@tempa{#1}\ifx\@tempa\@empty \href
  {http://dx.doi.org/#2} {doi:#2}\else \href {http://dx.doi.org/#2} {#1}\fi
  \endgroup}
\def\mn@eprint#1#2{\mn@eprint@#1:#2::\@nil}
\def\mn@eprint@arXiv#1{\href {http://arxiv.org/abs/#1} {{\tt arXiv:#1}}}
\def\mn@eprint@dblp#1{\href {http://dblp.uni-trier.de/rec/bibtex/#1.xml}
  {dblp:#1}}
\def\mn@eprint@#1:#2:#3:#4\@nil{\def\@tempa {#1}\def\@tempb {#2}\def\@tempc
  {#3}\ifx \@tempc \@empty \let \@tempc \@tempb \let \@tempb \@tempa \fi \ifx
  \@tempb \@empty \def\@tempb {arXiv}\fi \@ifundefined
  {mn@eprint@\@tempb}{\@tempb:\@tempc}{\expandafter \expandafter \csname
  mn@eprint@\@tempb\endcsname \expandafter{\@tempc}}}

\bibitem[\protect\citeauthoryear{Adamson, Whittet, Chrysostomou, Hough, Aitken,
  Wright  \& Roche}{Adamson et~al.}{1999}]{adamson1999spectropolarimetric}
Adamson A.~J.,  Whittet D.~C.~B.,  Chrysostomou A.,  Hough J.~H.,  Aitken
  D.~K.,  Wright G.~S.,   Roche P.~F.,  1999, \apj, 512, 224

\bibitem[\protect\citeauthoryear{Aitken}{Aitken}{1989}]{aitken1989spectropolarimetry}
Aitken D.~K.,  1989, in {B{\"o}hm-Vitense} E.,  ed.,  ESA Special Publication
  Vol. 290, Infrared Spectroscopy in Astronomy. pp 99--107

\bibitem[\protect\citeauthoryear{Aitken, Roche, Bailey, Briggs, Hough
  et~al.}{Aitken et~al.}{1986}]{aitken1986infrared}
Aitken D.~K.,  Roche P.~F.,  Bailey J.~A.,  Briggs G.~P.,  Hough J.~H.,
  et~al., 1986, \mnras, 218, 363

\bibitem[\protect\citeauthoryear{Aitken, Roche, Smith, James  \& Hough}{Aitken
  et~al.}{1988}]{aitken1988infrared}
Aitken D.~K.,  Roche P.~F.,  Smith C.~H.,  James S.~D.,   Hough J.~H.,  1988,
  \mnras, 230, 629

\bibitem[\protect\citeauthoryear{Aitken, Smith  \& Roche}{Aitken
  et~al.}{1989}]{aitken198910}
Aitken D.~K.,  Smith C.~H.,   Roche P.~F.,  1989, \mnras, 236, 919

\bibitem[\protect\citeauthoryear{{Aitken}, {Moore}, {Roche}, {Smith}  \&
  {Wright}}{{Aitken} et~al.}{1993}]{aitken1993bpic}
{Aitken} D.~K.,  {Moore} T.~J.~T.,  {Roche} P.~F.,  {Smith} C.~H.,   {Wright}
  C.~M.,  1993, \mnras, \href
  {http://adsabs.harvard.edu/abs/1993MNRAS.265L..41A} {265, L41}

\bibitem[\protect\citeauthoryear{Alexander, Casali, Andr{\'e}, Persi  \&
  Eiroa}{Alexander et~al.}{2003}]{alexander2003isocam}
Alexander R.~D.,  Casali M.~M.,  Andr{\'e} P.,  Persi P.,   Eiroa C.,  2003,
  \aap, 401, 613

\bibitem[\protect\citeauthoryear{Allamandola, Sandford, Tielens  \&
  Herbst}{Allamandola et~al.}{1992}]{allamandola1992infrared}
Allamandola L.~J.,  Sandford S.~A.,  Tielens A.~G.~G.~M.,   Herbst T.~M.,
  1992, \apj, 399, 134

\bibitem[\protect\citeauthoryear{Allen \& Burton}{Allen \&
  Burton}{1993}]{allen1993explosive}
Allen D.~A.,  Burton M.~G.,  1993, Nature, 363, 54

\bibitem[\protect\citeauthoryear{Becklin, Matthews, Neugebauer  \&
  Willner}{Becklin et~al.}{1978}]{becklin1978infrared}
Becklin E.~E.,  Matthews K.,  Neugebauer G.,   Willner S.~P.,  1978, \apj, 219,
  121

\bibitem[\protect\citeauthoryear{Beckwith, Persson, Neugebauer  \&
  Becklin}{Beckwith et~al.}{1978}]{beckwith1978observations}
Beckwith S.,  Persson S.~E.,  Neugebauer G.,   Becklin E.~E.,  1978, \apj, 223,
  464

\bibitem[\protect\citeauthoryear{Bernstein, Sandford  \& Allamandola}{Bernstein
  et~al.}{2005}]{bernstein2005mid}
Bernstein M.~P.,  Sandford S.~A.,   Allamandola L.~J.,  2005, \apj Supplement
  Series, 161, 53

\bibitem[\protect\citeauthoryear{Bohren \& Huffman}{Bohren \&
  Huffman}{1983}]{bohren1983light}
Bohren C.~F.,  Huffman D.~R.,  1983, Light Scattering and Absorption by Small
  Particles

\bibitem[\protect\citeauthoryear{{Boley} et~al.,}{{Boley}
  et~al.}{2013}]{boley2013midir}
{Boley} P.~A.,  et~al., 2013, \mn@doi [\aap] {10.1051/0004-6361/201321539},
  \href {http://adsabs.harvard.edu/abs/2013A%26A...558A..24B} {558, A24}

\bibitem[\protect\citeauthoryear{Boogert et~al.,}{Boogert
  et~al.}{2004}]{boogert2004spitzer}
Boogert A.~C.~A.,  et~al., 2004, \apj Supplement Series, 154, 359

\bibitem[\protect\citeauthoryear{Boogert et~al.,}{Boogert
  et~al.}{2008}]{boogert2008c2d}
Boogert A.~C.~A.,  et~al., 2008, \apj, 678, 985

\bibitem[\protect\citeauthoryear{Boogert, Gerakines  \& Whittet}{Boogert
  et~al.}{2015}]{boogert2015observations}
Boogert A.~C.~A.,  Gerakines P.,   Whittet D.,  2015, \araa, 53, 541

\bibitem[\protect\citeauthoryear{Bowey, Adamson  \& Yates}{Bowey
  et~al.}{2003}]{bowey2003galactic}
Bowey J.~E.,  Adamson A.~J.,   Yates J.~A.,  2003, \mnras, 340, 1173

\bibitem[\protect\citeauthoryear{{Bradley}}{{Bradley}}{2010}]{bradley2010astromineralogy}
{Bradley} J.~P.,  2010, in {Henning} T.,  ed.,  Lecture Notes in Physics,
  Berlin Springer Verlag Vol. 815, Lecture Notes in Physics, Berlin Springer
  Verlag. pp 259--276, \mn@doi{10.1007/978-3-642-13259-9_6}

\bibitem[\protect\citeauthoryear{Bradley, Humecki  \& Germani}{Bradley
  et~al.}{1992}]{bradley1992combined}
Bradley J.~P.,  Humecki H.~J.,   Germani M.~S.,  1992, \apj, 394, 643

\bibitem[\protect\citeauthoryear{{Bregman}, {Witteborn}, {Allamandola},
  {Campins}, {Wooden}, {Rank}, {Cohen}  \& {Tielens}}{{Bregman}
  et~al.}{1987}]{bregman1987halley}
{Bregman} J.~D.,  {Witteborn} F.~C.,  {Allamandola} L.~J.,  {Campins} H.,
  {Wooden} D.~H.,  {Rank} D.~M.,  {Cohen} M.,   {Tielens} A.~G.~G.~M.,  1987,
  \aap, \href {http://adsabs.harvard.edu/abs/1987A%26A...187..616B} {187, 616}

\bibitem[\protect\citeauthoryear{Bregman, Hayward  \& Sloan}{Bregman
  et~al.}{2000}]{bregman2000discovery}
Bregman J.~D.,  Hayward T.~L.,   Sloan G.~C.,  2000, \apj Letters, 544, L75

\bibitem[\protect\citeauthoryear{Bringa et~al.,}{Bringa
  et~al.}{2007}]{bringa2007energetic}
Bringa E.~M.,  et~al., 2007, \apj, 662, 372

\bibitem[\protect\citeauthoryear{Bromley, Goumans, Herbst, Jones  \&
  Slater}{Bromley et~al.}{2014}]{bromley2014challenges}
Bromley S.~T.,  Goumans T.~P.~M.,  Herbst E.,  Jones A.~P.,   Slater B.,  2014,
  Physical Chemistry Chemical Physics, 16, 18623

\bibitem[\protect\citeauthoryear{Brooke, Sellgren  \& Smith}{Brooke
  et~al.}{1996}]{brooke1996study}
Brooke T.~Y.,  Sellgren K.,   Smith R.~G.,  1996, \apj, 459, 209

\bibitem[\protect\citeauthoryear{Brooke, Sellgren  \& Geballe}{Brooke
  et~al.}{1999}]{brooke1999new}
Brooke T.~Y.,  Sellgren K.,   Geballe T.~R.,  1999, \apj, 517, 883

\bibitem[\protect\citeauthoryear{{Calvet}, {Magris}, {Patino}  \&
  {D'Alessio}}{{Calvet} et~al.}{1992}]{calvet1992rtdisk}
{Calvet} N.,  {Magris} G.~C.,  {Patino} A.,   {D'Alessio} P.,  1992, \rmxaa,
  \href {http://adsabs.harvard.edu/abs/1992RMxAA..24...27C} {24, 27}

\bibitem[\protect\citeauthoryear{Campins \& Ryan}{Campins \&
  Ryan}{1989}]{campins1989identification}
Campins H.,  Ryan E.~V.,  1989, \apj, 341, 1059

\bibitem[\protect\citeauthoryear{Ceccarelli, Caux, Tielens, Kemper, Waters  \&
  Phillips}{Ceccarelli et~al.}{2002}]{ceccarelli2002discovery}
Ceccarelli C.,  Caux E.,  Tielens A.~G.~G.~M.,  Kemper F.,  Waters L.~B.~F.~M.,
    Phillips T.,  2002, \aap, 395, L29

\bibitem[\protect\citeauthoryear{Cesarsky, Jones, Lequeux  \&
  Verstraete}{Cesarsky et~al.}{2000}]{cesarsky2000silicate}
Cesarsky D.,  Jones A.~P.,  Lequeux J.,   Verstraete L.,  2000, \aap, 358, 708

\bibitem[\protect\citeauthoryear{{Chiang} \& {Goldreich}}{{Chiang} \&
  {Goldreich}}{1997}]{chiang1997rtdisk}
{Chiang} E.~I.,  {Goldreich} P.,  1997, \apj, \href
  {http://adsabs.harvard.edu/abs/1997ApJ...490..368C} {490, 368}

\bibitem[\protect\citeauthoryear{Chiar, Tielens, Whittet, Schutte, Boogert,
  Lutz, Van~Dishoeck  \& Bernstein}{Chiar et~al.}{2000}]{chiar2000composition}
Chiar J.~E.,  Tielens A.~G.~G.~M.,  Whittet D.~C.~B.,  Schutte W.~A.,  Boogert
  A.~C.~A.,  Lutz D.,  Van~Dishoeck E.~F.,   Bernstein M.~P.,  2000, \apj, 537,
  749

\bibitem[\protect\citeauthoryear{Chiar, Adamson, Pendleton, Whittet, Caldwell
  \& Gibb}{Chiar et~al.}{2002}]{chiar2002hydrocarbons}
Chiar J.~E.,  Adamson A.~J.,  Pendleton Y.~J.,  Whittet D.~C.~B.,  Caldwell
  D.~A.,   Gibb E.~L.,  2002, \apj, 570, 198

\bibitem[\protect\citeauthoryear{Chiar et~al.,}{Chiar
  et~al.}{2006}]{chiar2006spectropolarimetry}
Chiar J.~E.,  et~al., 2006, \apj, 651, 268

\bibitem[\protect\citeauthoryear{Chiar, Tielens, Adamson  \& Ricca}{Chiar
  et~al.}{2013}]{chiar2013structure}
Chiar J.~E.,  Tielens A.~G.~G.~M.,  Adamson A.~J.,   Ricca A.,  2013, \apj,
  770, 78

\bibitem[\protect\citeauthoryear{Cl{\'e}ment, Mutschke, Klein  \&
  Henning}{Cl{\'e}ment et~al.}{2003}]{clement2003new}
Cl{\'e}ment D.,  Mutschke H.,  Klein R.,   Henning T.,  2003, \apj, 594, 642

\bibitem[\protect\citeauthoryear{Cox}{Cox}{1989}]{cox1989line}
Cox P.,  1989, \aap, 225, L1

\bibitem[\protect\citeauthoryear{Dartois \& d'Hendecourt}{Dartois \&
  d'Hendecourt}{2001}]{dartois2001search}
Dartois E.,  d'Hendecourt L.,  2001, \aap, 365, 144

\bibitem[\protect\citeauthoryear{Dartois et~al.,}{Dartois
  et~al.}{1998}]{dartois1998detection}
Dartois E.,  et~al., 1998, \aap, 338, L21

\bibitem[\protect\citeauthoryear{Dartois, d'Hendecourt, Thi, Pontoppidan  \&
  Van~Dishoeck}{Dartois et~al.}{2002}]{dartois2002combined}
Dartois E.,  d'Hendecourt L.,  Thi W.,  Pontoppidan K.~M.,   Van~Dishoeck
  E.~F.,  2002, \aap, 394, 1057

\bibitem[\protect\citeauthoryear{Day}{Day}{1979}]{day1979mid}
Day K.~L.,  1979, \apj, 234, 158

\bibitem[\protect\citeauthoryear{Day}{Day}{1981}]{day1981infrared}
Day K.~L.,  1981, \apj, 246, 110

\bibitem[\protect\citeauthoryear{Demyk, Jones, Dartois, Cox  \&
  d'Hendecourt}{Demyk et~al.}{1999}]{demyk1999chemical}
Demyk K.,  Jones A.~P.,  Dartois E.,  Cox P.,   d'Hendecourt L.,  1999, \aap,
  349, 267

\bibitem[\protect\citeauthoryear{Demyk, Dartois, Wiesemeyer, Jones  \&
  d'Hendecourt}{Demyk et~al.}{2000}]{demyk2000structure}
Demyk K.,  Dartois E.,  Wiesemeyer H.,  Jones A.~P.,   d'Hendecourt L.,  2000,
  \aap, 364, 170

\bibitem[\protect\citeauthoryear{Dorschner, Begemann, Henning, Jaeger  \&
  Mutschke}{Dorschner et~al.}{1995}]{dorschner1995steps}
Dorschner J.,  Begemann B.,  Henning T.,  Jaeger C.,   Mutschke H.,  1995,
  \aap, 300, 503

\bibitem[\protect\citeauthoryear{{Draine}}{{Draine}}{2003a}]{draine2003interstellar}
{Draine} B.~T.,  2003a, \araa, 41, 241

\bibitem[\protect\citeauthoryear{Draine}{Draine}{2003b}]{draine2003scattering}
Draine B.~T.,  2003b, \apj, 598, 1026

\bibitem[\protect\citeauthoryear{Draine}{Draine}{2009}]{draine2009asp}
Draine B.~T.,  2009

\bibitem[\protect\citeauthoryear{Draine \& Allaf-Akbari}{Draine \&
  Allaf-Akbari}{2006}]{draine2006x}
Draine B.~T.,  Allaf-Akbari K.,  2006, \apj, 652, 1318

\bibitem[\protect\citeauthoryear{Draine \& Lee}{Draine \&
  Lee}{1984}]{draine1984optical}
Draine B.~T.,  Lee H.~M.,  1984, \apj, 285, 89

\bibitem[\protect\citeauthoryear{Dyck \& Lonsdale}{Dyck \&
  Lonsdale}{1980}]{dyck1980ice}
Dyck H.~M.,  Lonsdale C.~J.,  1980, The Astronomical Journal, 85, 1077

\bibitem[\protect\citeauthoryear{Fabian, Henning, J{\"a}ger, Mutschke,
  Dorschner  \& Wehrhan}{Fabian et~al.}{2001}]{fabian2001steps}
Fabian D.,  Henning T.,  J{\"a}ger C.,  Mutschke H.,  Dorschner J.,   Wehrhan
  O.,  2001, \aap, 378, 228

\bibitem[\protect\citeauthoryear{{Fritz} et~al.,}{{Fritz}
  et~al.}{2011}]{fritz2011line}
{Fritz} T.~K.,  et~al., 2011, \mn@doi [\apj] {10.1088/0004-637X/737/2/73},
  \href {http://adsabs.harvard.edu/abs/2011ApJ...737...73F} {737, 73}

\bibitem[\protect\citeauthoryear{Fujiyoshi, Wright  \& Moore}{Fujiyoshi
  et~al.}{2015}]{fujiyoshi2015mid}
Fujiyoshi T.,  Wright C.~M.,   Moore T. J.~T.,  2015, \mnras, 451, 3371

\bibitem[\protect\citeauthoryear{Geballe \& Oka}{Geballe \&
  Oka}{1989}]{geballe1989infrared}
Geballe T.~R.,  Oka T.,  1989, \apj, 342, 855

\bibitem[\protect\citeauthoryear{{Gibb} et~al.,}{{Gibb}
  et~al.}{2000}]{gibb2000w33a}
{Gibb} E.~L.,  et~al., 2000, \mn@doi [\apj] {10.1086/308940}, \href
  {http://adsabs.harvard.edu/abs/2000ApJ...536..347G} {536, 347}

\bibitem[\protect\citeauthoryear{Gibb, Whittet  \& Chiar}{Gibb
  et~al.}{2001}]{gibb2001searching}
Gibb E.~L.,  Whittet D.~C.~B.,   Chiar J.~E.,  2001, \apj, 558, 702

\bibitem[\protect\citeauthoryear{Gibb, Whittet, Boogert  \& Tielens}{Gibb
  et~al.}{2004}]{gibb2004interstellar}
Gibb E.~L.,  Whittet D.~C.~B.,  Boogert A.~C.~A.,   Tielens A.~G.~G.~M.,  2004,
  \apj Supplement Series, 151, 35

\bibitem[\protect\citeauthoryear{Gielen, Van~Winckel, Waters, Min  \&
  Dominik}{Gielen et~al.}{2007}]{gielen2007dust}
Gielen C.,  Van~Winckel H.,  Waters L.~B.~F.~M.,  Min M.,   Dominik C.,  2007,
  \aap, 475, 629

\bibitem[\protect\citeauthoryear{Gillett, Forrest, Merrill, Soifer  \&
  Capps}{Gillett et~al.}{1975}]{gillett19758}
Gillett F.~C.,  Forrest W.~J.,  Merrill K.~M.,  Soifer B.~T.,   Capps R.~W.,
  1975, \apj, 200, 609

\bibitem[\protect\citeauthoryear{{Glass}, {Matsumoto}, {Carter}  \&
  {Sekiguchi}}{{Glass} et~al.}{2001}]{glass2001ohir}
{Glass} I.~S.,  {Matsumoto} S.,  {Carter} B.~S.,   {Sekiguchi} K.,  2001,
  \mn@doi [\mnras] {10.1046/j.1365-8711.2001.03971.x}, \href
  {http://adsabs.harvard.edu/abs/2001MNRAS.321...77G} {321, 77}

\bibitem[\protect\citeauthoryear{{Glasse}, {Atad-Ettedgui}  \&
  {Harris}}{{Glasse} et~al.}{1997}]{glasse1997michelle}
{Glasse} A.~C.,  {Atad-Ettedgui} E.~I.,   {Harris} J.~W.,  1997, in {Ardeberg}
  A.~L.,  ed.,  Society of Photo-Optical Instrumentation Engineers (SPIE)
  Conference Series Vol. 2871, Optical Telescopes of Today and Tomorrow. pp
  1197--1203

\bibitem[\protect\citeauthoryear{Goto, McCall, Geballe, Usuda, Kobayashi,
  Terada  \& Oka}{Goto et~al.}{2002}]{goto2002absorption}
Goto M.,  McCall B.~J.,  Geballe T.~R.,  Usuda T.,  Kobayashi N.,  Terada H.,
  Oka T.,  2002, Publications of the Astronomical Society of Japan, 54, 951

\bibitem[\protect\citeauthoryear{Goto et~al.,}{Goto
  et~al.}{2008}]{goto2008absorption}
Goto M.,  et~al., 2008, \apj, 688, 306

\bibitem[\protect\citeauthoryear{Goto, Geballe, Indriolo, Yusef-Zadeh, Usuda,
  Henning  \& Oka}{Goto et~al.}{2014}]{goto2014infrared}
Goto M.,  Geballe T.~R.,  Indriolo N.,  Yusef-Zadeh F.,  Usuda T.,  Henning T.,
    Oka T.,  2014, \apj, 786, 96

\bibitem[\protect\citeauthoryear{Greenberg \& Li}{Greenberg \&
  Li}{1996}]{greenberg1996true}
Greenberg J.~M.,  Li A.,  1996, \aap, 309, 258

\bibitem[\protect\citeauthoryear{Hanner, Brooke  \& Tokunaga}{Hanner
  et~al.}{1995}]{hanner199510}
Hanner M.~S.,  Brooke T.~Y.,   Tokunaga A.~T.,  1995, \apj, 438, 250

\bibitem[\protect\citeauthoryear{Hanner, Brooke  \& Tokunaga}{Hanner
  et~al.}{1998}]{hanner19988}
Hanner M.~S.,  Brooke T.~Y.,   Tokunaga A.~T.,  1998, \apj, 502, 871

\bibitem[\protect\citeauthoryear{{Helmich} et~al.,}{{Helmich}
  et~al.}{1996}]{helmich1996hotwater}
{Helmich} F.~P.,  et~al., 1996, \aap, \href
  {http://adsabs.harvard.edu/abs/1996A%26A...315L.173H} {315, L173}

\bibitem[\protect\citeauthoryear{Henning}{Henning}{2010}]{henning2010cosmic}
Henning T.,  2010, \araa, 48, 21

\bibitem[\protect\citeauthoryear{Henning \& Stognienko}{Henning \&
  Stognienko}{1993}]{henning1993porous}
Henning T.,  Stognienko R.,  1993, \aap, 280, 609

\bibitem[\protect\citeauthoryear{{Herman} \& {Habing}}{{Herman} \&
  {Habing}}{1985}]{herman1985ohir}
{Herman} J.,  {Habing} H.~J.,  1985, \physrep, \href
  {http://adsabs.harvard.edu/abs/1985PhR...124..257H} {124, 257}

\bibitem[\protect\citeauthoryear{Hildebrand}{Hildebrand}{1988}]{hildebrand1988magnetic}
Hildebrand R.~H.,  1988, \qjras, 29, 327

\bibitem[\protect\citeauthoryear{Hildebrand \& Dragovan}{Hildebrand \&
  Dragovan}{1995}]{hildebrand1995shapes}
Hildebrand R.~H.,  Dragovan M.,  1995, \apj, 450, 663

\bibitem[\protect\citeauthoryear{Holloway, Chrysostomou, Aitken, Hough  \&
  McCall}{Holloway et~al.}{2002}]{holloway2002spectropolarimetry}
Holloway R.~P.,  Chrysostomou A.,  Aitken D.~K.,  Hough J.~H.,   McCall A.,
  2002, \mnras, 336, 425

\bibitem[\protect\citeauthoryear{Hough, Whittet, Sato, Yamashita, Tamura,
  Nagata, Aitken  \& Roche}{Hough et~al.}{1989}]{hough1989spectropolarimetry}
Hough J.~H.,  Whittet D.~C.~B.,  Sato S.,  Yamashita T.,  Tamura M.,  Nagata
  T.,  Aitken D.~K.,   Roche P.~F.,  1989, \mnras, 241, 71

\bibitem[\protect\citeauthoryear{Jager, Mutschke  \& Henning}{Jager
  et~al.}{1998a}]{jager1998optical}
Jager C.,  Mutschke H.,   Henning T.,  1998a, \aap, 332, 291

\bibitem[\protect\citeauthoryear{J\"{a}ger, Molster, Dorschner, Henning,
  Mutschke  \& Waters}{J\"{a}ger et~al.}{1998b}]{jager1998steps}
J\"{a}ger C.,  Molster F.~J.,  Dorschner J.,  Henning T.,  Mutschke H.,
  Waters L.~B.~F.~M.,  1998b, \aap, 339, 904

\bibitem[\protect\citeauthoryear{J{\"a}ger, Dorschner, Mutschke, Posch  \&
  Henning}{J{\"a}ger et~al.}{2003}]{jager2003steps}
J{\"a}ger C.,  Dorschner J.,  Mutschke H.,  Posch T.,   Henning T.,  2003,
  \aap, 408, 193

\bibitem[\protect\citeauthoryear{Jiang, Zhang, Li  \& Lisse}{Jiang
  et~al.}{2013}]{jiang2013crystalline}
Jiang B.~W.,  Zhang K.,  Li A.,   Lisse C.~M.,  2013, \apj, 765, 72

\bibitem[\protect\citeauthoryear{{Jim{\'e}nez-Esteban}, {Garc{\'{\i}}a-Lario},
  {Engels}  \& {Perea Calder{\'o}n}}{{Jim{\'e}nez-Esteban}
  et~al.}{2006}]{jiminez2006ohir}
{Jim{\'e}nez-Esteban} F.~M.,  {Garc{\'{\i}}a-Lario} P.,  {Engels} D.,   {Perea
  Calder{\'o}n} J.~V.,  2006, \mn@doi [\aap] {10.1051/0004-6361:20053268},
  \href {http://adsabs.harvard.edu/abs/2006A%26A...446..773J} {446, 773}

\bibitem[\protect\citeauthoryear{Jones}{Jones}{2014}]{jones2014physical}
Jones A.~P.,  2014, arXiv preprint arXiv:1411.6666

\bibitem[\protect\citeauthoryear{Jones \& Nuth}{Jones \&
  Nuth}{2011}]{jones2011dust}
Jones A.~P.,  Nuth J.~A.,  2011, \aap, 530, A44

\bibitem[\protect\citeauthoryear{{Jones}, {Frey}, {Verstraete}, {Cox}  \&
  {Demyk}}{{Jones} et~al.}{1999}]{jones1999infrared}
{Jones} A.~P.,  {Frey} V.,  {Verstraete} L.,  {Cox} P.,   {Demyk} K.,  1999, in
  {Cox} P.,  {Kessler} M.,  eds,  ESA Special Publication Vol. 427, The
  Universe as Seen by ISO. p.~679

\bibitem[\protect\citeauthoryear{Jones, Fanciullo, K{\"o}hler, Verstraete,
  Guillet, Bocchio  \& Ysard}{Jones et~al.}{2013}]{jones2013evolution}
Jones A.~P.,  Fanciullo L.,  K{\"o}hler M.,  Verstraete L.,  Guillet V.,
  Bocchio M.,   Ysard N.,  2013, \aap, 558, A62

\bibitem[\protect\citeauthoryear{Juh{\'a}sz et~al.,}{Juh{\'a}sz
  et~al.}{2010}]{juhasz2010dust}
Juh{\'a}sz A.,  et~al., 2010, \apj, 721, 431

\bibitem[\protect\citeauthoryear{Keane, Tielens, Boogert, Schutte  \&
  Whittet}{Keane et~al.}{2001}]{keane2001ice}
Keane J.~V.,  Tielens A.~G.~G.~M.,  Boogert A.~C.~A.,  Schutte W.~A.,   Whittet
  D.~C.~B.,  2001, \aap, 376, 254

\bibitem[\protect\citeauthoryear{Kemper, Molster, J\"{a}ger  \& Waters}{Kemper
  et~al.}{2002a}]{kemper2002mineral}
Kemper F.,  Molster F.~J.,  J\"{a}ger C.,   Waters L.~B.~F.~M.,  2002a, \aap,
  394, 679

\bibitem[\protect\citeauthoryear{Kemper, J\"{a}ger, Waters, Henning, Molster,
  Barlow, Lim  \& De~Koter}{Kemper et~al.}{2002b}]{kemper2002detection}
Kemper F.,  J\"{a}ger C.,  Waters L.~B.~F.~M.,  Henning T.,  Molster F.~J.,
  Barlow M.~J.,  Lim T.,   De~Koter A.,  2002b, Nature, 415, 295

\bibitem[\protect\citeauthoryear{Kemper, Vriend  \& Tielens}{Kemper
  et~al.}{2004}]{kemper2004absence}
Kemper F.,  Vriend W.~J.,   Tielens A.~G.~G.~M.,  2004, \apj, 609, 826

\bibitem[\protect\citeauthoryear{Kemper, Vriend  \& Tielens}{Kemper
  et~al.}{2005}]{kemper2005erratum}
Kemper F.,  Vriend W.~J.,   Tielens A.~G.~G.~M.,  2005, \apj, 633, 534

\bibitem[\protect\citeauthoryear{Kessler-Silacci, Hillenbrand, Blake  \&
  Meyer}{Kessler-Silacci et~al.}{2005}]{kessler20058}
Kessler-Silacci J.~E.,  Hillenbrand L.~A.,  Blake G.~A.,   Meyer M.~R.,  2005,
  \apj, 622, 404

\bibitem[\protect\citeauthoryear{Knacke, Fajardo-Acosta, Telesco, Hackwell,
  Lynch  \& Russell}{Knacke et~al.}{1993}]{knacke1993silicates}
Knacke R.~F.,  Fajardo-Acosta S.~B.,  Telesco C.~M.,  Hackwell J.~A.,  Lynch
  D.~K.,   Russell R.~W.,  1993, \apj, 418, 440

\bibitem[\protect\citeauthoryear{Kobayashi, Kawara, Sato  \& Okuda}{Kobayashi
  et~al.}{1980}]{kobayashi1980narrow}
Kobayashi Y.,  Kawara K.,  Sato S.,   Okuda H.,  1980, Publications of the
  Astronomical Society of Japan, 32, 295

\bibitem[\protect\citeauthoryear{Koike, Chihara, Tsuchiyama, Suto, Sogawa  \&
  Okuda}{Koike et~al.}{2003}]{koike2003compositional}
Koike C.,  Chihara H.,  Tsuchiyama A.,  Suto H.,  Sogawa H.,   Okuda H.,  2003,
  \aap, 399, 1101

\bibitem[\protect\citeauthoryear{{Lacy}, {Faraji}, {Sandford}  \&
  {Allamandola}}{{Lacy} et~al.}{1998}]{lacy1998ammonia}
{Lacy} J.~H.,  {Faraji} H.,  {Sandford} S.~A.,   {Allamandola} L.~J.,  1998,
  \mn@doi [\apjl] {10.1086/311452}, \href
  {http://adsabs.harvard.edu/abs/1998ApJ...501L.105L} {501, L105}

\bibitem[\protect\citeauthoryear{Lane \& Christensen}{Lane \&
  Christensen}{1997}]{lane1997thermal}
Lane M.~D.,  Christensen P.~R.,  1997, \jgr, 102, 25581

\bibitem[\protect\citeauthoryear{Laor \& Draine}{Laor \&
  Draine}{1993}]{laor1993spectroscopic}
Laor A.,  Draine B.~T.,  1993, \apj, 402, 441

\bibitem[\protect\citeauthoryear{Lazarian}{Lazarian}{2007}]{lazarian2007tracing}
Lazarian A.,  2007, \jqsrt, 106, 225

\bibitem[\protect\citeauthoryear{Lee \& Draine}{Lee \&
  Draine}{1985}]{lee1985infrared}
Lee H.~M.,  Draine B.~T.,  1985, \apj, 290, 211

\bibitem[\protect\citeauthoryear{{Leech} et~al.,}{{Leech}
  et~al.}{2003}]{leech2003iso}
{Leech} K.,  et~al., 2003, {The ISO Handbook, Volume V - SWS - The Short
  Wavelength Spectrometer}

\bibitem[\protect\citeauthoryear{Li, Zhao  \& Li}{Li
  et~al.}{2007}]{li2007crystallinity}
Li M.~P.,  Zhao G.,   Li A.,  2007, \mnras, 382, L26

\bibitem[\protect\citeauthoryear{Li, Liang  \& Li}{Li
  et~al.}{2014}]{li2014spectropolarimetric}
Li Q.,  Liang S.~L.,   Li A.,  2014, \mnras, 440, L56

\bibitem[\protect\citeauthoryear{Lutz et~al.,}{Lutz et~al.}{1996}]{lutz1996sws}
Lutz D.,  et~al., 1996, \aap, 315, L269

\bibitem[\protect\citeauthoryear{Maldoni, Smith, Robinson  \& Rookyard}{Maldoni
  et~al.}{1998}]{maldoni1998study}
Maldoni M.~M.,  Smith R.~G.,  Robinson G.,   Rookyard V.~L.,  1998, \mnras,
  298, 251

\bibitem[\protect\citeauthoryear{{Maldoni}, {Egan}, {Smith}, {Robinson}  \&
  {Wright}}{{Maldoni} et~al.}{2003}]{maldoni2003ohir}
{Maldoni} M.~M.,  {Egan} M.~P.,  {Smith} R.~G.,  {Robinson} G.,   {Wright}
  C.~M.,  2003, \mn@doi [\mnras] {10.1046/j.1365-8711.2003.07013.x}, \href
  {http://adsabs.harvard.edu/abs/2003MNRAS.345..912M} {345, 912}

\bibitem[\protect\citeauthoryear{{Maldoni}, {Egan}, {Robinson}, {Smith}  \&
  {Wright}}{{Maldoni} et~al.}{2004}]{maldoni2004ohir}
{Maldoni} M.~M.,  {Egan} M.~P.,  {Robinson} G.,  {Smith} R.~G.,   {Wright}
  C.~M.,  2004, \mn@doi [\mnras] {10.1111/j.1365-2966.2004.07532.x}, \href
  {http://adsabs.harvard.edu/abs/2004MNRAS.349..665M} {349, 665}

\bibitem[\protect\citeauthoryear{{Maldoni}, {Ireland}, {Smith}  \&
  {Robinson}}{{Maldoni} et~al.}{2005}]{maldoni2005ohir}
{Maldoni} M.~M.,  {Ireland} T.~R.,  {Smith} R.~G.,   {Robinson} G.,  2005,
  \mn@doi [\mnras] {10.1111/j.1365-2966.2005.09314.x}, \href
  {http://adsabs.harvard.edu/abs/2005MNRAS.362..872M} {362, 872}

\bibitem[\protect\citeauthoryear{Malfait, Waelkens, Waters, Vandenbussche,
  Huygen  \& De~Graauw}{Malfait et~al.}{1998}]{malfait1998spectrum}
Malfait K.,  Waelkens C.,  Waters L.~B.~F.~M.,  Vandenbussche B.,  Huygen E.,
  De~Graauw M.~S.,  1998, Astronomy and astrophysics, 332, L25

\bibitem[\protect\citeauthoryear{Martin}{Martin}{1975}]{martin1975some}
Martin P.~G.,  1975, \apj, 202, 393

\bibitem[\protect\citeauthoryear{Mastrapa, Sandford, Roush, Cruikshank  \&
  Dalle~Ore}{Mastrapa et~al.}{2009}]{mastrapa2009optical}
Mastrapa R.~M.,  Sandford S.~A.,  Roush T.~L.,  Cruikshank D.~P.,   Dalle~Ore
  C.~M.,  2009, \apj, 701, 1347

\bibitem[\protect\citeauthoryear{{Mathis}}{{Mathis}}{1996}]{mathis1996dustmodels}
{Mathis} J.~S.,  1996, \mn@doi [\apj] {10.1086/178094}, \href
  {http://adsabs.harvard.edu/abs/1996ApJ...472..643M} {472, 643}

\bibitem[\protect\citeauthoryear{{Mathis}}{{Mathis}}{1998}]{mathis1998silicates}
{Mathis} J.~S.,  1998, \mn@doi [\apj] {10.1086/305477}, \href
  {http://adsabs.harvard.edu/abs/1998ApJ...497..824M} {497, 824}

\bibitem[\protect\citeauthoryear{{Mathis} \& {Whiffen}}{{Mathis} \&
  {Whiffen}}{1989}]{mathis1989porous}
{Mathis} J.~S.,  {Whiffen} G.,  1989, \mn@doi [\apj] {10.1086/167538}, \href
  {http://adsabs.harvard.edu/abs/1989ApJ...341..808M} {341, 808}

\bibitem[\protect\citeauthoryear{{Mathis}, {Rumpl}  \& {Nordsieck}}{{Mathis}
  et~al.}{1977}]{mathis1977mrn}
{Mathis} J.~S.,  {Rumpl} W.,   {Nordsieck} K.~H.,  1977, \mn@doi [\apj]
  {10.1086/155591}, \href {http://adsabs.harvard.edu/abs/1977ApJ...217..425M}
  {217, 425}

\bibitem[\protect\citeauthoryear{Meeus, Waters, Bouwman, Van Den~Ancker,
  Waelkens  \& Malfait}{Meeus et~al.}{2001}]{meeus2001iso}
Meeus G.,  Waters L.~B.~F.~M.,  Bouwman J.,  Van Den~Ancker M.~E.,  Waelkens
  C.,   Malfait K.,  2001, \aap, 365, 476

\bibitem[\protect\citeauthoryear{Min, Hovenier  \& de Koter}{Min
  et~al.}{2003}]{min2003shape}
Min M.,  Hovenier J.~W.,   de Koter A.,  2003, \aap, 404, 35

\bibitem[\protect\citeauthoryear{Min, Waters, De~Koter, Hovenier, Keller  \&
  Markwick-Kemper}{Min et~al.}{2007}]{min2007shape}
Min M.,  Waters L.~B.~F.~M.,  De~Koter A.,  Hovenier J.~W.,  Keller L.~P.,
  Markwick-Kemper F.,  2007, \aap, 462, 667

\bibitem[\protect\citeauthoryear{{Molster} \& {Kemper}}{{Molster} \&
  {Kemper}}{2005}]{molster2005crystalline}
{Molster} F.~J.,  {Kemper} C.,  2005, \mn@doi [\ssr]
  {10.1007/s11214-005-8066-x}, \href
  {http://adsabs.harvard.edu/abs/2005SSRv..119....3M} {119, 3}

\bibitem[\protect\citeauthoryear{Molster, Waters, Tielens  \& Barlow}{Molster
  et~al.}{2002}]{molster2002crystalline}
Molster F.~J.,  Waters L.~B.~F.~M.,  Tielens A.~G.~G.~M.,   Barlow M.~J.,
  2002, \aap, 382, 184

\bibitem[\protect\citeauthoryear{Molster, Waters  \& Kemper}{Molster
  et~al.}{2010}]{molster2010mineralogy}
Molster F.~J.,  Waters L.~B.~F.~M.,   Kemper F.,  2010, in , Astromineralogy.
Springer, pp 143--201

\bibitem[\protect\citeauthoryear{Monnier, Tuthill, Ireland, Cohen, Tannirkulam
  \& Perrin}{Monnier et~al.}{2009}]{monnier2009mid}
Monnier J.~D.,  Tuthill P.~G.,  Ireland M.,  Cohen R.,  Tannirkulam A.,
  Perrin M.~D.,  2009, \apj, 700, 491

\bibitem[\protect\citeauthoryear{Moultaka, Eckart, Viehmann, Mouawad,
  Straubmeier, Ott  \& Sch{\"o}del}{Moultaka et~al.}{2004}]{moultaka2004dust}
Moultaka J.,  Eckart A.,  Viehmann T.,  Mouawad N.,  Straubmeier C.,  Ott T.,
  Sch{\"o}del R.,  2004, \aap, 425, 529

\bibitem[\protect\citeauthoryear{Moultaka, Eckart, Sch{\"o}del, Viehmann  \&
  Najarro}{Moultaka et~al.}{2005}]{moultaka2005vlt}
Moultaka J.,  Eckart A.,  Sch{\"o}del R.,  Viehmann T.,   Najarro F.,  2005,
  \aap, 443, 163

\bibitem[\protect\citeauthoryear{{Mouri} \& {Taniguchi}}{{Mouri} \&
  {Taniguchi}}{2000}]{mouri2000survival}
{Mouri} H.,  {Taniguchi} Y.,  2000, \mn@doi [\apjl] {10.1086/312633}, \href
  {http://adsabs.harvard.edu/abs/2000ApJ...534L..63M} {534, L63}

\bibitem[\protect\citeauthoryear{Mukai \& Koike}{Mukai \&
  Koike}{1990}]{mukai1990optical}
Mukai T.,  Koike C.,  1990, Icarus, 87, 180

\bibitem[\protect\citeauthoryear{Nagata, Kobayashi  \& Sato}{Nagata
  et~al.}{1994}]{nagata1994spectropolarimetry}
Nagata T.,  Kobayashi N.,   Sato S.,  1994, \apj, 423, L113

\bibitem[\protect\citeauthoryear{O'Donnell}{O'Donnell}{1994}]{o1994effect}
O'Donnell J.~E.,  1994, \apj, 437, 262

\bibitem[\protect\citeauthoryear{Oka, Geballe, Goto, Usuda  \& McCall}{Oka
  et~al.}{2005}]{oka2005hot}
Oka T.,  Geballe T.~R.,  Goto M.,  Usuda T.,   McCall B.~J.,  2005, \apj, 632,
  882

\bibitem[\protect\citeauthoryear{{Okada} et~al.,}{{Okada}
  et~al.}{2003}]{okada2003mid}
{Okada} Y.,  et~al., 2003, \mn@doi [Astronomische Nachrichten Supplement]
  {10.1002/asna.200385079}, \href
  {http://adsabs.harvard.edu/abs/2003ANS...324..567O} {324, 567}

\bibitem[\protect\citeauthoryear{{Ossenkopf}}{{Ossenkopf}}{1991}]{ossenkopf1991emt}
{Ossenkopf} V.,  1991, \aap, \href
  {http://adsabs.harvard.edu/abs/1991A%26A...251..210O} {251, 210}

\bibitem[\protect\citeauthoryear{{Ossenkopf}, {Henning}  \&
  {Mathis}}{{Ossenkopf} et~al.}{1992}]{ossenkopf1992silicates}
{Ossenkopf} V.,  {Henning} T.,   {Mathis} J.~S.,  1992, \aap, \href
  {http://adsabs.harvard.edu/abs/1992A%26A...261..567O} {261, 567}

\bibitem[\protect\citeauthoryear{{Palik}}{{Palik}}{1985}]{palik1985hocs}
{Palik} E.~D.,  1985, {Handbook of optical constants of solids}

\bibitem[\protect\citeauthoryear{Peeters et~al.,}{Peeters
  et~al.}{2002}]{peeters2002iso}
Peeters E.,  et~al., 2002, \aap, 381, 571

\bibitem[\protect\citeauthoryear{{Peeters}, {Mart{\'{\i}}n-Hern{\'a}ndez},
  {Rodr{\'{\i}}guez-Fern{\'a}ndez}  \& {Tielens}}{{Peeters}
  et~al.}{2005}]{peeters2005high}
{Peeters} E.,  {Mart{\'{\i}}n-Hern{\'a}ndez} N.~L.,
  {Rodr{\'{\i}}guez-Fern{\'a}ndez} N.~J.,   {Tielens} X.,  2005, \mn@doi [\ssr]
  {10.1007/s11214-005-8070-1}, \href
  {http://adsabs.harvard.edu/abs/2005SSRv..119..273P} {119, 273}

\bibitem[\protect\citeauthoryear{P{\'e}gouri{\'e}}{P{\'e}gouri{\'e}}{1988}]{pegourie1988optical}
P{\'e}gouri{\'e} B.,  1988, \aap, 194, 335

\bibitem[\protect\citeauthoryear{Pendleton, Sandford, Allamandola, Tielens  \&
  Sellgren}{Pendleton et~al.}{1994}]{pendleton1994near}
Pendleton Y.~J.,  Sandford S.~A.,  Allamandola L.~J.,  Tielens A.~G.~G.~M.,
  Sellgren K.,  1994, \apj, 437, 683

\bibitem[\protect\citeauthoryear{{Perrin} \& {Lamy}}{{Perrin} \&
  {Lamy}}{1990}]{perrin1990emt}
{Perrin} J.-M.,  {Lamy} P.~L.,  1990, \mn@doi [\apj] {10.1086/169395}, \href
  {http://adsabs.harvard.edu/abs/1990ApJ...364..146P} {364, 146}

\bibitem[\protect\citeauthoryear{{Persi}, {Ferrari-Toniolo}  \&
  {Spinoglio}}{{Persi} et~al.}{1986}]{persi1986infrared}
{Persi} P.,  {Ferrari-Toniolo} M.,   {Spinoglio} L.,  1986, \aap, 157, 29

\bibitem[\protect\citeauthoryear{Pitman, Hofmeister, Corman  \& Speck}{Pitman
  et~al.}{2008}]{pitman2008optical}
Pitman K.~M.,  Hofmeister A.~M.,  Corman A.~B.,   Speck A.~K.,  2008, \aap,
  483, 661

\bibitem[\protect\citeauthoryear{Pitman, Dijkstra, Hofmeister  \& Speck}{Pitman
  et~al.}{2010}]{pitman2010infrared}
Pitman K.~M.,  Dijkstra C.,  Hofmeister A.~M.,   Speck A.~K.,  2010, \mnras,
  406, 460

\bibitem[\protect\citeauthoryear{Poteet et~al.,}{Poteet
  et~al.}{2011}]{poteet2011spitzer}
Poteet C.~A.,  et~al., 2011, \apjl, 733, L32

\bibitem[\protect\citeauthoryear{Pott, Eckart, Glindemann, Sch{\"o}del,
  Viehmann  \& Robberto}{Pott et~al.}{2008}]{pott2008enigma}
Pott J.-U.,  Eckart A.,  Glindemann A.,  Sch{\"o}del R.,  Viehmann T.,
  Robberto M.,  2008, \aap, 480, 115

\bibitem[\protect\citeauthoryear{Riaz, Mart{\'\i}n, Bouy  \& Tata}{Riaz
  et~al.}{2009}]{riaz20092mass}
Riaz B.,  Mart{\'\i}n E.~L.,  Bouy H.,   Tata R.,  2009, \apj, 700, 1541

\bibitem[\protect\citeauthoryear{Robinson}{Robinson}{2013}]{robinson2013models}
Robinson G.,  2013, \mnras, p. stt2115

\bibitem[\protect\citeauthoryear{Robinson \& Maldoni}{Robinson \&
  Maldoni}{2010}]{robinson2010librational}
Robinson G.,  Maldoni M.~M.,  2010, \mnras, 408, 1956

\bibitem[\protect\citeauthoryear{Robinson, Smith  \& Maldoni}{Robinson
  et~al.}{2012}]{robinson2012water}
Robinson G.,  Smith R.~G.,   Maldoni M.~M.,  2012, \mnras, 424, 1530

\bibitem[\protect\citeauthoryear{Roche \& Aitken}{Roche \&
  Aitken}{1984a}]{roche1984investigation}
Roche P.~F.,  Aitken D.~K.,  1984a, \mnras, 208, 481

\bibitem[\protect\citeauthoryear{Roche \& Aitken}{Roche \&
  Aitken}{1984b}]{roche1984oh}
Roche P.~F.,  Aitken D.~K.,  1984b, \mnras, 209, 33P

\bibitem[\protect\citeauthoryear{Roche \& Aitken}{Roche \&
  Aitken}{1985}]{roche1985investigation}
Roche P.~F.,  Aitken D.~K.,  1985, \mnras, 215, 425

\bibitem[\protect\citeauthoryear{{Roche}, {Packham}, {Telesco}, {Radomski},
  {Alonso-Herrero}, {Aitken}, {Colina}  \& {Perlman}}{{Roche}
  et~al.}{2006}]{roche2006circinus}
{Roche} P.~F.,  {Packham} C.,  {Telesco} C.~M.,  {Radomski} J.~T.,
  {Alonso-Herrero} A.,  {Aitken} D.~K.,  {Colina} L.,   {Perlman} E.,  2006,
  \mn@doi [\mnras] {10.1111/j.1365-2966.2006.10072.x}, \href
  {http://adsabs.harvard.edu/abs/2006MNRAS.367.1689R} {367, 1689}

\bibitem[\protect\citeauthoryear{{Roche}, {Alonso-Herrero}  \&
  {Gonzalez-Martin}}{{Roche} et~al.}{2015}]{roche2015ngc4418}
{Roche} P.~F.,  {Alonso-Herrero} A.,   {Gonzalez-Martin} O.,  2015, \mn@doi
  [\mnras] {10.1093/mnras/stv495}, \href
  {http://adsabs.harvard.edu/abs/2015MNRAS.449.2598R} {449, 2598}

\bibitem[\protect\citeauthoryear{Rosenthal, Bertoldi  \& Drapatz}{Rosenthal
  et~al.}{2000}]{rosenthal2000iso}
Rosenthal D.,  Bertoldi F.,   Drapatz S.,  2000, \aap, 356, 705

\bibitem[\protect\citeauthoryear{Ruan, Wang, Qi  \& Wang}{Ruan
  et~al.}{2011}]{ruan2011aerosol}
Ruan L.~M.,  Wang X.~Y.,  Qi H.,   Wang S.~G.,  2011, Journal of aerosol
  science, 42, 759

\bibitem[\protect\citeauthoryear{{Sako} et~al.,}{{Sako}
  et~al.}{2003}]{sako2003detector}
{Sako} S.,  et~al., 2003, \mn@doi [\pasp] {10.1086/379748}, \href
  {http://adsabs.harvard.edu/abs/2003PASP..115.1407S} {115, 1407}

\bibitem[\protect\citeauthoryear{Sandford}{Sandford}{1986}]{sandford1986acid}
Sandford S.~A.,  1986, Science, 231, 1540

\bibitem[\protect\citeauthoryear{Sandford \& Walker}{Sandford \&
  Walker}{1985}]{sandford1985laboratory}
Sandford S.~A.,  Walker R.~M.,  1985, \apj, 291, 838

\bibitem[\protect\citeauthoryear{Sandford, Allamandola, Tielens, Sellgren,
  Tapia  \& Pendleton}{Sandford et~al.}{1991}]{sandford1991interstellar}
Sandford S.~A.,  Allamandola L.~J.,  Tielens A.~G.~G.~M.,  Sellgren K.,  Tapia
  M.,   Pendleton Y.,  1991, \apj, 371, 607

\bibitem[\protect\citeauthoryear{Sargent et~al.,}{Sargent
  et~al.}{2009}]{sargent2009dust}
Sargent B.~A.,  et~al., 2009, \apj Supplement Series, 182, 477

\bibitem[\protect\citeauthoryear{Sch{\"o}del, Najarro, Muzic  \&
  Eckart}{Sch{\"o}del et~al.}{2010}]{schodel2010peering}
Sch{\"o}del R.,  Najarro F.,  Muzic K.,   Eckart A.,  2010, \aap, 511, A18

\bibitem[\protect\citeauthoryear{{Schutte} \& {Khanna}}{{Schutte} \&
  {Khanna}}{2003}]{schutte2003origin}
{Schutte} W.~A.,  {Khanna} R.~K.,  2003, \mn@doi [\aap]
  {10.1051/0004-6361:20021705}, \href
  {http://adsabs.harvard.edu/abs/2003A%26A...398.1049S} {398, 1049}

\bibitem[\protect\citeauthoryear{{Schutte}, {Gerakines}, {Geballe}, {van
  Dishoeck}  \& {Greenberg}}{{Schutte} et~al.}{1996}]{schutte1996discovery}
{Schutte} W.~A.,  {Gerakines} P.~A.,  {Geballe} T.~R.,  {van Dishoeck} E.~F.,
  {Greenberg} J.~M.,  1996, \aap, \href
  {http://adsabs.harvard.edu/abs/1996A%26A...309..633S} {309, 633}

\bibitem[\protect\citeauthoryear{Scott \& Duley}{Scott \&
  Duley}{1996}]{scott1996ultraviolet}
Scott A.,  Duley W.~W.,  1996, \apjs, 105, 401

\bibitem[\protect\citeauthoryear{Scoville, Stolovy, Rieke, Christopher  \&
  Yusef-Zadeh}{Scoville et~al.}{2003}]{scoville2003hubble}
Scoville N.~Z.,  Stolovy S.~R.,  Rieke M.,  Christopher M.,   Yusef-Zadeh F.,
  2003, \apj, 594, 294

\bibitem[\protect\citeauthoryear{Serabyn, Lacy  \& Achtermann}{Serabyn
  et~al.}{1991}]{serabyn1991gaseous}
Serabyn E.,  Lacy J.~H.,   Achtermann J.~M.,  1991, \apj, 378, 557

\bibitem[\protect\citeauthoryear{Siebenmorgen, Voshchinnikov  \&
  Bagnulo}{Siebenmorgen et~al.}{2014}]{siebenmorgen2014dust}
Siebenmorgen R.,  Voshchinnikov N.~V.,   Bagnulo S.,  2014, \aap, 561, A82

\bibitem[\protect\citeauthoryear{Sihvola}{Sihvola}{1994}]{sihvola1994anisotropic}
Sihvola A.~H.,  1994, Journal of electromagnetic waves and applications, 8, 115

\bibitem[\protect\citeauthoryear{Sihvola \& Pekonen}{Sihvola \&
  Pekonen}{1994}]{sihvola1994dielectric}
Sihvola A.~H.,  Pekonen O. P.~M.,  1994, Journal of electromagnetic waves and
  applications, 8, 1605

\bibitem[\protect\citeauthoryear{Skinner, Tielens, Barlow  \&
  Justtanont}{Skinner et~al.}{1992}]{skinner1992methanol}
Skinner C.~J.,  Tielens A.~G.~G.~M.,  Barlow M.~J.,   Justtanont K.,  1992,
  \apj, 399, L79

\bibitem[\protect\citeauthoryear{Slavin, Dwek  \& Jones}{Slavin
  et~al.}{2015}]{slavin2015destruction}
Slavin J.~D.,  Dwek E.,   Jones A.~P.,  2015, \apj, 803, 7

\bibitem[\protect\citeauthoryear{{Smith}}{{Smith}}{2003}]{smith2003ohir}
{Smith} B.~J.,  2003, \mn@doi [\aj] {10.1086/376743}, \href
  {http://adsabs.harvard.edu/abs/2003AJ....126..935S} {126, 935}

\bibitem[\protect\citeauthoryear{Smith \& Herman}{Smith \&
  Herman}{1990}]{smith1990absorption}
Smith R.~G.,  Herman J.,  1990, \aap, 227, 147

\bibitem[\protect\citeauthoryear{Smith \& Wright}{Smith \&
  Wright}{2011}]{smith2011librational}
Smith R.~G.,  Wright C.~M.,  2011, \mnras, 414, 3764

\bibitem[\protect\citeauthoryear{Smith, Sellgren  \& Tokunaga}{Smith
  et~al.}{1989}]{smith1989absorption}
Smith R.~G.,  Sellgren K.,   Tokunaga A.~T.,  1989, \apj, 344, 413

\bibitem[\protect\citeauthoryear{Smith, Wright, Aitken, Roche  \& Hough}{Smith
  et~al.}{2000}]{smith2000studies}
Smith C.~H.,  Wright C.~M.,  Aitken D.~K.,  Roche P.~F.,   Hough J.~H.,  2000,
  \mnras, 312, 327

\bibitem[\protect\citeauthoryear{Sogawa, Koike, Chihara, Suto, Tachibana,
  Tsuchiyama  \& Kozasa}{Sogawa et~al.}{2006}]{sogawa2006infrared}
Sogawa H.,  Koike C.,  Chihara H.,  Suto H.,  Tachibana S.,  Tsuchiyama A.,
  Kozasa T.,  2006, \aap, 451, 357

\bibitem[\protect\citeauthoryear{Soifer, Willner, Rudy  \& Capps}{Soifer
  et~al.}{1981}]{soifer19814}
Soifer B.~T.,  Willner S.~P.,  Rudy R.~J.,   Capps R.~W.,  1981, \apj, 250, 631

\bibitem[\protect\citeauthoryear{{Sokolik} \& {Toon}}{{Sokolik} \&
  {Toon}}{1999}]{sokolik1999aerosol}
{Sokolik} I.~N.,  {Toon} O.~B.,  1999, \mn@doi [\jgr] {10.1029/1998JD200048},
  \href {http://adsabs.harvard.edu/abs/1999JGR...104.9423S} {104, 9423}

\bibitem[\protect\citeauthoryear{{Somsikov} \& {Voshchinnikov}}{{Somsikov} \&
  {Voshchinnikov}}{1999}]{somsikov1999rayleigh}
{Somsikov} V.~V.,  {Voshchinnikov} N.~V.,  1999, \aap, \href
  {http://adsabs.harvard.edu/abs/1999A%26A...345..315S} {345, 315}

\bibitem[\protect\citeauthoryear{Spoon et~al.,}{Spoon
  et~al.}{2006}]{spoon2006detection}
Spoon H.~W.~W.,  et~al., 2006, \apj, 638, 759

\bibitem[\protect\citeauthoryear{Stognienko, Henning  \& Ossenkopf}{Stognienko
  et~al.}{1995}]{stognienko1995optical}
Stognienko R.,  Henning T.,   Ossenkopf V.,  1995, \aap, 296, 797

\bibitem[\protect\citeauthoryear{Stroud}{Stroud}{1975}]{stroud1975generalized}
Stroud D.,  1975, Physical Review B, 12, 3368

\bibitem[\protect\citeauthoryear{Suto et~al.,}{Suto et~al.}{2006}]{suto2006low}
Suto H.,  et~al., 2006, \mnras, 370, 1599

\bibitem[\protect\citeauthoryear{{Sutton}, {Danchi}, {Jaminet}  \&
  {Masson}}{{Sutton} et~al.}{1990}]{sutton1990co}
{Sutton} E.~C.,  {Danchi} W.~C.,  {Jaminet} P.~A.,   {Masson} C.~R.,  1990,
  \mn@doi [\apj] {10.1086/168259}, \href
  {http://adsabs.harvard.edu/abs/1990ApJ...348..503S} {348, 503}

\bibitem[\protect\citeauthoryear{{Sylvester}, {Kemper}, {Barlow}, {de Jong},
  {Waters}, {Tielens}  \& {Omont}}{{Sylvester} et~al.}{1999}]{sylvester19992}
{Sylvester} R.~J.,  {Kemper} F.,  {Barlow} M.~J.,  {de Jong} T.,  {Waters}
  L.~B.~F.~M.,  {Tielens} A.~G.~G.~M.,   {Omont} A.,  1999, \aap, \href
  {http://adsabs.harvard.edu/abs/1999A%26A...352..587S} {352, 587}

\bibitem[\protect\citeauthoryear{Tamanai, Mutschke, Blum  \& Meeus}{Tamanai
  et~al.}{2006}]{tamanai200610}
Tamanai A.,  Mutschke H.,  Blum J.,   Meeus G.,  2006, \apj Letters, 648, L147

\bibitem[\protect\citeauthoryear{{Telesco}, {Pina}, {Hanna}, {Julian}, {Hon}
  \& {Kisko}}{{Telesco} et~al.}{1998}]{telesco1998gatircam}
{Telesco} C.~M.,  {Pina} R.~K.,  {Hanna} K.~T.,  {Julian} J.~A.,  {Hon} D.~B.,
   {Kisko} T.~M.,  1998, in {Fowler} A.~M.,  ed.,  Society of Photo-Optical
  Instrumentation Engineers (SPIE) Conference Series Vol. 3354, Infrared
  Astronomical Instrumentation. pp 534--544

\bibitem[\protect\citeauthoryear{{Telesco} et~al.,}{{Telesco}
  et~al.}{2003}]{telesco2003canaricam}
{Telesco} C.~M.,  et~al., 2003, in {Iye} M.,  {Moorwood} A.~F.~M.,  eds,
  Society of Photo-Optical Instrumentation Engineers (SPIE) Conference Series
  Vol. 4841, Instrument Design and Performance for Optical/Infrared
  Ground-based Telescopes. pp 913--922, \mn@doi{10.1117/12.458979}

\bibitem[\protect\citeauthoryear{Thi, van Dishoeck, Dartois, Pontoppidan,
  Schutte, Ehrenfreund, d'Hendecourt  \& Fraser}{Thi et~al.}{2006}]{thi2006vlt}
Thi W.-F.,  van Dishoeck E.~F.,  Dartois E.,  Pontoppidan K.~M.,  Schutte
  W.~A.,  Ehrenfreund P.,  d'Hendecourt L.,   Fraser H.~J.,  2006, \aap, 449,
  251

\bibitem[\protect\citeauthoryear{Tielens, Wooden, Allamandola, Bregman  \&
  Witteborn}{Tielens et~al.}{1996}]{tielens1996infrared}
Tielens A.~G.~G.~M.,  Wooden D.~H.,  Allamandola L.~J.,  Bregman J.,
  Witteborn F.~C.,  1996, \apj, 461, 210

\bibitem[\protect\citeauthoryear{Van~den Ancker, Tielens  \& Wesselius}{Van~den
  Ancker et~al.}{2000}]{vandenancker2000iso}
Van~den Ancker M.~E.,  Tielens A.~G.~G.~M.,   Wesselius P.~R.,  2000, \aap,
  358, 1035

\bibitem[\protect\citeauthoryear{Verstraete et~al.,}{Verstraete
  et~al.}{2001}]{verstraete2001aromatic}
Verstraete L.,  et~al., 2001, \aap, 372, 981

\bibitem[\protect\citeauthoryear{Viehmann, Eckart, Sch{\"o}del, Moultaka,
  Straubmeier  \& Pott}{Viehmann et~al.}{2005}]{viehmann2005and}
Viehmann T.,  Eckart A.,  Sch{\"o}del R.,  Moultaka J.,  Straubmeier C.,   Pott
  J.-U.,  2005, \aap, 433, 117

\bibitem[\protect\citeauthoryear{Viehmann, Eckart, Sch{\"o}del, Pott  \&
  Moultaka}{Viehmann et~al.}{2006}]{viehmann2006dusty}
Viehmann T.,  Eckart A.,  Sch{\"o}del R.,  Pott J.-U.,   Moultaka J.,  2006,
  \apj, 642, 861

\bibitem[\protect\citeauthoryear{{Voshchinnikov} \& {Henning}}{{Voshchinnikov}
  \& {Henning}}{2010}]{voshchinnikov2010accretion}
{Voshchinnikov} N.~V.,  {Henning} T.,  2010, \mn@doi [\aap]
  {10.1051/0004-6361/200912817}, \href
  {http://adsabs.harvard.edu/abs/2010A%26A...517A..45V} {517, A45}

\bibitem[\protect\citeauthoryear{Voshchinnikov, Il'in, Henning  \&
  Dubkova}{Voshchinnikov et~al.}{2006}]{voshchinnikov2006dust}
Voshchinnikov N.~V.,  Il'in V.~B.,  Henning T.,   Dubkova D.~N.,  2006, \aap,
  445, 167

\bibitem[\protect\citeauthoryear{{Voshchinnikov}, {Videen}  \&
  {Henning}}{{Voshchinnikov} et~al.}{2007}]{voshchinnikov2007emt}
{Voshchinnikov} N.~V.,  {Videen} G.,   {Henning} T.,  2007, \mn@doi [\ao]
  {10.1364/AO.46.004065}, \href
  {http://adsabs.harvard.edu/abs/2007ApOpt..46.4065V} {46, 4065}

\bibitem[\protect\citeauthoryear{Watson et~al.,}{Watson
  et~al.}{2009}]{watson2009crystalline}
Watson D.~M.,  et~al., 2009, \apjs, 180, 84

\bibitem[\protect\citeauthoryear{{Whittet}, {Duley}  \& {Martin}}{{Whittet}
  et~al.}{1990}]{whittet1990sicism}
{Whittet} D.~C.~B.,  {Duley} W.~W.,   {Martin} P.~G.,  1990, \mnras, \href
  {http://adsabs.harvard.edu/abs/1990MNRAS.244..427W} {244, 427}

\bibitem[\protect\citeauthoryear{Whittet et~al.,}{Whittet
  et~al.}{1997}]{whittet1997infrared}
Whittet D.~C.~B.,  et~al., 1997, \apj, 490, 729

\bibitem[\protect\citeauthoryear{{Willner} \& {Pipher}}{{Willner} \&
  {Pipher}}{1982}]{willner1982three}
{Willner} S.~P.,  {Pipher} J.~L.,  1982, in {Riegler} G.~R.,  {Blandford}
  R.~D.,  eds,  American Institute of Physics Conference Series Vol. 83, The
  Galactic Center. pp 77--81, \mn@doi{10.1063/1.33505}

\bibitem[\protect\citeauthoryear{Willner et~al.,}{Willner
  et~al.}{1982}]{willner1982infrared}
Willner S.~P.,  et~al., 1982, \apj, 253, 174

\bibitem[\protect\citeauthoryear{Witteborn, Sandford, Bregman, Allamandola,
  Cohen, Wooden  \& Graps}{Witteborn et~al.}{1989}]{witteborn1989new}
Witteborn F.~C.,  Sandford S.~A.,  Bregman J.~D.,  Allamandola L.~J.,  Cohen
  M.,  Wooden D.~H.,   Graps A.~L.,  1989, \apj, 341, 270

\bibitem[\protect\citeauthoryear{{Wright}}{{Wright}}{1994}]{wright1994thesis}
{Wright} C.~M.,  1994, PhD thesis, University of New South Wales

\bibitem[\protect\citeauthoryear{{Wright} \& {Glasse}}{{Wright} \&
  {Glasse}}{2005}]{wright2005mid}
{Wright} C.~M.,  {Glasse} A.~C.~H.,  2005, in {Adamson} A.,  {Aspin} C.,
  {Davis} C.,   {Fujiyoshi} T.,  eds,  Astronomical Society of the Pacific
  Conference Series Vol. 343, Astronomical Polarimetry: Current Status and
  Future Directions. p.~316

\bibitem[\protect\citeauthoryear{{Wright}, {Aitken}, {Smith}  \&
  {Roche}}{{Wright} et~al.}{1999}]{wright1999mid}
{Wright} C.~M.,  {Aitken} D.~K.,  {Smith} C.~H.,   {Roche} P.~F.,  1999, in
  {Greenberg} J.~M.,  {Li} A.,  eds,  NATO Advanced Science Institutes (ASI)
  Series C Vol. 523, NATO Advanced Science Institutes (ASI) Series C. pp 77--83

\bibitem[\protect\citeauthoryear{{Wright}, {van Dishoeck}, {Black},
  {Feuchtgruber}, {Cernicharo}, {Gonz{\'a}lez-Alfonso}  \& {de
  Graauw}}{{Wright} et~al.}{2000}]{wright2000iso}
{Wright} C.~M.,  {van Dishoeck} E.~F.,  {Black} J.~H.,  {Feuchtgruber} H.,
  {Cernicharo} J.,  {Gonz{\'a}lez-Alfonso} E.,   {de Graauw} T.,  2000, \aap,
  358, 689

\bibitem[\protect\citeauthoryear{{Wright}, {Aitken}, {Smith}, {Roche}  \&
  {Laureijs}}{{Wright} et~al.}{2002}]{wright2002mineralogy}
{Wright} C.~M.,  {Aitken} D.~K.,  {Smith} C.~H.,  {Roche} P.~F.,   {Laureijs}
  R.~J.,  2002, in {Alves} J.~F.,  {McCaughrean} M.~J.,  eds, The Origin of
  Stars and Planets: The VLT View. p.~85, \mn@doi{10.1007/10856518_10}

\bibitem[\protect\citeauthoryear{{Wright}, {Siebenmorgen}, {Stecklum},
  {Sterzik}  \& {K{\"a}ufl}}{{Wright} et~al.}{2008}]{wright2008mid}
{Wright} C.~M.,  {Siebenmorgen} R.,  {Stecklum} B.,  {Sterzik} M.,
  {K{\"a}ufl} H.-U.,  2008, in Society of Photo-Optical Instrumentation
  Engineers (SPIE) Conference Series. p.~29, \mn@doi{10.1117/12.787899}

\bibitem[\protect\citeauthoryear{Yusef-Zadeh \& Melia}{Yusef-Zadeh \&
  Melia}{1992}]{yusef1992bow}
Yusef-Zadeh F.,  Melia F.,  1992, \apj, 385, L41

\bibitem[\protect\citeauthoryear{Yusef-Zadeh \& Morris}{Yusef-Zadeh \&
  Morris}{1991}]{yusef1991windswept}
Yusef-Zadeh F.,  Morris M.,  1991, \apj, 371, L59

\bibitem[\protect\citeauthoryear{{Zinner}}{{Zinner}}{2013}]{zinner2013laboratory}
{Zinner} E.,  2013, Analytical Chemistry, 85, 1264

\bibitem[\protect\citeauthoryear{{de Muizon}, {D'Hendecourt}  \& {Perrier}}{{de
  Muizon} et~al.}{1986}]{demuizon1986evidence}
{de Muizon} M.,  {D'Hendecourt} L.~B.,   {Perrier} C.,  1986, in {Israel}
  F.~P.,  ed.,  Astrophysics and Space Science Library Vol. 124, Light on Dark
  Matter. pp 221--224

\bibitem[\protect\citeauthoryear{de Vries, Min, Waters, Blommaert  \&
  Kemper}{de~Vries et~al.}{2010}]{de2010determining}
de Vries B.~L.,  Min M.,  Waters L.~B.~F.~M.,  Blommaert J.~A.~D.~L.,   Kemper
  F.,  2010, \aap, 516, A86

\bibitem[\protect\citeauthoryear{de Vries, Blommaert, Waters, Waelkens, Min,
  Lombaert  \& Van~Winckel}{de~Vries et~al.}{2014}]{de2014problematically}
de Vries B.~L.,  Blommaert J.~A.~D.~L.,  Waters L.~B.~F.~M.,  Waelkens C.,  Min
  M.,  Lombaert R.,   Van~Winckel H.,  2014, \aap, 561, A75

\bibitem[\protect\citeauthoryear{van Boekel et~al.,}{van Boekel
  et~al.}{2004}]{van2004building}
van Boekel R.~J.~H.~M.,  et~al., 2004, Nature, 432, 479

\bibitem[\protect\citeauthoryear{{van Dishoeck} \& {Helmich}}{{van Dishoeck} \&
  {Helmich}}{1996}]{vandishoeck1996hotwater}
{van Dishoeck} E.~F.,  {Helmich} F.~P.,  1996, \aap, \href
  {http://adsabs.harvard.edu/abs/1996A%26A...315L.177V} {315, L177}

\bibitem[\protect\citeauthoryear{van Dishoeck, Wright, Cernicharo,
  Gonz{\'a}lez-Alfonso, de Graauw, Helmich  \& Vandenbussche}{van Dishoeck
  et~al.}{1998}]{vandishoeck1998iso}
van Dishoeck E.~F.,  Wright C.~M.,  Cernicharo J.,  Gonz{\'a}lez-Alfonso E.,
  de Graauw T.,  Helmich F.~P.,   Vandenbussche B.,  1998, \apjl, 502, L173

\bibitem[\protect\citeauthoryear{{van der Tak}, {Tuthill}  \& {Danchi}}{{van
  der Tak} et~al.}{2005}]{vandertak2005subarc}
{van der Tak} F.~F.~S.,  {Tuthill} P.~G.,   {Danchi} W.~C.,  2005, \mn@doi
  [\aap] {10.1051/0004-6361:20041595}, \href
  {http://adsabs.harvard.edu/abs/2005A%26A...431..993V} {431, 993}

\makeatother
\end{thebibliography}




\appendix

\section[]{Locating SgrA~IRS3}

\begin{figure*}
\includegraphics[scale=0.855]{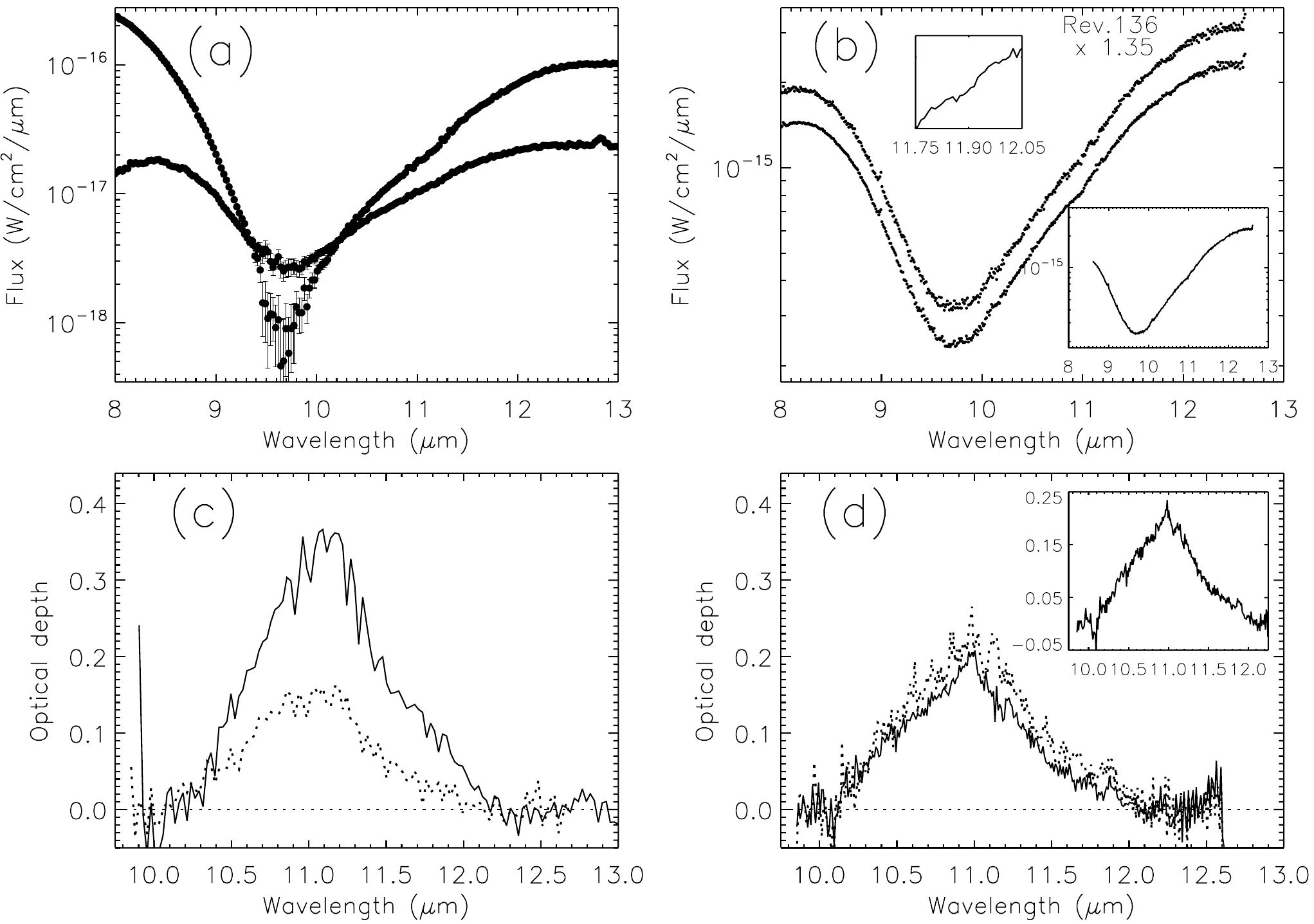}
\caption{Comparison between spectra of SgrA IRS3 and IRSX obtained simultaneously 
with Michelle (a), as well as with the two ISO spectra (b). Panels (c) and (d) show
their extracted optical depths for the 11.1~$\mu$m feature. The emission line at 
12.8~$\mu$m in SgrA~IRSX is from [Ne~II], which is clearly absent in SgrA~IRS3.
An 11.85~$\mu$m feature can be seen in both ISO data sets, but is most obvious in
the combined spectrum and its extracted optical depth (insets in (b) and (d) respectively).
It affirms the identification of crystalline silicates in the ISM path to the Galactic Centre.}
\label{IRS3X-ISOstar}
\end{figure*}

The fact that -- projected onto the plane-of-the-sky -- IRS3 lies only about 5~arcsec 
(0.2~pc) NW of the dynamical centre of the Galaxy, Sgr~A$^{*}$, makes it likely that 
the two are also close in 3-dimensional space. But its lack of association with the 
ionised gas and warm dust of the Galactic Centre (GC) mini-spiral which orbits 
Sgr~A$^{*}$, evidenced for instance by the non-detection of a 12.8~$\mu$m [Ne II] 
emission line in Figure~\ref{FIG-Spectra} -- unlike the case for IRSX in 
Figure\ref{IRS3X-ISOstar}-a -- could be used to argue against a physical 
association. Or instead it could be that IRS3 lies within the innermost region of 
the cavity encompassed by the $\sim$~1~pc radius cicumnuclear disk or ring (CND) 
around Sgr~A$^{*}$. 

This latter option finds some support in the M-band imaging of 
\cite{viehmann2005and}. They find extended emission with a bow-shock-like shape 
around IRS3, which they interpret to be due to strong winds from the nuclear 
star cluster within about half an arcsecond ($\leq$~0.02~pc) of Sgr~A$^{*}$. The 
8.6~$\mu$m images in \cite{pott2008enigma} show a similar morphology. On the 
other hand, if such a wind was influencing IRS3 then it might be expected to 
produce similar phenomena to that observed for IRS7 (only an arcsecond or so away), 
where a cometary-like tail at least 5~arcsec in length is observed in both radio 
continuum (\citet{yusef1992bow}, \cite{yusef1991windswept}) and [Ne II] 
(\citet{serabyn1991gaseous}). 

IRS3 must be further distant than about 4-5~kpc since absorption line 
spectroscopy of several molecules show narrow features at LSR velocities of 
-3, -31 and -53 km~s$^{-1}$. Examples include infrared lines of H$_{3}^{+}$ 
and CO in \cite{goto2014infrared}, \cite{goto2008absorption}, \cite{oka2005hot}, 
\cite{goto2002absorption} and \cite{geballe1989infrared}. These are also seen 
in mm-wave rotational transitions, such as CO 3--2 in \cite{sutton1990co}. 
The -3, -31 and -53 km~s$^{-1}$ components respectively arise from molecular 
clouds local to the solar neighborhood, the 3~kpc arm and the 4.5~kpc arm. 

Further, a discrete feature at -140~km~s$^{-1}$ and a 'trough' of absorption 
over a velocity extent of minus a few hundred to a few tens of km~s$^{-1}$ in 
the IR H$_{3}^{+}$ and CO transitions toward IRS3 has been identified with 
clouds of the so-called central molecular zone (CMZ). The CMZ is a region of 
radius about 200~pc around the Galactic Centre (\citet{oka2005hot}). Another 
component in the IRS3 spectrum at +60~km~s$^{-1}$ has recently been identified 
by \cite{goto2014infrared} with a compact cloud within the CND, although a 
previous analysis by \cite{goto2008absorption} associated it with the well 
known 50~km~s$^{-1}$ cloud on the far side of SgrA. The location of clouds at 
-72 and +45~km~s$^{-1}$ in the IRS3 spectrum remain undetermined 
(\citet{goto2014infrared}).  

From these considerations IRS3 must lie behind at least the front side of the 
CMZ as well as the CND, and probably also behind the dynamical centre. But 
how far behind remains an open question. 

\subsection[]{Enhanced silicate absorption toward SgrA~IRS3}

Perhaps consistent with IRS3 lying on the far side of the GC there is additional 
silicate absorption toward IRS3 compared to other sources within the GC. This was 
first noted by \cite{becklin1978infrared}, and then by 
\cite{roche1985investigation}, based on observations of multiple points within 
the N-S arm and E-W bar of the SgrA mini-spiral. It can also be seen in 
Figure~\ref{IRS3X-ISOstar}-a where the IRS3 spectrum is compared with that of a 
region about 5 arcsec south (we call it IRSX, just east of IRS2). This was obtained 
from the same Gemini--Michelle observation, thus is immune to possible artefacts 
from the standard star division, such as the telluric 9.6~$\mu$m ozone band mimicking 
deeper silicate absorption. Clearly the depths of the respective silicate bands 
toward IRS3 and IRSX are very different. 

Whilst IRSX almost certainly has underlying silicate emission, even accounting for 
this the silicate optical depth $\tau_{9.7}$ was found by \cite{roche1985investigation} 
to be confined to a narrow range of 3.6$\pm$0.3 for almost all of the compact mid-IR 
sources within the mini-spiral. A value about 0.8--1.3 higher was found for IRS3 by 
\cite{roche1985investigation} and \cite{becklin1978infrared} respectively. On the 
other hand, the almost 3 orders of magnitude extinction at 9.7~$\mu$m for IRS3 evident 
in our spectrum suggests $\tau_{9.7}$ is around 7, double that for the mini-spiral. 
This is in very good agreement with the figure of 7.0$\pm$0.5 inferred by 
\cite{pott2008enigma} from their spectrum obtained with mid-IR interferometry. 

An enhanced silicate optical depth toward IRS3 holds true even when more recent 
data on the extinction to the GC is considered. For instance, it is higher than 
the value of around 4.7 to the ionised gas and warm dust of the mini-spiral,
inferred from the data of \cite{fritz2011line} in their Figure 11. They used ISO 
and other hydrogen recombination line data to determine the extinction from 
1 to 19~$\mu$m and subsequently 'reconstructed' the underlying continuum 
emission. 

The origin of the extra silicate absorption toward IRS3 was suggested by 
\cite{roche1985investigation} and \cite{becklin1978infrared} to lie in a dust 
shell intrinsic (or local) to IRS3 itself, possibly a result of mass loss 
assuming IRS3 to be an O-rich star. However, the recent mid-IR observations of 
\cite{pott2008enigma} have cast a question over this model. They find that IRS3 
consists of up to three components. The largest is a diffuse emission region found 
from direct 8.6~$\mu$m imaging and extending over a few arcseconds (and also seen 
in the L- and M-bands by \cite{viehmann2005and}). Embedded within this is a compact 
component, unresolved in the direct images but resolved by interferometry into 
two additional components, the most compact of which has a wavelength-independent 
FWHM of 18$\pm$1~mas (144~AU at 8~kpc) surrounded by another component whose 
FWHM increases with wavelength from $\leq$~40~mas to $\geq$~50~mas (320--400~AU) 
between 8 and 13~$\mu$m.

Interestingly, the additional silicate absorption toward IRS3 cannot occur at 
spatial scales less than several hundred AU since the mid-IR visibilities of 
\cite{pott2008enigma} are spectrally featureless for all baselines. Combined 
with radiative transfer modelling this leads Pott et al. to conclude that IRS3 
is in fact a cool ($\sim$~3000~K) C-rich star with a carbonaceous circumstellar 
dust shell. The additional silicate absorption therefore probably occurs at radii 
of tens of thousands of AU from IRS3, presumably exterior even to the few arcsec 
extended dust region which is still hot enough to be emitting at L, M and 
8--13~$\mu$m (\citet{viehmann2006dusty} and \citet{viehmann2005and}). Whether 
this is physically associated with IRS3 is an open question, but if so then it 
would likely be a detached dust shell which resulted from a previous (O-rich) 
phase of IRS3's evolution. 

Whatever the origin of the dust producing the enhanced silicate absorption 
toward IRS3, we believe that the bulk of the 11.1~$\mu$m absorption band, and 
thus the putative crystalline silicates, occur in the ISM. This is evident in 
Figure~\ref{IRS3X-ISOstar} where we show our spectrum of IRSX against both 
ISO SWS01 spectra, as well as the extracted optical depths which are 
essentially equal. Even though $\tau_{11.1}$ for IRS3 is roughly twice that
for IRSX and the two ISO-observed positions, when $\tau_{9.7}$ is considered 
then the relative strengths of the amorphous and putative crystalline silicate 
bands in all four spectra are essentially equivalent.

\begin{figure}
\includegraphics[scale=0.55]{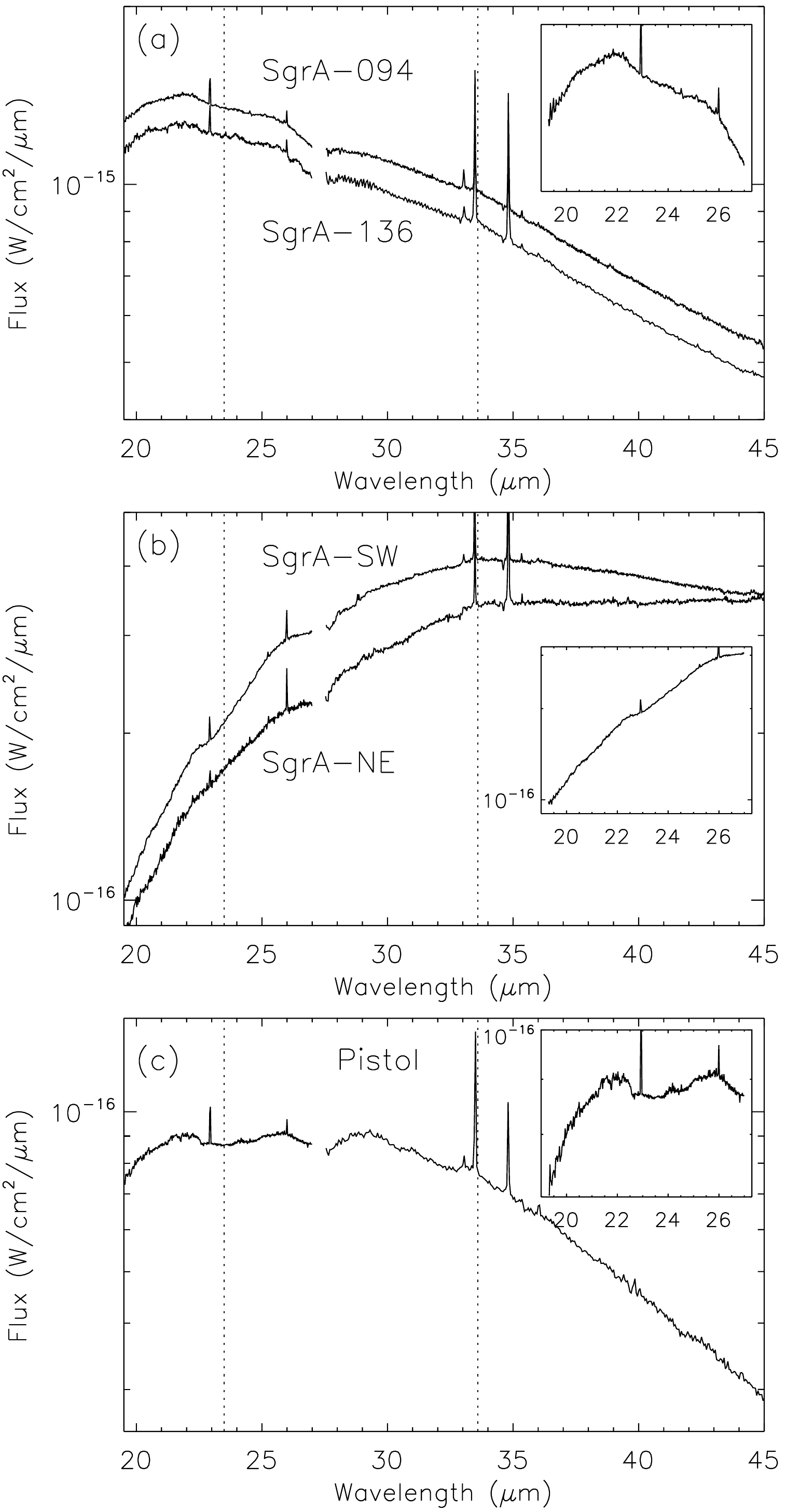}
\caption{ISO--SWS Galactic Centre data in the 20 to 45 $\mu$m interval. (a) two spectra
centred on the mini-spiral, with the revolution 94 and 136 band 4 data scaled by 
factors of 0.52 and 0.43; (b) two spectra at approximately 40--45 arcsec to the SSW
and NNE, with the band 4 SSW data scaled by 0.58, band 3D NNE by 1.48 and band 4 
NNE by 0.70; (c) a spectrum toward the Pistol star, with the band 4 data scaled by 0.42. 
Insets are zooms of the band 3D data, showing the clear presence of a feature around 
23.5~$\mu$m, further strengthening the identification of crystalline silicates in the ISM 
toward the Galactic Centre.}
\label{ISO-SgrA-20to45um}
\end{figure}

\subsection{$\tau_{9.7}$ vs A$_{\rm V}$}

Another issue on which the SgrA~IRS3 spectrum may feasibly bear, pointed out 
by \cite{pott2008enigma}, is the long-accepted ratio of visual extinction 
to the 9.7~$\mu$m optical depth, i.e. $A_{\rm V}/\tau_{9.7}$. For the local 
ISM this is thought to be around 18 (\citet{roche1984investigation}) but toward 
the GC it is more like around 9 (\citet{roche1985investigation}). 

From their mid-IR interferometric spectrum of IRS3 \cite{pott2008enigma} suggested 
that $A_{\rm V}/\tau_{9.7}$ could be as low as 4. This was based on an assumption 
that A$_{\rm V}$ was only around 25, an often quoted 'average' toward the GC. But 
other work (e.g. \citet{schodel2010peering}; \citet{scoville2003hubble}) has shown 
that the visual and near-IR extinction varies on relatively small spatial scales -- 
an arcsecond or less -- across SgrA. We believe it is more likely that A$_{\rm V}$ 
of IRS3 is higher than the 'average' GC value. 

Obviously a direct measure of A$_{\rm V}$ for IRS3 -- or indeed any of 
A$_{\rm J}$, A$_{\rm H }$ or A$_{\rm K}$ -- would be required to decide 
whether A$_{\rm V}$/$\tau_{9.7}$ toward IRS3 is abnormally low. This however 
is a very difficult task given IRS3 is optically invisible and its spectral 
type remains uncertain. So at this stage we do not believe the data is 
sufficient to say with any confidence that A$_{\rm V}$/$\tau_{9.7}$ of IRS3 
is lower than the commonly-accepted value of around 9 toward the GC.

\section[]{11~$\mu$m band in W28~A2 and IRAS19110+1045, and speculation on 
crystalline enstatite}

\begin{figure*}
\includegraphics[scale=0.68]{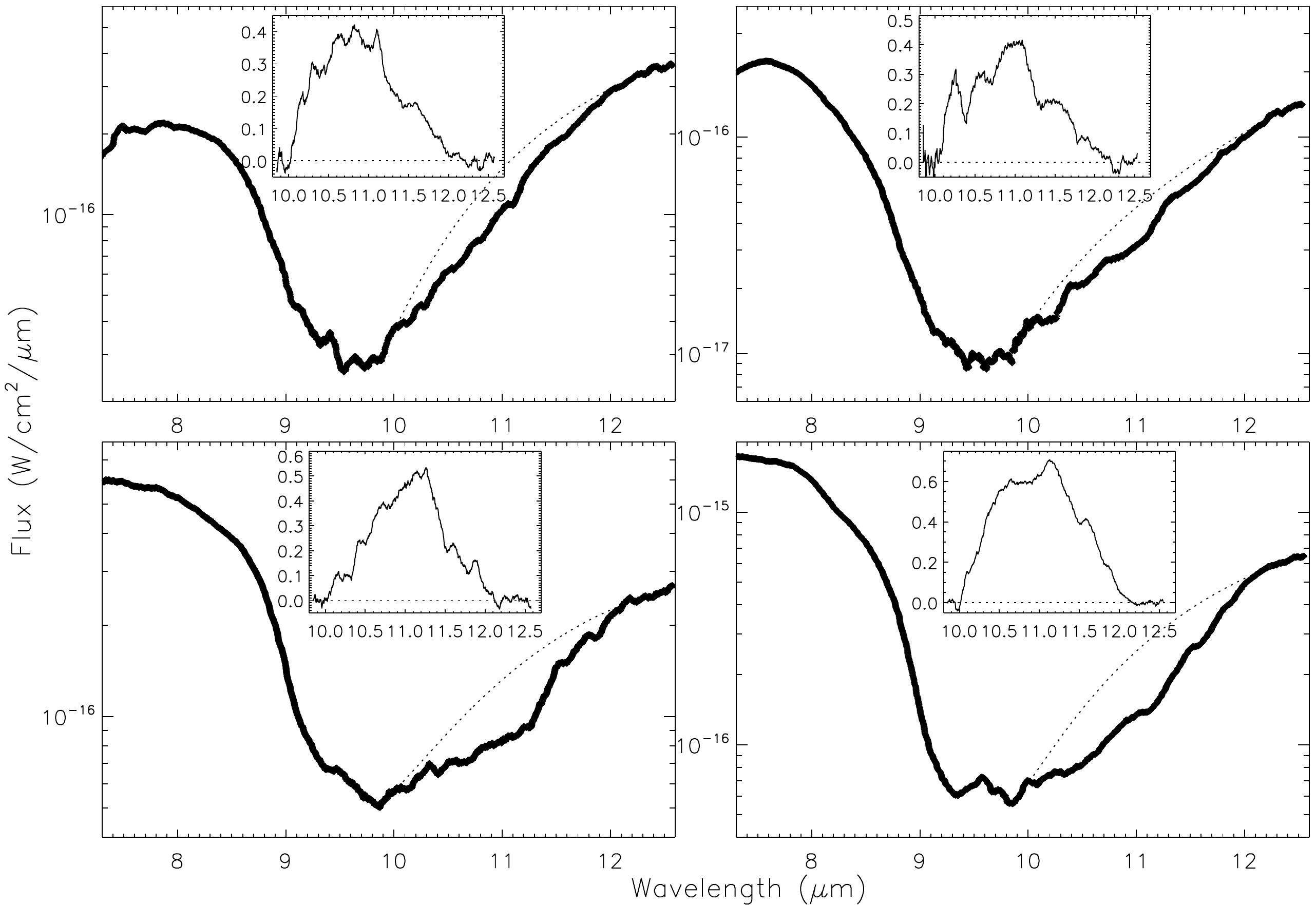}
\caption{ISO--SWS01 7--13~$\mu$m spectra (main), and extracted optical depth (insets), of 
the YSOs W28~A2 (G5.89-0.39, left) and IRAS19110+1045 (G45.07+0.13, right) on the top row, 
and the OH/IR stars AFGL~2403 (left) and OH26.5+0.6 (right) on the bottom row. The spectra 
have been sigma clipped and smoothed. The dotted line in the main plots is the polynomial 
fit between about 10~$\mu$m and 12.1--12.7~$\mu$m used to extract the optical depth spectra 
in the insets. The reader is reminded that narrow features (FWHM $\sim$ 0.1--0.3~$\mu$m) 
may be present at $\sim$ 9.35, 10.1 and 11.05~$\mu$m due to structure in the RSRF of the 
SWS band 2C (\protect\citet{leech2003iso}), and which can perturb the shape of the spectrum 
given their overlap with crystalline enstatite and/or forsterite bands. Also, the OH/IR star 
spectra, and thus the overall shape of their extracted optical depth spectra and/or relative 
strengths of features, will likely be more influenced by radiative transfer effects 
(\citet{maldoni2005ohir}, \citet{maldoni2004ohir}, \citet{maldoni2003ohir}). Nevertheless, 
the similarity of the YSO and OH/IR spectra here are strongly suggestive of similar dust 
mineralogies, specifically the presence of both crystalline forsterite and enstatite.}
\label{ISO-W28-IRAS19110}
\end{figure*}

Since we used the YSOs W28~A2 and IRAS19110+1045 in Figure~\ref{Spectra_20to45um} to 
demonstrate the existence of a 23.5~$\mu$m crystalline silicate band we show here that
they also possess an 11.1~$\mu$m feature. Figure~\ref{ISO-W28-IRAS19110} shows the 
sigma-clipped and smoothed ISO SWS01 spectra from 7.3 to 12.6~$\mu$m, along with the 
extracted 11~$\mu$m optical depths (insets), of W28~A2 (left) and IRAS19110 (right) in 
the top row, and for comparison the OH/IR stars AFGL~2403 (left) and OH26.5+0.6 (right) 
in the bottom row.  

\begin{figure*}
\hspace*{-0.5cm}
\includegraphics[scale=0.67]{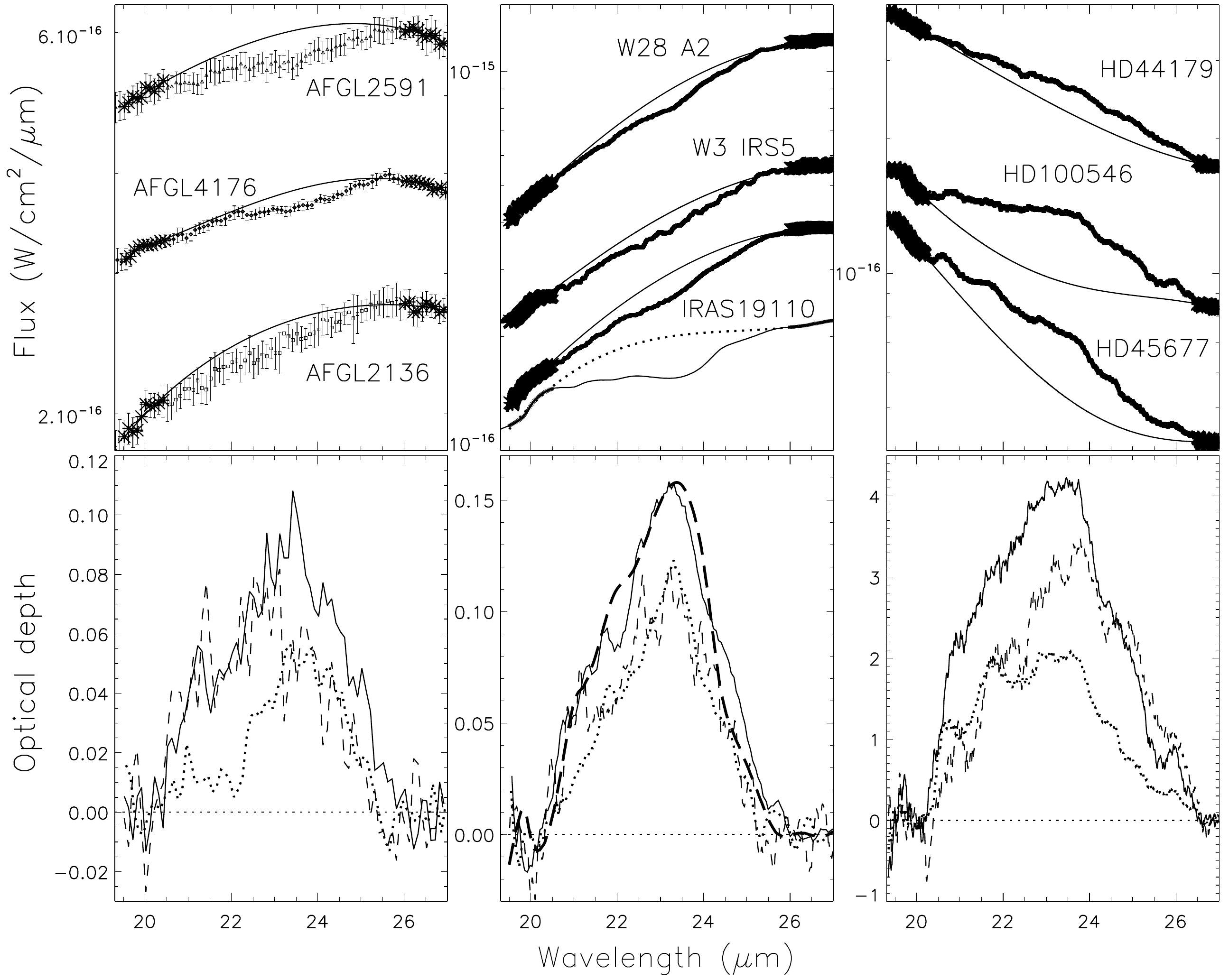}
\caption{ISO--SWS01 19--28~$\mu$m spectra of most of the targets previously 
presented in Figure~\ref{Spectra_20to45um}, aiming to demonstrate both the
reality of the 23.5~$\mu$m features in the YSOs and its identification with crystalline
silicate. Solid lines in the top panels show polynomial fits across the observed feature
(dotted line for the model feature in the lowermost spectrum of the middle top panel). 
The bottom panels show the extracted feature for each source, being optical depths 
for the absorption cases and continuum subtracted excesses -- in units of 10$^{-17}$
W/cm$^2$/$\mu$m -- for the emission cases. In the bottom left panel the solid, 
dotted and dashed spectra are for AFGL~2591, AFGL~4176 and AFGL~2136 
respectively. In the bottom middle panel the solid, dotted and light short dashed 
spectra are for IRAS19110+1045, W28~A2 and W3~IRS5 respectively, and have 
been averaged in 10 pixel wide bins for plotting purposes. The heavy long dashed 
line is the model, scaled to match the peak optical depth of IRAS19110+1045. The 
model uses optical constants of amorphous olivine MgFeSiO$_4$ and crystalline 
forsterite respectively from \protect\cite{dorschner1995steps} and
\protect\cite{fabian2001steps}. Effective medium theory has been used, with a 
crystalline volume fraction of 0.1 and Maxwell-Garnett mixing rule. In the bottom 
right panel the solid, dotted and dashed spectra are for HD100546, HD45677 and 
HD44179 respectively, and have been averaged in 10 pixel wide bins for plotting 
purposes.}
\label{ISO-23p5obs-mods}
\end{figure*}

Generally speaking the extracted feature of the two YSOs is consistent with that 
of the other objects presented here. However, in both cases there appears to be 
another shallow feature, centred around 11.5--11.6~$\mu$m, more clearly seen in 
the optical depth ($\tau$) insets. Looking back at Figure~\ref{FIG-Obs-XMods}, 
possible corresponding features can be seen in the YSOs W3~IRS5 and AFGL~2136, 
as well as the ISM path toward SgrA~IRS3, in the sense that the extracted optical 
depths lie significantly above the crystalline forsterite model.

For the IR sources W28~A2 and IRAS19110, being close to or coincident with 
ultracompact HII regions, it might be thought that extended 11.25~$\mu$m PAH 
emission within the large ISO beam could mimic the dip in $\tau$ between the 
11.1 and 11.6~$\mu$m features. However, these two sources are dominated by 
solid-state absorption bands, of both silicates and ices, with only very faint 
3.3 and possibly 6.2~$\mu$m PAH emission bands in W28~A2, and barely detectable 
3.3~$\mu$m emission in IRAS19110. In neither case are the intrinsically
stronger 7.7 and 8.6~$\mu$m PAH bands discernible. This is unlike the case for 
other embedded YSOs which we have looked at in the ISO archive (e.g. W3~IRS5,
NGC7538~IRS1, MonR2~IRS3), which, due to their proximity to HII regions, have 
obvious PAH features. Also, the extracted 11~$\mu$m feature, obtained in precisely
the same way, is markedly different between YSOs with and without 11.25~$\mu$m
PAH emission. 

Furthermore, as well as a probable 11.85~$\mu$m crystalline forsterite band in 
both IRAS19110+1045 and W28~A2, there appears to be another feature around 
10.5--10.6~$\mu$m, and/or a distinct broadening of the extracted optical depth 
spectrum on the short wavelength side of the peak, compared to the YSOs in 
Figure~\ref{FIG-Obs-XMods} and Figure~\ref{FIG-IRAS13481}. This apparent difference 
is particularly noticeable for AFGL~2136 and AFGL~2591, also measured with ISO 
and processed in precisely the same way, thus minimising any doubt over its reality. 
Once again there is a manifest similarity to the OH/IR spectra.

Overall, the correspondence between the $\tau$ spectral shapes of W28~A2 and IRAS19110  
and those of the OH/IR stars, which certainly have no PAH emission, is striking and 
strongly suggests that absorption bands are responsible for the YSO features. Finally, 
as one more indication of this we point out the likely existence of a feature at 
$\sim$ 9.3~$\mu$m in W28~A2, also seen in the two OH/IR stars. As noted throughout
this paper, we caution that a known band 2C RSRF artefact potentially exists at 
9.35~$\mu$m. But the fact that neither of the other two known band 2C RSRF features 
(prominently) appear provides confidence that the W28~A2 spectrum contains a real 
astronomical feature at 9.3~$\mu$m. 

We speculate that these features can be identified with a crystalline 
pyroxene, as has been reported for OH/IR stars (\citet{sylvester19992}). We further 
speculate that it would be a magnesium rich variant, i.e. enstatite-like, as the 
wavelengths are in pretty good agreement with the laboratory values for the 
MgSiO$_3$ aerosol sample  of \cite{tamanai200610}. To our knowledge there is little 
precedent for the presence of cold crystalline pyroxene in the near environs of 
embedded YSOs, as opposed to its fairly common presence in emission from the disks 
of older, though still pre-main sequence, stars (e.g. \citet{juhasz2010dust}). This 
is apart from the example of SVS13, recently reported by \cite{fujiyoshi2015mid}.  

\subsection[]{The 23.5~$\mu$m feature in DEYSOs}

Finally, in Figure~\ref{ISO-23p5obs-mods} we show zooms of the 19--28~$\mu$m
portion of most of the spectra previously presented in Figure~\ref{Spectra_20to45um}. 
This includes the six YSOs with 23.5~$\mu$m absorption bands (e.g. W28~A2 and 
IRAS19110+1045), plus for comparison the objects with 23.5~$\mu$m emission, being
the two Herbig Be stars HD100546 and HD45677 and the pre-planetary nebula HD44179.
Their 23.5~$\mu$m bands have been extracted in a manner completely analogous to 
that used previously for the 11.1~$\mu$m feature, as has a representative model (the
lowermost spectrum in the top middle panel and heavy dashed line in the bottom middle
panel). 

As even a cursory examination of the laboratory data for crystalline silicates
(both olivine and pyroxene) shows, it is difficult if not impossible to find any wavelength
interval in the 20-30~$\mu$m region free of a discrete feature (\citet{pitman2010infrared}; 
\citet{sogawa2006infrared}; \citet{suto2006low}; \citet{koike2003compositional}; 
\citet{jager1998steps}; \citet{mukai1990optical}). However, the 23.5~$\mu$m
band is intrinsically the strongest and so our choice of 'continuum' in 
Figure~\ref{ISO-23p5obs-mods} should have only a minor influence on the extracted 
optical depths or excesses.

There is a clear correspondence between the absorption and emission cases, and
the model. This is inclusive of the central wavelength (no shift has been applied) and
the overall feature shape, e.g. the shallow and steep slopes on the short and long
wavelength sides of the peak. For the emission cases at least some of the short
wavelength 'wing' probably arises from crystalline enstatite. 

At these wavelengths, and with a dust temperature around 100~K, the competition 
between emission and absorption is likely to be more important than at 11~$\mu$m, 
and thus a more sophisticated modelling approach is called for. Even so, our confidence 
in the detection of crystalline forsterite in deeply embedded YSOs, and that it is (perhaps 
unexpectedly) common, is increased by consideration of the data and preliminary model 
presented in Figure~\ref{ISO-23p5obs-mods}.

\section[]{Correcting ISO--SWS band 4 responsivity/memory effects of the up/down 
scans}

Briefly described here is the technique we used to correct the ISO band 4 
(28--45~$\mu$m) data for memory/responsivity effects in the spectra presented 
in Figure~\ref{Spectra_20to45um}, Figure~\ref{ISO-HII-20to45um} and 
Figure~\ref{ISO-SgrA-20to45um}.

\begin{figure*}
\includegraphics[scale=0.60]{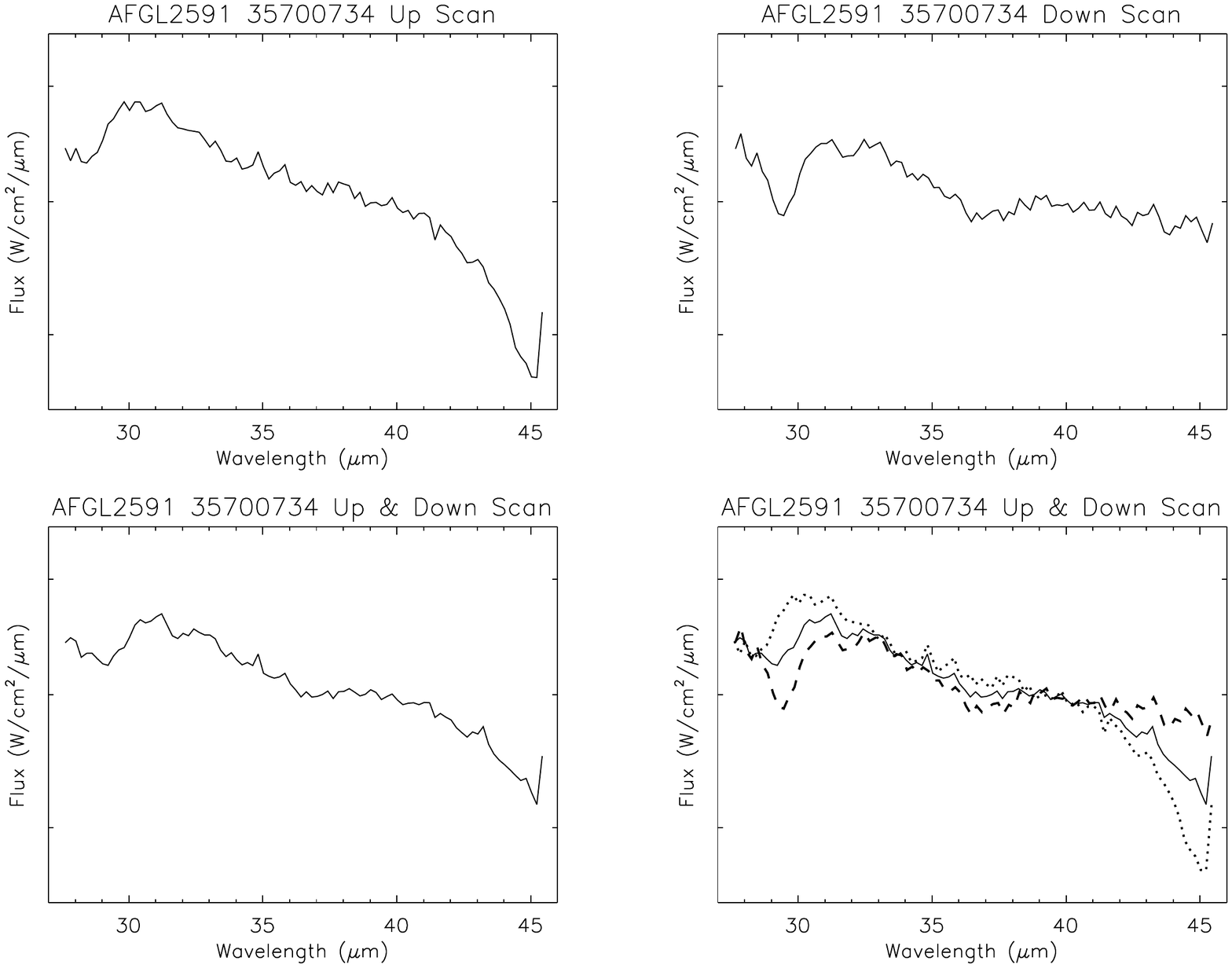}
\caption{Sigma-clipped and binned uncorrected up and down scans for the band 4 ISO observation of AFGL2591 (tdt=35700734).}
\label{GL2591_UPDO_UNCO}
\end{figure*}

The SWS01 mode observed the full spectral range 2.4--45.2~$\mu$m of the ISO--SWS 
in two directions, denoted the 'up' and 'down' scans. These refer to grating steps, 
such that an 'up' scan actually goes from longer to shorter wavelengths, and 
vice-versa for the 'down' scan. For band 4 the scan starts at 45.2~$\mu$m, where for 
many of the sources considered here -- i.e. massive YSOs, HII regions, Orion, Galactic 
Centre -- the flux is hundreds or even many thousands of Jansky. The band 4 detectors
take a finite time to respond to such a sudden signal increase, and so there can be
a notable 'droop' in the shape of the spectrum from 40--45~$\mu$m. This can be
seen in Figure~\ref{GL2591_UPDO_UNCO}, which shows the uncorrected up and down 
scans for the AFGL~2591 SWS01 (speed 3) observation of revolution 357. Conversely, 
when the down scan begins the detectors 'remember' the illumination history and again 
take time to respond to the new signal level. The resultant spectral shape is not 
as predictable as the 'droop' of the up scan, but could potentially mimic a spectral 
feature. Given the possible existence of a 28~$\mu$m crystalline silicate band it 
is important to account for these effects.

We thus assume that the up (down) scan most accurately represents the true spectral 
shape in the 28--33~$\mu$m (40--45~$\mu$m) interval. These wavelength intervals are
selected based on experience, depending somewhat on the speed of the SWS01. From 
Figure~\ref{GL2591_UPDO_UNCO} it can be seen that between 33 and 40~$\mu$m the 
up/down scans overlay quite well for this speed 3 SWS01. This is also the case for 
most speed 2's, whilst for speed 4 (slowest) the two scan directions overlay quite 
well from about 30--43~$\mu$m.

\begin{figure*}
\includegraphics[scale=0.60]{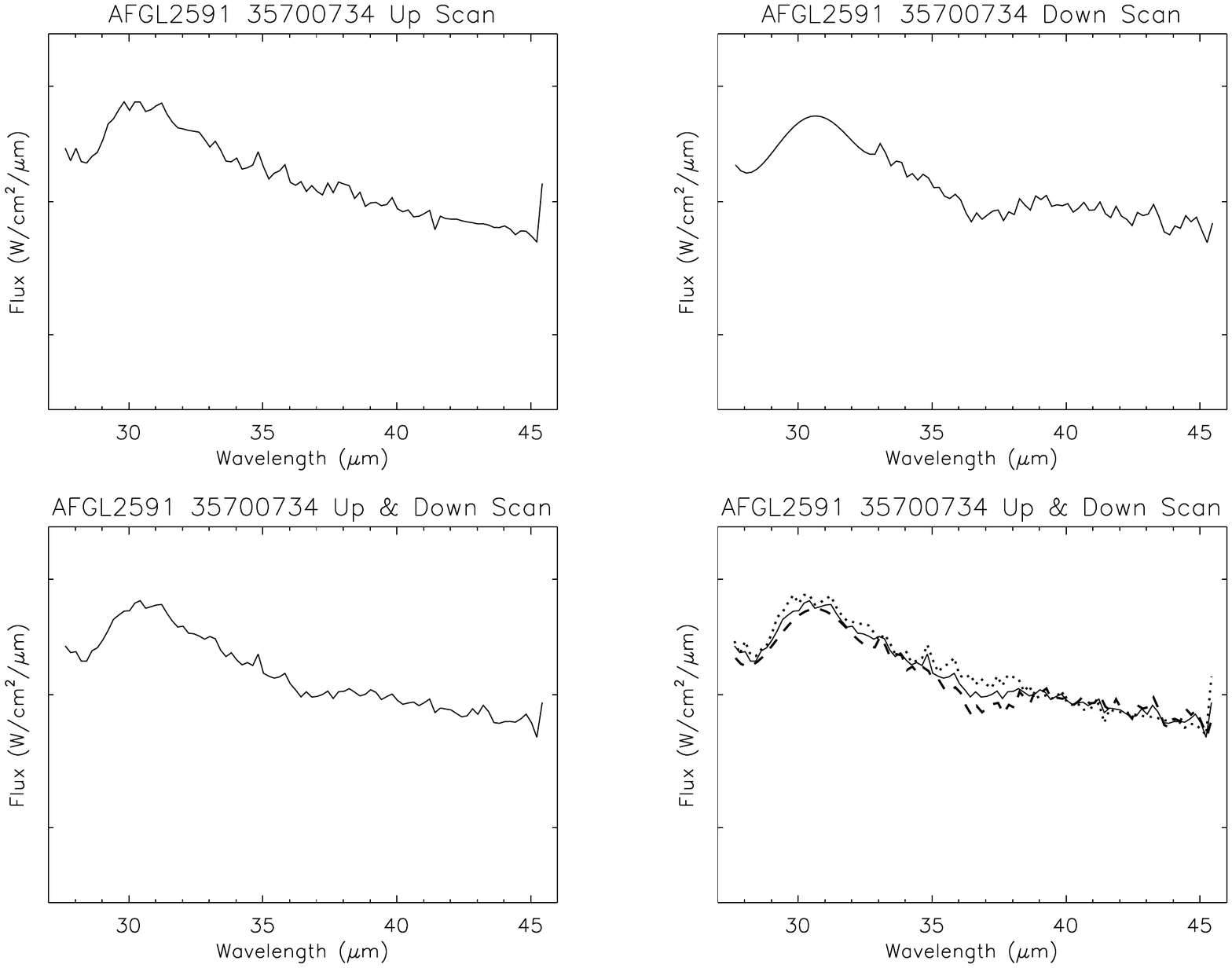}
\caption{Sigma-clipped and binned corrected up and down scans for the band 4 ISO observation of AFGL2591 (tdt=35700734).}
\label{GL2591_UPDO_CORR}
\end{figure*}

With this assumption we then fit a low-order polynomial to the two spectral regions,
28--33~$\mu$m of the up scan and 40--45~$\mu$m of the down scan, and force the other
scan to take on that shape. The correction is done at the detector level, i.e. before
sigma clipping and averaging, although the same correction is applied to all 12 detectors. 
Figure~\ref{GL2591_UPDO_CORR} shows the result of this procedure for the same data set 
as in Figure~\ref{GL2591_UPDO_UNCO}. The up and down scans now overlay well throughout 
the entire band 4 spectral range.

\bsp	
\label{lastpage}
\end{document}